\documentclass[twocolumn,english,superscriptaddress,citeautoscript,showpacs,preprintnumbers,amsmath,amssymb,prd,floatfix,nofootinbib]{revtex4-1}
\pdfoutput=1
\usepackage{amsmath, amsthm, amssymb, graphicx,bm,amstext,subfigure, mathdots}
\usepackage{hyperref}
\usepackage{color}

\begin{document}
\title{How the huge energy of quantum vacuum gravitates to drive the slow accelerating expansion of the Universe}
\author{Qingdi Wang}
\author{Zhen Zhu}
\author{William G. Unruh}
\affiliation{Department of Physics and Astronomy, 
The University of British Columbia,
Vancouver, Canada V6T 1Z1}
\begin{abstract}
We investigate the gravitational property of the quantum vacuum by treating its large energy density predicted by quantum field theory seriously and assuming that it does gravitate to obey the equivalence principle of general relativity. We find that the quantum vacuum would gravitate differently from what people previously thought. The consequence of this difference is an accelerating universe with a small Hubble expansion rate $H\propto \Lambda e^{-\beta\sqrt{G}\Lambda}\to 0$ instead of the previous prediction $H=\sqrt{8\pi G\rho^{vac}/3}\propto\sqrt{G}\Lambda^2\to\infty$ which was unbounded, as the high energy cutoff $\Lambda$ is taken to infinity. In this sense, at least the ``old'' cosmological constant problem would be resolved. Moreover, it gives the observed slow rate of the accelerating expansion as $\Lambda$ is taken to be some large value of the order of Planck energy or higher. This result suggests that there is no necessity to introduce the cosmological constant, which is required to be fine tuned to an accuracy of $10^{-120}$, or other forms of dark energy, which are required to have peculiar negative pressure, to explain the observed accelerating expansion of the Universe.
\end{abstract}
\maketitle

\section{Introduction}
The two pillars that much of modern physics is based on are Quantum Mechanics (QM) and General Relativity (GR). QM is the most successful scientific theory in history, which has never been found to fail in repetitive experiments. GR is also a successful theory which has so far managed to survive every test \cite{lrr-2006-3}. In particular, the last major prediction of GR--the gravitational waves, has finally been directly detected on Sept 2015 \cite{Abbott:2016blz}. However, these two theories seem to be incompatible at a fundamental level (see e.g. \cite{Wald:1984rg}). The unification of both theories is a big challenge to modern theoretical physicists. 

While the test of the combination of QM and GR is still difficult in lab, our Universe already provides one of the biggest confrontations between both theories: the Cosmological Constant Problem \cite{RevModPhys.61.1}. Quantum field theory (QFT) predicts a huge vacuum energy density from various sources. Meanwhile, the equivalence principle of GR requires that every form of energy gravitates in the same way. When combining these concepts together, it is widely supposed that the vacuum energy gravitates as a cosmological constant. However, the observed effective cosmological constant is so small compared with the QFT's prediction that an unknown bare cosmological constant \eqref{contributions to effective cc} has to cancel this huge contribution from the vacuum to better than at least 50 to 120 decimal places! It  is an extremely difficult fine-tune problem that gets even worse when the higher loop corrections are included \cite{Padilla:2015aaa}.

In 1998, the discovery of the accelerating expansion of the Universe  \cite{1538-3881-116-3-1009, 0004-637X-517-2-565} has further strengthened the importance of this problem. Before this, one only needs to worry about the ``old'' cosmological constant problem of explaining why the effective cosmological constant is not large. Now,  one also has to face the challenge of the ``new'' cosmological constant problem of explaining why it has such a specific small value from the observation, which is the same order of magnitude as the present mass density of the Universe (coincidence problem).

This problem is widely regarded as one of the major obstacles to further progress in fundamental physics (for example, see Witten 2001 \cite{witten}). Its importance has been emphasized by various authors from different aspects. For example, it has been described as a ``veritable crisis'' (Weinberg 1989, \cite{RevModPhys.61.1} p.1), an ``unexplained puzzle'' (Kolb and Turner 1993 , \cite{Kolb:EU90} p.198), ``the most striking problem in contemporary physics'' (Dolgov 1997 \cite{Dolgov:1997za} p.1) and even ``the mother of all physics problems'' , ``the worst prediction ever''(Susskind 2015 \cite{citeulike:430764} chapter two). While it might be possible that people working on a particular problem tend to emphasize or even exaggerate its importance, those authors all agree that this is a problem that needs to be solved, although there is little agreement on what is the right direction to find the solution \cite{pittphilsci398}.

In this paper, we make a proposal for addressing the cosmological constant problem. We treat the divergent vacuum energy density predicted by QFT seriously without trying renormalization and assume that it does gravitate to obey the equivalence principle of GR. We notice that the magnitude of the vacuum fluctuation itself also fluctuates, which leads to a constantly fluctuating and extremely inhomogeneous vacuum energy density. As a result, the quantum vacuum gravitates differently from a cosmological constant. Instead, at each spatial point, the spacetime sourced by the vacuum oscillates alternatively between expansion and contraction, and the phases of the oscillations at neighboring points are different. In this manner of vacuum gravitation, although the gravitational effect produced by the vacuum energy is still huge at sufficiently small scales (Planck scale), its effect at macroscopic scales is largely canceled. Moreover, due to the weak parametric resonance of those oscillations, the expansion outweighs contraction a little bit during each oscillation. This effect accumulates at sufficiently large scales (cosmological scale), resulting in an observable effect---the slow accelerating expansion of the Universe.

Our proposal harkens back to Wheeler's spacetime foam \cite{Wheeler:1957mu, misner1973gravitation} and suggests that it is this foamy structure which leads to the cosmological constant we see today.

The paper is organized as follows: in section \ref{section ii}, we first review several key steps in formulating the cosmological constant problem; in section \ref{sec iii}, we point out that the vacuum energy density is not a constant but is constantly fluctuating and extremely inhomogeneous; in section \ref{sec iv}, we investigate the differences made by the extreme inhomogeneity of the quantum vacuum by introducing a simple model; in section \ref{section v}, we give the solutions to this model by solving the Einstein field equations and show how metric fluctuations leads to the slow accelerating expansion of the Universe; in section \ref{sec vi}, we explain the meaning of our results; in section \ref{back reaction}, we investigate the back reaction effect of the resulting spacetime on the matter fields propagating on it; in section \ref{sec viii}, we generalize our results to more general metrics; in section \ref{sec ix}, we discuss some questions raised and a couple of new concepts suggested by the different way of vacuum gravitation.

The units and metric signature are set to be $c=\hbar=1$ and $(-, +, +,+)$ throughout except otherwise specified.

\section{The formulation of the cosmological constant problem}\label{section ii}
The cosmological constant problem arises when trying to combine GR and QFT to investigate the gravitational property of the vacuum:
\begin{equation}\label{Einstein equation with cosmological constant}
G_{\mu\nu}+\lambda_bg_{\mu\nu}=8\pi G T_{\mu\nu}^{\mathrm{vac}},
\end{equation}
where $G_{\mu\nu}\equiv R_{\mu\nu}-\frac{1}{2} Rg_{\mu\nu}$ is the Einstein tensor and the parameter $\lambda_b$ is the bare cosmological constant.

One crucial step in formulating the cosmological constant problem is assuming that the vacuum energy density is equivalent to a cosmological constant. First, it is argued that the vacuum is Lorentz invariant and thus every observer would see the same vacuum. In Minkowski spacetime, $\eta_{\mu\nu}$ is the only Lorentz invariant $(0,2)$ tensor up to a constant. Thus the vacuum stress-energy tensor must be proportional to $\eta_{\mu\nu}$ (see, e.g.  \cite{carroll2004spacetime}, \cite{pittphilsci398}) 
\begin{equation}\label{flat vacuum equation of state}
T_{\mu\nu}^{\mathrm{vac}}(t, \mathbf{x})=-\rho^{\mathrm{vac}} \eta_{\mu\nu}.
\end{equation}
This relation is then straightforwardly generalized to curved spacetime:
\begin{equation}\label{vacuum equation of state}
T_{\mu\nu}^{\mathrm{vac}}(t, \mathbf{x})=-\rho^{\mathrm{vac}} g_{\mu\nu}(t, \mathbf{x}).
\end{equation}
If $T_{\mu\nu}^{\mathrm{vac}}$ does take the above form \eqref{vacuum equation of state}, the vacuum energy density $\rho^{\mathrm{vac}}$ has to be a constant, which is the requirement of the conservation of the stress-energy tensor
\begin{equation}\label{constancey of vacuum energy density}
\nabla^{\mu}T_{\mu\nu}^{\mathrm{vac}}=0.
\end{equation}

The effect of a stress-energy tensor of the form \eqref{vacuum equation of state} is equivalent to that of a cosmological constant, as can be seen by moving the term $8\pi G T_{\mu\nu}^{\mathrm{vac}}$ in \eqref{Einstein equation with cosmological constant} to the left-hand side
\begin{equation}\label{Einstein equation for effective cc}
G_{\mu\nu}+\lambda_{\mathrm{eff}}g_{\mu\nu}=0,
\end{equation}
where,
\begin{equation}\label{contributions to effective cc}
\lambda_{\mathrm{eff}}=\lambda_b+8\pi G\rho^{\mathrm{vac}};
\end{equation}
Or equivalently by moving the term $\lambda_bg_{\mu\nu}$ in \eqref{Einstein equation with cosmological constant} to the right-hand side
\begin{equation}\label{Einstein equatin for effective vac}
G_{\mu\nu}=-8\pi G \rho_{\mathrm{eff}}^{\mathrm{vac}}g_{\mu\nu},
\end{equation}
where,
\begin{equation}
\rho_{\mathrm{eff}}^{\mathrm{vac}}=\rho^{\mathrm{vac}}+\frac{\lambda_b}{8\pi G}.
\end{equation}
So anything that contributes to the energy density of the vacuum acts like a cosmological constant and thus contributes to the effective cosmological constant. Or equivalently we can say that the bare cosmological constant acts like a source of vacuum energy and thus contributes to the total effective vacuum energy density. This equivalence is the origin of the identification of the cosmological constant with the vacuum energy density. 

Following the above formulations, the effective cosmological constant $\lambda_{\mathrm{eff}}$ or the total effective vacuum energy density $\rho_{\mathrm{eff}}^{\mathrm{vac}}$ are the quantities that can be constrained and measured through experiments. While solar system and galactic observations have placed a small upper bound on $\lambda_{\mathrm{eff}}$, large scale cosmological observations provide the most accurate measurement. It is interpreted as a form of ``dark energy'', which drives the observed accelerating expansion of the Universe \cite{1538-3881-116-3-1009, 0004-637X-517-2-565}.

Based on the assumption of homogeneity and isotropy of the Universe, the metric has the cosmology's standard FLRW form, which is, for the spatially flat case,
\begin{equation}\label{FRW metric}
ds^2=-dt^2+a^2(t)\left(dx^2+dy^2+dz^2\right).
\end{equation}
Then by applying the equations \eqref{Einstein equation for effective cc} or \eqref{Einstein equatin for effective vac} for the above special metric \eqref{FRW metric}, one obtains the contributions to the Hubble expansion rate $H=\dot{a}/a$ and the acceleration of the scale factor $\ddot{a}$ from $\lambda_{\mathrm{eff}}$ and/or $\rho_{\mathrm{eff}}^{\mathrm{vac}}$ are
\begin{eqnarray}
&3&H^2=\lambda_{\mathrm{eff}}=8\pi G \rho_{\mathrm{eff}}^{\mathrm{vac}},\label{vac 1}\\
&\ddot{a}&=\frac{\lambda_{\mathrm{eff}}}{3}a_{\mathrm{vac}}=\frac{8\pi G\rho_{\mathrm{eff}}^{\mathrm{vac}}}{3}a.\label{vac 2}
\end{eqnarray}
The solution to the dynamic equation \eqref{vac 2} is
\begin{equation}\label{cmc prediction}
a(t)=a(0)e^{H t},
\end{equation}
where $H$ is determined by the initial value constraint equation \eqref{vac 1}.

According to the Lambda-CDM model of the big bang cosmology, the effective cosmological constant is responsible for the accelerating expansion of the Universe as shown in \eqref{vac 2} and contributes about $69\%$ to the current Hubble expansion rate \cite{Ade:2015xua}:
\begin{equation}
\lambda_{\mathrm{eff}}=3\Omega_{\lambda} H_0^2\approx 4.32\times10^{-84}(\mathrm{GeV})^2,
\end{equation}
or
\begin{equation}\label{observed vacuum energy}
\rho_{\mathrm{eff}}^{\mathrm{vac}}=\Omega_{\lambda}\rho_{\mathrm{crit}}\approx 2.57\times 10^{-47} (\mathrm{GeV})^4,
\end{equation}
where $\Omega_{\lambda}=0.69$ is the dark energy density parameter, $H_0$ is the current observed Hubble constant and $\rho_{\mathrm{crit}}=\frac{3H_0^2}{8\pi G}$ is the critical density.

Unfortunately the predicted energy density of the vacuum from QFT is much larger than this. It receives contributions from various sources, including the zero point energies ($\sim10^{72} (\mathrm{GeV})^4$) of all fundamental quantum fields due to vacuum fluctuations, the phase transitions due to the spontaneous symmetry breaking of electroweak theory ($\sim 10^9 (\mathrm{GeV})^4$) and any other known and unknown phase transitions in the early Universe (e.g. from chiral symmetry breaking in QCD ($\sim 10^{-2} (\mathrm{GeV})^4$), grand unification ($\sim 10^{64} (\mathrm{GeV})^4$) etc)\cite{pittphilsci398, Carroll:2000fy}. Each contribution is larger than the observed value \eqref{observed vacuum energy} by $50$ to $120$ orders of magnitude. There is no mechanism in the standard model which suggests any relations between the individual contributions, so it is customary to assume that the total vacuum energy density is at least as large as any of the individual contributions \cite{pittphilsci398}. One thus has to fine tune the unknown bare cosmological constant $\lambda_b$ to a precision of at least 50 decimal places to cancel the excess vacuum energy density.

\section{The fluctuating quantum vacuum energy density}\label{sec iii}
The vacuum energy density is treated as a constant in the usual formulation of the cosmological constant problem. While this is true for the expectation value, it is not true for the actual energy density. 

That's because the vacuum is not an eigenstate of the local energy density operator $T_{00}$, although it is an eigenstate of the global Hamiltonian operator $\operatorname{H}=\int d^3x\, T_{00}$. This implies that the total vacuum energy all over the space is constant but its density fluctuates at individual points. 

To see this more clearly, consider a quantized real massless scalar field $\phi$ in Minkowski spacetime as an example:
\begin{equation}\label{field expansion}
\phi(t,\mathbf{x})=\int\frac{d^3k}{(2\pi)^{3/2}}\frac{1}{\sqrt{2\omega}}\left(a_{\mathbf{k}}e^{-i(\omega t-\mathbf{k}\cdot\mathbf{x})}+a_{\mathbf{k}}^{\dag}e^{+i(\omega t-\mathbf{k}\cdot\mathbf{x})}\right), 
\end{equation}
where the temporal frequency $\omega$ and the spatial frequency $\mathbf{k}$ in \eqref{field expansion} are related to each other by $\omega=|\mathbf{k}|$.

The vacuum state $|0\rangle$, which is defined as
\begin{equation}
a_{\mathbf{k}}|0\rangle=0, \quad\mbox{for all}\,\, \mathbf{k},
\end{equation}
is an eigenstate of the Hamiltonian operator
\begin{equation}
\operatorname{H}=\int d^3x\, T_{00}=\frac{1}{2}\int d^3k\,\omega\left(a_{\mathbf{k}}a_{\mathbf{k}}^{\dag}+a_{\mathbf{k}}^{\dag}a_{\mathbf{k}}\right),
\end{equation}
where $T_{00}$ is defined as
\begin{equation}\label{energy density definition}
T_{00}=\frac{1}{2}\left(\dot{\phi}^2+(\nabla\phi)^2\right).
\end{equation}

But, $|0\rangle$ is not an eigenstate of the energy density operator
\begin{widetext}
\begin{eqnarray}
&&T_{00}(t,\mathbf{x})=\frac{1}{2}\int\frac{d^3kd^3k'}{(2\pi)^3}\frac{1}{2}\left(\sqrt{|\mathbf{k}||\mathbf{k}'|}+\frac{\mathbf{k}\cdot\mathbf{k}'}{\sqrt{|\mathbf{k}||\mathbf{k}'|}}\right)\Bigg(a_{\mathbf{k}}a_{\mathbf{k}'}^{\dag}e^{-i\left[\left(|\mathbf{k}|-|\mathbf{k}'|\right)t-\left(\mathbf{k}-\mathbf{k}'\right)\cdot\mathbf{x}\right]}\\
&&+a_{\mathbf{k}}^{\dag}a_{\mathbf{k}'}e^{+i\left[\left(|\mathbf{k}|-|\mathbf{k}'|\right)t-\left(\mathbf{k}-\mathbf{k}'\right)\cdot\mathbf{x}\right]}\nonumber-a_{\mathbf{k}}a_{\mathbf{k}'}e^{-i\left[\left(|\mathbf{k}|+|\mathbf{k}'|\right)t-\left(\mathbf{k}+\mathbf{k}'\right)\cdot\mathbf{x}\right]}-a_{\mathbf{k}}^{\dag}a_{\mathbf{k}'}^{\dag}e^{+i\left[\left(|\mathbf{k}|+|\mathbf{k}'|\right)t-\left(\mathbf{k}+\mathbf{k}'\right)\cdot\mathbf{x}\right]}\Bigg),\nonumber
\end{eqnarray}
\end{widetext}
because of the terms of the form $a_{\mathbf{k}}a_{\mathbf{k}'}$ and $a_{\mathbf{k}}^{\dag}a_{\mathbf{k}'}^{\dag}$.

Direct calculation shows the magnitude of the fluctuation of the vacuum energy density diverges as the same order as the energy density itself,
\begin{equation}\label{variance}
\left\langle\big(T_{00}-\left\langle T_{00}\right\rangle\big)^2\right\rangle=\frac{2}{3}\langle T_{00} \rangle^2,
\end{equation}
where 
\begin{equation}
\langle T_{00} \rangle=\frac{\Lambda^4}{16\pi^2},
\end{equation}
where $\Lambda$ is the effective QFT's high energy cutoff. (For more details on this calculation, see equation \eqref{magnitude of mean squared energy density} in Appendix \ref{appendix 1}.) Thus the energy density fluctuates as violently as its own magnitude. With such huge fluctuations, the vacuum energy density $\rho^{\mathrm{vac}}$ is not a constant in space or time.

\begin{figure}
\includegraphics[scale=0.65]{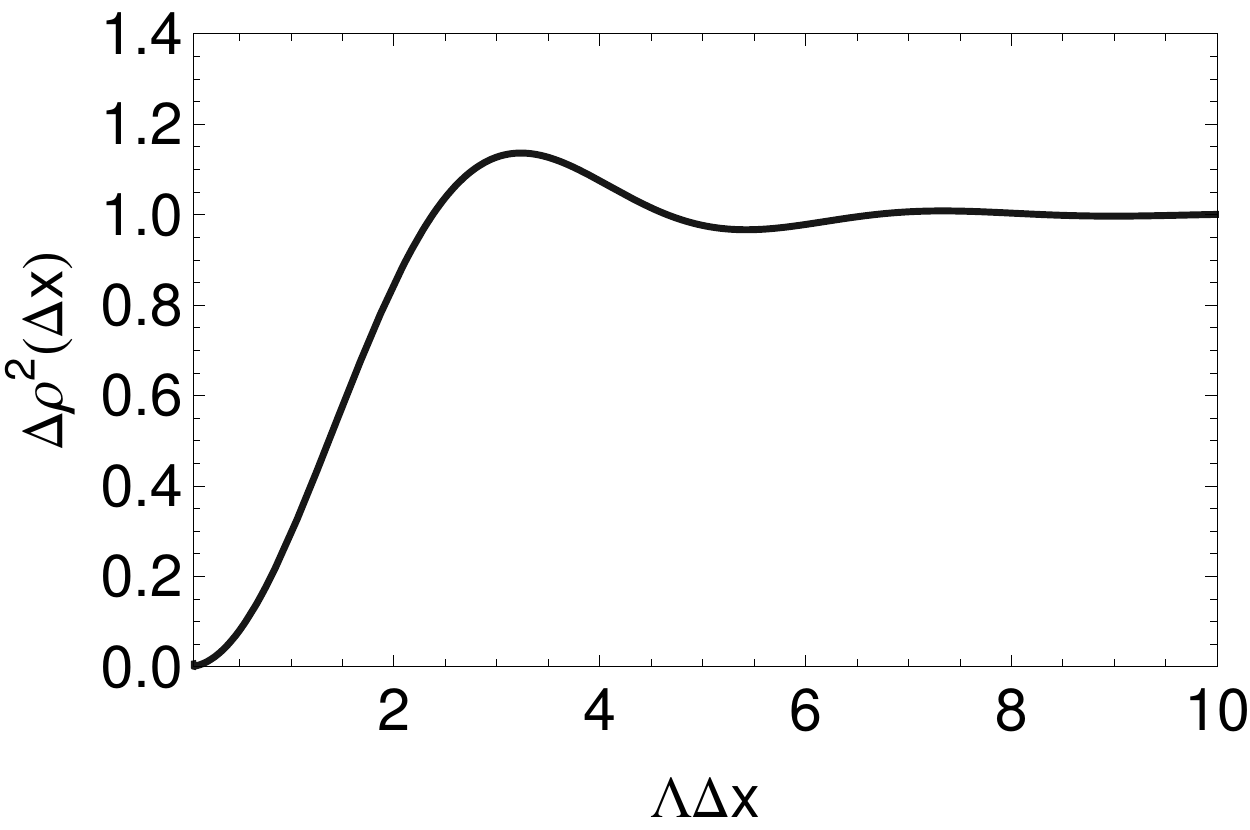}
\caption{\label{density difference}Plot of the expectation value of the square of the energy density difference as a function of spacial separation $\Lambda\Delta x$.}
\end{figure}

Furthermore,  the energy density of the vacuum is not only not a constant in time at a fixed spatial point, it also varies from place to place. In other words, the energy density of vacuum is varying wildly at every spatial point and the variation is not in phase for different spatial points. This results in an extremely inhomogeneous vacuum. The extreme inhomogeneity can be illustrated by directly calculating the expectation value of the square of difference between energy density at different spatial points,
\begin{equation}\label{spatial inhomogeneous rho}
\Delta\rho^2\left(\Delta x\right)=\frac{\left\langle\left\{\big(T_{00}\left(t,\mathbf{x}\right)-T_{00}\left(t,\mathbf{x}'\right)\big)^2\right\}\right\rangle}{\frac{4}{3}\left\langle T_{00}(t,\mathbf{x})\right\rangle ^2},
\end{equation}
where $\Delta x=|\mathbf{x}-\mathbf{x}'|$ and we have normalized $\Delta\rho^2$ by dividing its asymptotic value $\frac{4}{3}\langle T_{00}\rangle ^2$ (the curly bracket $\{\}$ is the symmetrization operator which is defined by \eqref{symmetrization def}). The behavior of $\Delta\rho^2$ for the scalar field \eqref{field expansion} in Minkowski vacuum is plotted in FIG. \ref{density difference}, which shows that the magnitude of the energy density difference between two spacial points quickly goes up to the order of $\langle T_{00}\rangle$ itself as their distance increases by only the order of $1/\Lambda$. (For more details on the calculations and how the energy density fluctuates all over the spacetime, see Appendix \ref{appendix 1}.)

As the vacuum is clearly not homogeneous, equation \eqref{vac 1} is not valid as it depends on a homogeneous and isotropic matter field and metric. Therefore a new method of relating vacuum energy density to the observed Hubble expansion rate is required.

\section{Differences made by the inhomogeneous vacuum---a simple model}\label{sec iv}
The extreme inhomogeneity of the vacuum means its gravitational effect cannot be treated perturbatively, so another method is required. As solutions to the fully general Einstein equations are difficult to obtain, we will first look at a highly simplified model.

\subsection{Beyond the FLRW metric}\label{Beyond the FLRW metric}
To describe the gravitational property of the inhomogeneous quantum vacuum, we must allow inhomogeneity in the metric. This is accomplished by allowing the scale factor $a(t)$ in the FLRW metric \eqref{FRW metric} to have spatial dependence,
\begin{equation}\label{inhomogenous FLRW coordinate}
ds^2=-dt^2+a^2(t, \mathbf{x})(dx^2+dy^2+dz^2).
\end{equation}

The full Einstein field equations for the coordinate \eqref{inhomogenous FLRW coordinate} are
\begin{widetext}
\begin{eqnarray}
G_{00}&=&3\left(\frac{\dot{a}}{a}\right)^2+\frac{1}{a^2}\left(\frac{\nabla a}{a}\right)^2-\frac{2}{a^2}\left(\frac{\nabla^2 a}{a}\right)=8\pi G T_{00},\label{Einstein equations 00}\\
G_{ii}&=&-2a\ddot{a}-\dot{a}^2-\left(\frac{\nabla a}{a}\right)^2+\frac{\nabla^2 a}{a}+2\left(\frac{\partial_i a}{a}\right)^2-\frac{\partial_i^2a}{a}=8\pi G T_{ii},\label{Einstein equations ii}\\
G_{0i}&=&2\frac{\dot{a}}{a}\frac{\partial_i a}{a}-2\frac{\partial_i \dot{a}}{a}=8\pi G T_{0i},\label{Einstein equations 0i}\\
G_{ij}&=&2\frac{\partial_i a}{a}\frac{\partial_j a}{a}-\frac{\partial_i\partial_j a}{a}=8\pi G T_{ij},\label{Einstein equations ij}\quad i, j=1,2,3,\quad i\neq j,
\end{eqnarray}
\end{widetext}
where $\nabla=(\partial_1, \partial_2, \partial_3)$ is the ordinary gradient operator with respect to the spatial coordinates $x, y, z$.

By choosing the above simplest inhomogeneous metric \eqref{inhomogenous FLRW coordinate}, we are assuming a mini-superspace type model, and will choose which of these equations do apply later. This treatment might result in inconsistencies as general vacuum fluctuations of the matter fields posses rich structures that they may not produce spacetime described by the metric \eqref{inhomogenous FLRW coordinate}. To fully describe the resulting inhomogeneous spacetime, one needs a more general metric. However, as a first approximation, using \eqref{inhomogenous FLRW coordinate} is relatively easy to calculate and leads to interesting results. We are also going to do the calculations for a more general metric in section \ref{general coordinate}.

\subsection{The fluctuating spacetime}\label{fluctuating spacetime}
The role played by the value of vacuum energy density in the above equations \eqref{Einstein equations 00}, \eqref{Einstein equations ii}, \eqref{Einstein equations 0i} and \eqref{Einstein equations ij} is different from \eqref{vac 1}. The value of vacuum energy density is no longer directly related to the Hubble rate $H$ through the equation \eqref{vac 1}. This is evident from the $00$ component of the Einstein equation \eqref{Einstein equations 00}. The equation \eqref{vac 1} is only the special case of \eqref{Einstein equations 00} when the spatial derivatives $\nabla a$ and $\nabla^2 a$ are zero, which requires that the matter distribution is strictly homogeneous and isotropic. However, as shown in the last section, the quantum vacuum is extremely inhomogeneous and necessarily anisotropic, which requires $\nabla a$ and $\nabla^2 a$ be huge. This can be seen through the $ij$ component of the Einstein equation \eqref{Einstein equations ij}. In fact, due to symmetry properties of the quantum vacuum, we have the expectation value of shear stress $T_{ij}$ on the right side of \eqref{Einstein equations ij}
\begin{equation}\label{shear stress average}
\left\langle T_{ij}\right\rangle=0, \quad i, j=1,2,3,\quad i\neq j.
\end{equation}

Meanwhile, $T_{ij}$ must fluctuate since the quantum vacuum is not its eigenstate either,  and the magnitude of the fluctuation is on the same order of the vacuum energy density
\begin{equation}\label{shear stress square}
\left\langle T_{ij}^2\right\rangle\sim\left\langle T_{00}\right\rangle^2.
\end{equation}
This means that the $T_{ij}$ is constantly fluctuating around zero with a huge magnitude of the order of vacuum energy density. As a result, in \eqref{Einstein equations ij}, the spatial derivatives of $a(t, \mathbf{x})$ must also constantly fluctuate with huge magnitudes.

More importantly, since the scale factor $a(t, \mathbf{x})$ is spatially dependent, the physical distance $L$ between two spatial points with comoving coordinates $\mathbf{x}_1$ and $\mathbf{x}_2$ is no longer related to their comoving distance $\Delta x=|\mathbf{x}_1-\mathbf{x}_2|$ by the simple equation $L(t)=a(t)\Delta x$ and the observed global Hubble rate $H$ is no longer equal to the local Hubble rate $\dot{a}/a$. Instead, the physical distance and the global Hubble rate are defined as
\begin{equation}\label{length definition}
L(t)=\int_{\mathbf{x}_1}^{\mathbf{x}_2}\sqrt{a^2(t, \mathbf{x})}dl
\end{equation}
and
\begin{equation}\label{hubble expansion definition}
H(t)=\frac{\dot{L}}{L}=\frac{\int_{\mathbf{x}_1}^{\mathbf{x}_2}\frac{\dot{a}}{a}(t, \mathbf{x})\sqrt{a^2(t, \mathbf{x})}dl}{\int_{\mathbf{x}_1}^{\mathbf{x}_2}\sqrt{a^2(t, \mathbf{x})}dl},
\end{equation}
where the line element $dl=\sqrt{dx^2+dy^2+dz^2}$.

Equation \eqref{hubble expansion definition} shows the key difference between the gravitational behavior of quantum vacuum predicted by the homogeneous FLRW metric \eqref{FRW metric} and by the inhomogeneous metric \eqref{inhomogenous FLRW coordinate}.

For the homogeneous metric \eqref{FRW metric}, the scale factor $a$ is spatially independent and \eqref{hubble expansion definition} just reduces to
\begin{equation}
H(t)=\frac{\dot{a}}{a}(t).
\end{equation}
In this case, there are only two distinct choices for Hubble rates on a spatial slice $t=Const$ under the initial value constraint equation \eqref{vac 1} 
\begin{equation}
\frac{\dot{a}}{a}=\pm\sqrt{\frac{8\pi G \rho^{\mathrm{vac}}}{3}},
\end{equation}
which implies that all points in space have to be simultaneously expanding or contracting at the same constant rate (Here we do not include the cosmological constant $\lambda$).

But for the inhomogeneous metric \eqref{inhomogenous FLRW coordinate}, the scale factor $a$ is spatially dependent and there is much more freedom in choosing different local Hubble rates at different spatial points of the slice $t=Const$ under the corresponding initial value constraint equation \eqref{Einstein equations 00}. 

In fact, the local Hubble rates must be constantly changing over spatial directions within very small length scales. This can be seen from the initial value constraint equations \eqref{Einstein equations 0i}, which can be rewritten as
\begin{equation}\label{energy flux constraint}
\nabla\left(\frac{\dot{a}}{a}\right)=-4\pi G\mathbf{J},
\end{equation}
where $\mathbf{J}=\left(T_{01}, T_{02}, T_{03}\right)$ is vacuum energy flux\footnote{One might notice that \eqref{energy flux constraint} requires $\nabla\times\mathbf{J}=0$, which means that to produce the metric of the form \eqref{inhomogenous FLRW coordinate}, the energy flux of the matter field needs to be curl free. As mentioned in the last paragraph of section \ref{Beyond the FLRW metric}, this is not true for general matter fields, but here as a first approximation we will use \eqref{energy flux constraint} to estimate the magnitude of change in $\dot{a}/a$.}.

The solution to \eqref{energy flux constraint} or \eqref{Einstein equations 0i} is
\begin{equation}\label{hubble rate spatial difference}
\frac{\dot{a}}{a}(t, \mathbf{x})=\frac{\dot{a}}{a}(t, \mathbf{x}_0)-4\pi G\int_{\mathbf{x}_0}^{\mathbf{x}}\mathbf{J}\left(t, \mathbf{x}'\right)\cdot\mathbf{dl}',
\end{equation}
where $\mathbf{dl}'=(dx', dy', dz')$ and $\mathbf{x}_0$ is an arbitrary spatial point. The above solution \eqref{hubble rate spatial difference} shows that the difference in the local Hubble rates $\dot{a}/a$ between $\mathbf{x}_0$ and $\mathbf{x}_1$ is determined by the spatial accumulations (integral) of the vacuum energy flux $\mathbf{J}$. Similar to the shear stress, $\mathbf{J}$ has zero expectation value 
\begin{equation}\label{zero expectation value of energy flux}
\left\langle\mathbf{J}\right\rangle=\mathbf{0}
\end{equation}
but huge fluctuations
\begin{equation}
J=\sqrt{\left\langle\mathbf{J}^2\right\rangle}\sim\left\langle T_{00}\right\rangle\sim\Lambda^4\to+\infty,
\end{equation}
which implies that the local Hubble rates differ from point to point due to the fluctuations. The average of the absolute value of $\dot{a}/a$ can be estimated with the constraint equation \eqref{Einstein equations 00}
\begin{equation}
\sqrt{\left\langle\left(\frac{\dot{a}}{a}\right)^2\right\rangle}\sim\sqrt{G\left\langle T_{00}\right\rangle}\sim\sqrt{G}\Lambda^2.
\end{equation}
By using \eqref{hubble rate spatial difference}, we find that the difference in local Hubble rates becomes comparable with itself for points separated by only a distance of the order $\Delta x\sim \frac{1}{\sqrt{G}\Lambda^2}$ as $\Lambda\to+\infty$:
\begin{equation}\label{estimiation of hubble rate spatial difference}
\Delta\left(\frac{\dot{a}}{a}\right)\sim 4\pi G J\Delta x\sim\sqrt{G}\Lambda^2\sim\sqrt{\left\langle\left(\frac{\dot{a}}{a}\right)^2\right\rangle}.
\end{equation}

Up to this point, we have used the equations \eqref{Einstein equations 00}, \eqref{Einstein equations 0i} and \eqref{Einstein equations ij}. These equations are all initial value constraint equations which do not contain the scale factor's second order time derivative $\ddot{a}$. To get the information about the time evolution of $a(t, \mathbf{x})$, we also need to use \eqref{Einstein equations ii}. A linear combination of equations \eqref{Einstein equations 00} and \eqref{Einstein equations ii} gives,
\begin{equation}\label{00+ii}
G_{00}+\frac{1}{a^2}\left(G_{11}+G_{22}+G_{33}\right)=-\frac{6\ddot{a}}{a},
\end{equation}
where all the spatial derivatives of $a$ cancel and only the second order time derivative left. Therefore we reach the following dynamic evolution equation for $a(t, \mathbf{x})$:
\begin{equation}\label{hmo}
\ddot{a}+\Omega^2(t,\mathbf{x})a=0,
\end{equation}
where 
\begin{equation}\label{Omega definition}
\Omega^2=\frac{4\pi G}{3}\left(\rho+\displaystyle\sum_{i=1}^{3}P_i\right), \quad\rho=T_{00}, P_i=\frac{1}{a^2}T_{ii}. 
\end{equation}

If $\Omega^2>0$, which is true if the matter fields satisfy normal energy conditions, \eqref{hmo} describes a harmonic oscillator with time dependent frequency. The most basic behavior of a harmonic oscillator is that it oscillates back and forth around its equilibrium point, which implies that the local Hubble rates $\dot{a}/a$ are periodically changing signs over time. By using equation \eqref{estimiation of hubble rate spatial difference} you can find that $\dot{a}/a$ must also have this periodic sign change in a given spatial direction. 

Physically, these fluctuating features of $\dot{a}/a$ imply that, at any instant of time, if the space is expanding in a small region, it has to be contracting in neighboring regions; and at any spatial point, if the space is expanding now, it has to be contracting later.

These features result in huge cancellations when calculating the averaged $H$ through \eqref{hubble expansion definition}. The observable overall net Hubble rate can be small although the absolute value of the local Hubble rate $|\dot{a}/a|$ at each individual point has to be huge to satisfy the constraint equation \eqref{Einstein equations 00}. In other words, while the instantaneous rates of expansion or contraction at a fixed spatial point can be large, their effects can be canceled in a way that the physical distance \eqref{length definition} would not grow $10^{120}$ times larger than what is observed.

This picture of fluctuating spacetime is not completely new. It is similar to the concept of spacetime foam devised by John Wheeler \cite{Wheeler:1957mu, misner1973gravitation} that in a quantum theory of gravity spacetime would have a foamy, jittery nature and would consist of many small, ever-changing, regions in which spacetime are not definite, but fluctuates. His reason for this ``foamy'' picture is the same as ours---at sufficiently small scales the energy of vacuum fluctuations would be large enough to cause significant departures from the smooth spacetime we see at macroscopic scales.

The solution for $a(t, \mathbf{x})$ will be given by equations \eqref{expected form of solution}, \eqref{wkb} and \eqref{phase} in the next section \ref{section v} to describe this foamy structure more precisely.

\subsection{Methods and assumptions in solving the system}
In principle, we need a full quantum theory of gravity to solve the evolution details of this quantum gravitational system. Unfortunately, no satisfactory theory of quantum gravity exists yet.

In this paper, we are not trying to quantize gravity. Instead, we are still keeping the spacetime metric $a(t, \mathbf{x})$ as classical, but quantizing the fields propagating on it. The key difference from the usual semiclassical gravity is that we go one more step---instead of assuming the semiclassical Einstein equation, where the curvature of the spacetime is sourced by the expectation value of the quantum field stress energy tensor, we also take the huge fluctuations of the stress energy tensor into account. In our method, the sources of gravity are stochastic classical fields whose stochastic properties are determined by their quantum fluctuations.

The evolution details of the scale factor $a(t, \mathbf{x})$ described by equation\eqref{hmo} depends on the property of the time dependent
frequency $\Omega(t, \mathbf{x})$ given by \eqref{Omega definition}. For both simplicity and clarity, in the following sections we investigate the properties of $\Omega$ by considering the contribution from a real massless scalar field $\phi$. In this case, the stress energy tensor for a general spacetime metric $g_{\mu\nu}$ is
\begin{equation}\label{stress energy tensor for a massless scalar field}
T_{\mu\nu}=\nabla_{\mu}\phi\nabla_{\nu}\phi-\frac{1}{2}g_{\mu\nu}\nabla^{\lambda}\phi\nabla_{\lambda}\phi.
\end{equation}
Direct calculation using the inhomogeneous metric \eqref{inhomogenous FLRW coordinate} gives that
\begin{equation}\label{massless field contribution}
\rho+\displaystyle\sum_{i=1}^{3}P_i=2\dot{\phi}^2, 
\end{equation}
where all the spatial derivatives and explicit dependence on the metric $a$ are canceled. Thus we obtain 
\begin{equation}
\Omega^2=\frac{8\pi G \dot{\phi}^2}{3}>0,
\end{equation}
which is not explicitly dependent on the metric $a(t, \mathbf{x})$.

However, the resulting spacetime sourced by this massless scalar field $\phi$ does have back reaction effect on $\phi$ itself. This is because $\phi$ obeys the equation of motion in curved spacetime
\begin{equation}\label{wave eqaution in curved spacetime}
\nabla^{\mu}\nabla_{\mu}\phi=\frac{1}{\sqrt{-g}}\partial_{\mu}\left(\sqrt{-g}g^{\mu\nu}\partial_{\nu}\phi\right)=0,
\end{equation}
which reduces to
\begin{equation}\label{eq:fullwave}
\partial_t\left(a^3\partial_t \phi\right)-\nabla\cdot\left(a\nabla \phi\right)=0
\end{equation}
for the special metric \eqref{inhomogenous FLRW coordinate}.

Incorporating the back reaction effect by solving both the Einstein equations \eqref{Einstein equations 00}, \eqref{Einstein equations ii}, \eqref{Einstein equations 0i}, \eqref{Einstein equations ij} for the metric $a$ and the equation of motion \eqref{eq:fullwave} for the field $\phi$ at the same time is difficult. Fortunately, solving the system in this way is unnecessary. Physically, the quantum vacuum locally behaves as a huge energy reservoir, so that the back reaction effect on it should be small and can be neglected. In our method, we will first assume that the quantized field $\phi$ is still taking the flat spacetime form of \eqref{field expansion} for field modes below the effective QFT's high frequency cutoff $\Lambda$. We use \eqref{field expansion} to calculate the stochastic property of the time dependent frequency $\Omega$ and then solve \eqref{hmo} to get the resulting curved spacetime described by the metric $a(t, \mathbf{x})$. This will be done in the next section \ref{section v}.

We then investigate the back reaction effect in section \ref{back reaction}by quantizing the field $\phi$ in the resulting curved spacetime. It turns out that, while the resulting spacetime is fluctuating, the fluctuation happens at scales which are much smaller than the length scale $1/\Lambda$. Therefore the corrections to the field modes with frequencies below the cutoff $\Lambda$ is quite small and thus the flat spacetime quantization \eqref{field expansion} is valid to high precision. (See equations \eqref{toy model asymptotic mode solution} (or \eqref{general uk}), \eqref{parseval} and \eqref{Delta u} for quantitatively how high this precision is.) In this way we justify neglecting the aforementioned back reaction.

Empirically, this must be true since ordinary QFT has achieved great successes by assuming flat Minkowski background and using the expansion \eqref{field expansion}. So if our method is correct, \eqref{field expansion} has to be still valid even the background spacetime is no longer flat but wildly fluctuating at small scales. In other words, the resulting spacetime should still looks like Minkowskian for low frequency field modes. Long wavelength fields ride over the Wheeler's foam as if it is not there. This is similar to the behavior of very long wavelength water waves which do not notice the rapidly fluctuating atomic soup over which they ride.

\section{The solution for $a(t, \mathbf{x})$}\label{section v}
In this section we give the solution for the local scale factor $a(t, \mathbf{x})$.

\subsection{Parametric resonance}\label{The accelerating expansion from weak parametric resonance}
One important feature of a harmonic oscillator with time dependent frequency is that it may exhibit parametric resonance behavior.

If the $\Omega(t, \mathbf{x})$ is strictly periodic in time with a period $T$, the property of the solutions of \eqref{hmo} has been thoroughly studied by Floquet theory \cite{teschlordinary}. Under certain conditions (for example, the condition \eqref{parametric resonance condition}), the parametric resonance phenomenan occurs and the general solution of \eqref{hmo} is (see e.g. Eq(27.6) in Chapter V of \cite{LANDAU197658})
\begin{equation}\label{general floquet solution}
a(t, \mathbf{x})=c_1 e^{H_{\mathbf{x}}t}P_1(t, \mathbf{x})+c_2 e^{-H_{\mathbf{x}}t}P_2(t, \mathbf{x}),
\end{equation}
where $H_{\mathbf{x}}>0$, $c_1$ and $c_2$ are constants. The $P_1$ and $P_2$ are purely periodic functions of time with period $T$. They are in general functions oscillating around zero. The amplitude of the first term in \eqref{general floquet solution} increases exponentially with time while the second term decreases exponentially. Therefore the first term will become dominant and the solution will approach a pure exponential evolution
\begin{equation}\label{floquet solution}
a(t, \mathbf{x})\simeq e^{H_{\mathbf{x}}t}P(t, \mathbf{x}),
\end{equation}
where we have absorbed the constant $c_1$ into $P(t, \mathbf{x})$ by letting $P(t, \mathbf{x})=c_1P_1(t, \mathbf{x})$.

Physically, the exponential evolution of the amplitude of $a(t, \mathbf{x})$ is easy to understand. If $\Omega$ is strictly periodic, the system will finally reach a steady pattern of evolution (when the second term in \eqref{general floquet solution} has been highly suppressed). In this pattern, after each period of evolution of the system, $a$ increases by a fixed ratio, i.e. $a(t+T, \mathbf{x})=\mu_{\mathbf{x}} a(t, \mathbf{x})$, which results in the exponential increase since after $n$ cycles, $a(t+nT, \mathbf{x})=\mu_{\mathbf{x}}^n a(t, \mathbf{x})$. Here the $\mu_{\mathbf{x}}$ is related to the $H_{\mathbf{x}}$ by $H_{\mathbf{x}}=\frac{\ln\mu_{\mathbf{x}}}{T}$.

Due to the stochastic nature of quantum fluctuations, the $\Omega(t, \mathbf{x})$ in \eqref{hmo} is not strictly periodic. However, its behavior is still similar to a periodic function. In fact, $\Omega$ exhibits quasiperiodic behavior in the sense that it is always varying around its mean value back and forth on an approximately fixed time scale. To see this, we calculate the following normalized covariance:
\begin{eqnarray}\label{cov def}
\chi\left(\Delta t\right)&=&\operatorname{Cov}\left(\Omega^2(t_1,\mathbf{x}),\Omega^2(t_2,\mathbf{x})\right)\\
&=&\frac{\left\langle\left\{\left(\Omega^2(t_1)-\left\langle\Omega^2(t_1)\right\rangle\right)\left(\Omega^2(t_2)-\left\langle\Omega^2(t_2)\right\rangle\right)\right\}\right\rangle}{\left\langle \left(\Omega^2-\left\langle\Omega^2\right\rangle\right)^2\right\rangle}, \nonumber
\end{eqnarray}
where $\Delta t=t_1-t_2$ and we have dropped the label $\mathbf{x}$ in the second line of the above definition \eqref{cov def} since the final result is independent with $\mathbf{x}$.

\begin{figure}
\includegraphics[scale=0.65]{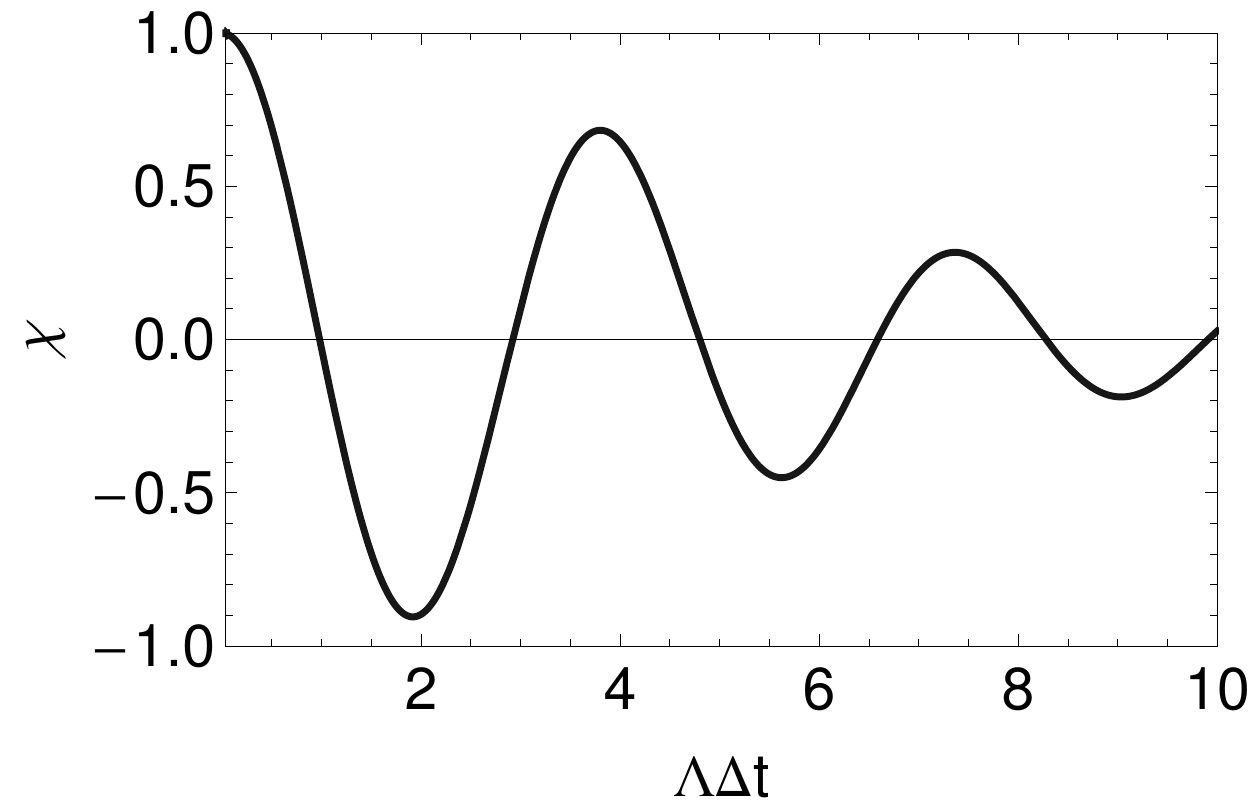}
\caption{\label{omega covariance}Plot of the normalized covariance $\chi$ as a function of temporal separation $\Lambda\Delta t$.}
\end{figure}

Explicit expression for $\chi$ as a function of $\Delta t$ is given by \eqref{phi dot square covariance}, which is plotted in FIG. \ref{omega covariance}. It describes how $\Omega^2$ at different times change around their mean values together. We say that two $\Omega^2$ separated by time difference $\Delta t$ are positively (negatively) correlated if $\chi(\Delta t)>0(<0)$, since it means that they are most likely to be at the same (opposite) side of their mean value $\langle\Omega^2\rangle$. 

FIG. \ref{omega covariance} and \eqref{phi dot square covariance} show that $\Omega^2$ at different times are strongly correlated at close range. Especially, the negative correlation is strongest when $\Delta t\sim 2/\Lambda$, which implies that if at $t=0$ the $\Omega^2$ is above its mean value $\left\langle\Omega^2\right\rangle$, then at $t\sim 2/\Lambda$, it is most likely below $\left\langle\Omega^2\right\rangle$. So basically, $\Omega^2$ varies around its mean value quasiperiodically on the time scale $T\sim 1/\Lambda$.

This quasiperiodic behavior of $\Omega$ should also lead to parametric resonance behavior seen in \eqref{floquet solution}, instead with a difference in that $H_{\mathbf{x}}$ would become time dependent, i.e. the solution would take the following form
\begin{equation}\label{expected form of solution}
a(t, \mathbf{x})\simeq e^{\int_0^t H_{\mathbf{x}}(t')dt'}P(t, \mathbf{x}),
\end{equation}
where $P(t, \mathbf{x})$ here is no longer a strictly periodic function as in \eqref{floquet solution} but a quasiperiodic function with the same quasiperiod of the order $1/\Lambda$ as the time dependent frequency $\Omega(t, \mathbf{x})$. (The solution \eqref{wkb} for $P(t, \mathbf{x})$ in the next subsection \ref{wkb approximation} reveals this property.)

The physical mechanism is similar. The system will also reach a final steady evolution pattern. In this pattern, after each quasiperiod of evolution of the system, $a$ will increase by an approximately fixed ratio.  Suppose that during the $i$th cycle of quasiperiod $T_i$, $a$ increases by a factor $\mu_{i\mathbf{x}}$, i.e. $a(t+T_i, \mathbf{x})=\mu_{i\mathbf{x}}a(t, \mathbf{x})$. Then after the $n$ cycles, we have $a(t+\displaystyle\sum_{i=1}^{n}T_i, \mathbf{x})=\left(\prod_{i=1}^{n}\mu_{i\mathbf{x}}\right)a(t, \mathbf{x})$. Because the quasiperiods $T_i$ and the factors $\mu_{i\mathbf{x}}$ are generally different from each other, the exponent in \eqref{expected form of solution} would need to take the form of integration.

The detailed oscillating behavior of $P(t, \mathbf{x})$ is not observable at macroscopic scales. However, the factor of the exponential increase $e^{\int_0^t H_{\mathbf{x}}(t')dt'}$ can be observed. In fact, inserting \eqref{expected form of solution} into \eqref{length definition}, the observable physical distance would become
\begin{equation}\label{new accelerating expansion}
L(t)=L(0) e^{H t},
\end{equation}
where
\begin{equation}
L(0)=\int_{\mathbf{x}_1}^{\mathbf{x}_2}\sqrt{P^2(t, \mathbf{x})}dl
\end{equation}
and the global Hubble expansion rate $H$ is
\begin{equation}\label{definition of expected H}
H=\frac{1}{t}\int_0^t H_{\mathbf{x}}(t')dt'.
\end{equation}

In the next two subsections, we are going to give the solution for $P(t, \mathbf{x})$ and the global Hubble expansion rate $H$.

\subsection{The solution for $P(t, \mathbf{x})$}\label{wkb approximation}
The magnitude of the time dependent frequency $\Omega$ is of the order $\sim \sqrt{G\left\langle T_{00}\right\rangle}\sim\sqrt{G}\Lambda^2$, while $\Omega$ itself varies roughly with a characteristic frequency $\Lambda$ (this has been shown by FIG. \ref{omega covariance}). Then according to \eqref{hmo}, the scale factor $a$ would oscillate with a period that roughly goes as $T=2\pi/\Omega\sim 1/\sqrt{G}\Lambda^2\ll1/\Lambda$, as $\Lambda\to\infty$, where $1/\Lambda$ is the time scale on which the $\Omega$ itself would change significantly.

So comparing to the oscillating period $T$ of the scale factor $a$, the variation of $\Omega$ itself is very slow, although the time $1/\Lambda$ is already very short for large $\Lambda$. Therefore, during one period of the oscillation of $a$, $\Omega$ is almost constant since it has not have a chance to change significantly during such a short time scale. In this sense the time dependent frequency $\Omega$ is slowly varying and the evolution of the scale factor $a$ is an adiabatic process.

The slow variation of $\Omega$ can be verified in a more formal way by calculating the expectation values of $\Omega^2=\frac{8\pi G}{3}\dot{\phi}^2$ and $\left(\frac{d\Omega}{dt}\right)^2=\frac{8\pi G}{3}\ddot{\phi}^2$. Using \eqref{field expansion}, we have
\begin{eqnarray}\label{omega square expectation value}
\left\langle\Omega^2\right\rangle&=&\frac{8\pi G}{3}\frac{1}{(2\pi)^3}\int d^3k \frac{1}{2}\omega\nonumber\\
&=&\frac{8\pi G}{3}\frac{1}{4\pi^2}\int_0^{\Lambda}k^3dk=\frac{1}{6\pi}G\Lambda^4,\label{omega expectation value}
\end{eqnarray}
\begin{eqnarray}\label{d omega expectation value}
\left\langle\left(\frac{d\Omega}{dt}\right)^2\right\rangle&=&\frac{8\pi G}{3}\frac{1}{(2\pi)^3}\int d^3k \frac{1}{2}\omega^3\nonumber\\
&=&\frac{8\pi G}{3}\frac{1}{4\pi^2}\int_0^{\Lambda}k^5dk=\frac{1}{9\pi}G\Lambda^6.\label{omega time derivative expectation value}
\end{eqnarray}

\eqref{omega square expectation value} just gives $\Omega\sim\sqrt{G}\Lambda^2$ as expected, \eqref{d omega expectation value} gives $d\Omega/dt\sim\sqrt{G}\Lambda^3$. Therefore, during one period of oscillation $T\sim 2\pi/\Omega\sim\frac{1}{\sqrt{G}\Lambda^2}$, we have, as $\Lambda\to+\infty$, the slow varying condition (see equation (49.1) in Chapter VII of \cite{LANDAU1976131})
\begin{equation}\label{slow varying condition}
T d\Omega/dt\ll\Omega,
\end{equation}
is satisfied. Thus the system varies adiabatically since $\Omega$ varies only slightly during the one period of oscillation time $T$. 

The leading order solution of the equation \eqref{hmo} for a harmonic oscillator with the slowly varying frequency $\Omega$ can be obtained by a first order WKB approximation. This adiabatic approximation neglects the small exponential factor in \eqref{expected form of solution}. It gives the solution $P(t, \mathbf{x})$ which is describing the oscillating behavior of $a(t, \mathbf{x})$. The result is,
\begin{equation}\label{wkb}
P(t,\mathbf{x})=\frac{A_0}{\sqrt{\Omega(t,\mathbf{x})}}\cos\left(\int_0^t\Omega(t',\mathbf{x}) dt'+\theta_{\mathbf{x}}\right).
\end{equation}
The $P(t, \mathbf{x})$ above is a quasiperiodic function with the same quasiperiod of the order $1/\Lambda$ as the time dependent frequency $\Omega(t, \mathbf{x})$ just as expected. The two constants of integration $A_0$ and $\theta_{\mathbf{x}}$ in \eqref{wkb} can be determined by the initial values $a(0, \mathbf{x})$ and $\dot{a}(0, \mathbf{x})$. 

The quantum vacuum is fluctuating everywhere, but its statistical property must be still the same everywhere. Correspondingly, the statistical property of $P(t, \mathbf{x})$ must also be the same everywhere, which requires that the constant $A_0$ to be independent with respect to the spatial coordinate $\mathbf{x}$. In addition, the constant $A_0$ can be chosen as any nonzero value since the scale factor $a$ multiplying by any nonzero constant describes physically equivalent spacetimes.

The initial phase $\theta_{\mathbf{x}}$ at different places must be dependent on $\mathbf{x}$. In applying the initial value constraint equation \eqref{hubble rate spatial difference}, neglecting the small exponential factor in \eqref{expected form of solution} and neglecting the relatively small time derivative terms of the slowly varying frequency $\Omega$, we obtain the result,
\begin{equation}\label{phase}
\tan\theta_{\mathbf{x}}=\frac{\Omega(0, \mathbf{x}_0)}{\Omega(0, \mathbf{x})}\tan\theta_{\mathbf{x}_0}+\frac{4\pi G}{\Omega(0, \mathbf{x})}\int_{\mathbf{x}_0}^{\mathbf{x}}\mathbf{J}(0, \mathbf{x}')\cdot\mathbf{dl}',
\end{equation}
where $\theta_{\mathbf{x}_0}$ is the initial phase of the scale factor $a$ at an arbitrary spatial point $\mathbf{x}_0$.

In solutions \eqref{wkb} and \eqref{phase} we see the fluctuating nature of spacetime at very small scales as described in the previous section \ref{fluctuating spacetime}. In particular, \eqref{phase} shows that the phases of $a(t, \mathbf{x})$ vary on a given initial Cauchy slice; some locations contract while others expand. In this new physical picture the catastrophic vacuum energy density is confined to very small scales.

\subsection{The global Hubble expansion rate $H$}\label{expoential increase}
As the system is adiabatic, the parametric resonance effect is weak. The adiabatic solution \eqref{wkb} in the last subsection does not include the parametric resonance and thus misses the small exponential factor expected in \eqref{expected form of solution}. In this subsection we go beyond the adiabatic approximation and investigate the exact strength of the weak parametric resonance. 

When considering the weak parametric resonance effect, the constant $A_0$ in \eqref{wkb} would become time and space dependent and take the following form
\begin{equation}\label{exponential increase of A}
A(t, \mathbf{x})=A_0 e^{\int_0^t H_{\mathbf{x}}(t')dt'}
\end{equation}
in order to satisfy \eqref{expected form of solution}. 

To determine how the $H_{\mathbf{x}}(t)$ depends on the spacetime dependent frequency $\Omega(t, \mathbf{x})$, we consider the adiabatic invariant of a harmonic oscillator with time dependent frequency, which is defined as
\begin{equation}\label{adiabatic invariance definition}
I(t, \mathbf{x})=\frac{E}{\Omega},
\end{equation}
where
\begin{equation}
E=\frac{1}{2}(\dot{a}^2+\Omega^2a^2)/\Omega.
\end{equation}

Replace the constant $A_0$ in \eqref{wkb} by $A(t, \mathbf{x})$ and then plug it into the above expression \eqref{adiabatic invariance definition} we get that
\begin{equation}\label{connection between I and A}
I(t, \mathbf{x})=\frac{1}{2}A^2(t, \mathbf{x}),
\end{equation}
where we have neglected the time derivatives of $A$ and $\Omega$ in the above equation \eqref{connection between I and A}, which are higher order infinitesimals. $I$ is invariant in the first order adiabatic approximation. When going to higher orders, $I$ will slowly change with time. Through the relation \eqref{connection between I and A} between $I$ and $A$ we can obtain how the $A(t, \mathbf{x})$ changes by investigating how accurately the adiabatic invariant is preserved and how it changes with time. 

It has been proved by Robnik and Romanovski \cite{adibaticinvariant, 0305-4470-39-1-L05} that, in full generality (no restrictions on the function $\Omega(t,\mathbf{x})$), the final value of the adiabatic invariant for the average energy $\bar{I}=\bar{E}/\Omega$ is always greater or equal to the initial value $I_0=E_0/\Omega_0$ (see the references \cite{adibaticinvariant, 0305-4470-39-1-L05} for precise definition about the average energy). In other words, the average value of the adiabatic invariant $\bar{I}=\bar{E}/\Omega$ for the mean value of the energy never decreases, which is a kind of irreversibility statement. It is conserved only for infinitely slow process, i.e. an ideal adiabatic process. 

Therefore, in the case of our quasiperiodic frequency $\Omega(t, \mathbf{x})$ in \eqref{hmo}, $\bar{I}$ will also always increase. Moreover, it will increase by a fixed factor after each quasiperiod of evolution, which results in an exponentially increasing $\bar{I}$. This is in fact evident because of the weak parametric resonance effect. In the following we investigate this exponential behavior in detail.

First we construct the evolution equation for the adiabatic invariant $I$. Do the canonical transformation
\begin{eqnarray}
a&=&\sqrt{2I/\Omega}\sin\varphi,\\
\dot{a}&=&\sqrt{2I\Omega}\cos\varphi.
\end{eqnarray}
Then the evolution equations for $a$ and its conjugate momentum $\dot{a}$ transfer to the evolution equation for the new action variable $I$ and the angle variable $\varphi$,
\begin{eqnarray}
\frac{dI}{dt}&=&-I\frac{\dot{\Omega}}{\Omega}\cos 2\varphi,\label{canonical equation 1}\\
\frac{d\varphi}{dt}&=&\Omega+\frac{\dot{\Omega}}{2\Omega}\sin 2\varphi. \label{canonical equation 2}
\end{eqnarray}
Integrating \eqref{canonical equation 1} yields
\begin{equation}
I(t)=I(0)\exp\left(2\int_0^t H_{\mathbf{x}}(t')dt'\right),
\end{equation}
where
\begin{equation}\label{dependence of H on Omega}
H_{\mathbf{x}}(t')=-\frac{\dot{\Omega}}{2\Omega}\cos 2\varphi.
\end{equation}
The $H_{\mathbf{x}}(t')$ in the above equation \eqref{dependence of H on Omega} is just the same with the $H_{\mathbf{x}}(t')$ defined in \eqref{expected form of solution} and \eqref{exponential increase of A}, which can be seen by applying equation \eqref{connection between I and A}. Thus equation \eqref{dependence of H on Omega} constructed the dependence of $H_{\mathbf{x}}(t')$ on the time dependent frequency $\Omega(t', \mathbf{x})$.

The observable global Hubble expansion rate $H$ is the average of $H_{\mathbf{x}}(t')$ over time, which was defined by equation \eqref{definition of expected H}. Plugging \eqref{dependence of H on Omega} into \eqref{definition of expected H} gives,
\begin{equation}\label{hubble constant}
H=\operatorname{Re}\left(-\frac{1}{t}\int_{0}^{t}\frac{\dot{\Omega}}{2\Omega}e^{2i\varphi}dt'\right).
\end{equation}
When the slow varying condition \eqref{slow varying condition} holds, from equation \eqref{canonical equation 2} we know that $d\varphi/dt$ is positive, i.e. $\varphi$ is a monotonic function in time. Thus we can change the integral in \eqref{hubble constant} from the integration over $t'$ to integration over $\varphi'$:
\begin{equation}\label{hubble constant phi}
H=\operatorname{Re}\left(-\frac{1}{t}\int_{\varphi_0}^{\varphi}\frac{\dot{\Omega}}{2\Omega}e^{2i\varphi}\frac{dt'}{d\varphi'}d\varphi'\right),
\end{equation}
where $\varphi_0=\varphi(0)$ and $\varphi=\varphi(t)$.

To evaluate $H$, we formally treat $\varphi$ as a complex variable and close the contour integral in the upper half plane. The integrand in \eqref{hubble constant phi} has no singularities for real $\varphi$ if the slow varying condition \eqref{slow varying condition} holds. Equation \eqref{canonical equation 2} implies that $\varphi\sim\Omega t\sim\sqrt{G}\Lambda^2 t$, so the length of the interval $\varphi-\varphi_0\sim\sqrt{G}\Lambda^2t$ goes to infinity as $\Lambda\to+\infty$. Hence the principle contribution to the integral in \eqref{hubble constant phi} comes from the residue values at singularities $\varphi_{(k)}$ inside the contour:
\begin{equation}\label{residue}
H=\frac{1}{t}\operatorname{Re}\left(2\pi i\displaystyle\sum_k\operatorname{Res}\left(-\frac{\dot{\Omega}}{2\Omega}e^{2i\varphi}\frac{dt}{d\varphi},\,\varphi_{(k)}\right)\right).
\end{equation}

Each term in \eqref{residue} gives a contribution containing a factor $\exp\left(-2\operatorname{Im}\varphi_{(k)}\right)$. So the dominant contribution in \eqref{residue} comes from the singularities near the real axis, i.e. those with the smallest positive imaginary part. To keep the calculation simple, we retain only those terms. Since $\Omega(t)$ varies quasiperiodically with a characteristic time $\tau\sim 1/\Lambda$, the number of singularities near the real axis would roughly be on the order $t/\tau\sim\Lambda t$. Therefore the $H$ in \eqref{residue} is roughly
\begin{equation}\label{hubble constant estimation 1}
H\sim\Lambda\exp\left(-2\operatorname{Im}\varphi_{(k)}\right).
\end{equation}

Let $t_{(k)}$ be the (complex) ``instant'' corresponding to the singularity $\varphi_{(k)}$: $\varphi_{(k)}=\varphi(t_{(k)})\sim\Omega\, t_{(k)}$. In general, $|t_{(k)}|$ has the same order of magnitude as the characteristic time $\tau\sim 1/\Lambda$ of variation of the $\Omega$. Remember that $\Omega\sim\sqrt{G}\Lambda^2$, thus the order of magnitude of the exponent in \eqref{hubble constant estimation 1} is
\begin{equation}\label{estimation of magnitude of im varphi}
\operatorname{Im}\varphi_{(k)}\sim\Omega\tau\sim\sqrt{G}\Lambda.
\end{equation}
Therefore, inserting \eqref{estimation of magnitude of im varphi} into \eqref{hubble constant estimation 1} gives
\begin{equation}\label{dependence of H on Lambda}
H=\alpha\Lambda e^{-\beta\sqrt{G}\Lambda},
\end{equation}
where $\alpha$ and $\beta$ are two dimensionless constants which depend on the variation details of the time dependent frequency $\Omega(t, \mathbf{x})$. Therefore $H$ becomes exponentially small in the limit of taking $\Lambda$ to infinity. This is a manifestation of the well-established result that the error in adiabatic invariant is exponentially small for analytic $\Omega$ \cite{LANDAU1976131, adibaticinvariant}. In fact, the technique we used in deriving \eqref{dependence of H on Lambda} is very similar to the one used in deriving the error in adiabatic invariant in the pages ``$160-161$" of \cite{LANDAU1976131}.

\subsection{A more intuitive explanation}
So far we have obtained our key result \eqref{dependence of H on Lambda} for the global Hubble expansion rate $H$. To understand the mechanism of weak parametric resonance better, we give a more intuitive explanation in this subsection.

Consider the following simplest parametric oscillator:
\begin{equation}\label{simplest hmo}
\ddot{x}+\omega^2(t)x=0,
\end{equation}
where
\begin{equation}\label{small omega}
\omega^2(t)=\omega_0^2\left(1+h\cos\gamma t\right).
\end{equation}
The behavior of the above harmonic oscillator with time dependent frequency has been thoroughly studied (see e.g. eq(27.7) in Chapter V of \cite{LANDAU197658}). The parametric resonance occurs when the frequency $\gamma$ with which $\omega(t)$ varies is close to any value $2\omega_0/n$, i.e.
\begin{equation}\label{parametric resonance condition}
\gamma\sim\frac{2\omega_0}{n},
\end{equation}
where $n$ is an integer. The strength of the parametric resonance is strongest if $\gamma$ is nearly twice $\omega_0$, i.e. if $n=1$. As $n$ increases to infinity, the strength of the parametric resonance decreases to zero. This is easy to understand since as $n$ increases, the varying frequency $\gamma$ of $\omega(t)$ becomes slower compared to the oscillator's natural frequency $\omega_0$ and as $n\to\infty$, \eqref{simplest hmo} reduces to an ordinary harmonic oscillator with constant frequency which has no parametric resonance behavior. 

\begin{figure}
\includegraphics[scale=0.65]{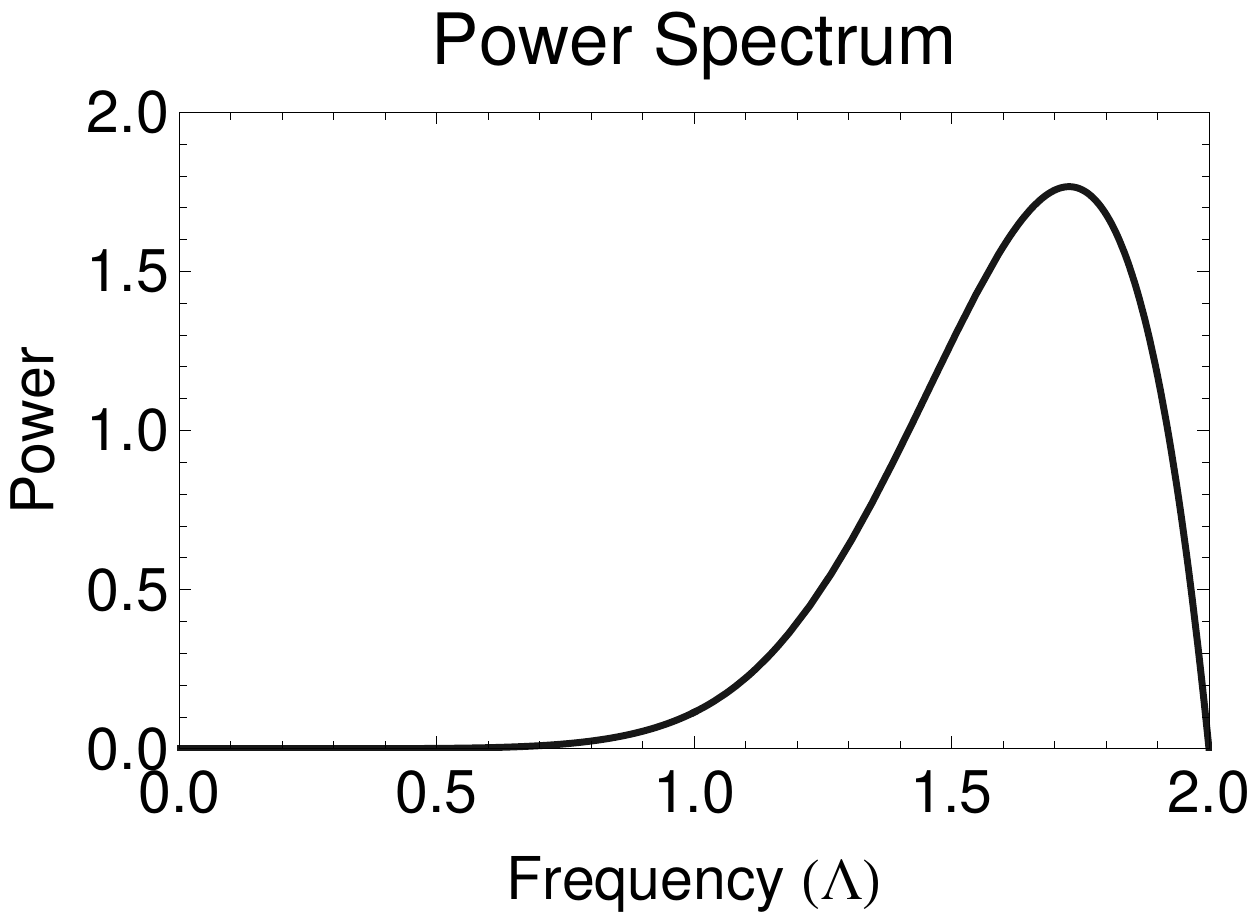}
\caption{\label{power spectrum}Plot of the power spectrum density of the varying part of $\Omega^2(t, \mathbf{0})$ (except for the constant $\Omega_0^2$ part).}
\end{figure}

Now let us go back to Eq.\eqref{hmo} for $a(t, \mathbf{x})$. The time dependent frequency $\Omega(t, \mathbf{x})$ in \eqref{hmo} is more complicated than the $\omega(t)$ given in our example \eqref{small omega}. However, it can be written in a similar form:
\begin{equation}\label{big Omega}
\Omega^2(t, \mathbf{0})=\Omega_0^2\left(1+\int_0^{2\Lambda}d\gamma\left(f\left(\gamma\right)\cos\gamma t+g\left(\gamma\right)\sin\gamma t\right)\right),
\end{equation}
where
\begin{equation}\label{expectation value of Omega square}
\Omega_0^2=\left\langle \Omega^2\right\rangle=\frac{G\Lambda^4}{6\pi},
\end{equation}
and $f(\gamma)$, $g(\gamma)$ are operator coefficients, whose exact form are given by \eqref{f gamma} and \eqref{g gamma} in Appendix \ref{appendix 1}. The behavior of $\Omega^2(t, \mathbf{x})$ for an arbitrary $\mathbf{x}$ is the same with $\Omega^2(t, \mathbf{0})$ except phase differences. The power spectrum density of the varying part of $\Omega^2(t, \mathbf{0})$ (except for the constant $\Omega_0^2$ part) given by \eqref{power spectrum calculation} is plotted in FIG. \ref{power spectrum}.

Unlike the case \eqref{small omega} where the $\omega(t)$ varies with a single frequency $\gamma$, the $\Omega(t, \mathbf{0})$ in \eqref{big Omega} varies with frequencies continuously distributed in the range $(0, 2\Lambda)$ with a peak around $1.7\Lambda$ (see FIG. \ref{power spectrum}). From \eqref{expectation value of Omega square} we have that, as taking the cutoff frequency $\Lambda$ to infinity, $\Omega_0\sim\sqrt{G}\Lambda^2\gg 2\Lambda$. Because of the continuity of the spectrum of $\Omega$, we can always find integers $n$ such that if
\begin{equation}\label{n inequality}
n\geq\sqrt{\frac{G}{6\pi}}\Lambda, \quad \Lambda\to+\infty,
\end{equation}
then
\begin{equation}
\frac{2\Omega_0}{n}\in\left(0, 2\Lambda\right).
\end{equation}
So $\Omega(t, \mathbf{x})$ always contains frequencies $2\Omega_0/n$ that may excite resonances. From \eqref{n inequality} we see that $n\to\infty$ as taking the cutoff $\Lambda$ to infinity. While as $n$ increases, the relative magnitude of the resonance frequency $2\Omega_0/n$ decreases comparing to the $a(t, \mathbf{x})$'s natural frequency $\Omega_0$. Then for reasons similar to the simplest parametric oscillator \eqref{simplest hmo}, the strength of the parametric resonance of \eqref{hmo} would also decrease to zero. This weak parametric resonance effect leads to the global Hubble expansion rate
\begin{equation}
H\to 0, \quad \text{as} \quad \Lambda\to+\infty.
\end{equation}

\subsection{Numerical verification}\label{numerical}
In this subsection, we do a numerical calculation for the evolution equation \eqref{hmo} to verify our result \eqref{dependence of H on Lambda}, which describes the dependence of $H$ on cutoff $\Lambda$.

In this subsection, Planck units will be used, so all instances of Newton's constant are set to unity, $G=1$.

\begin{figure}
\includegraphics[scale=0.35]{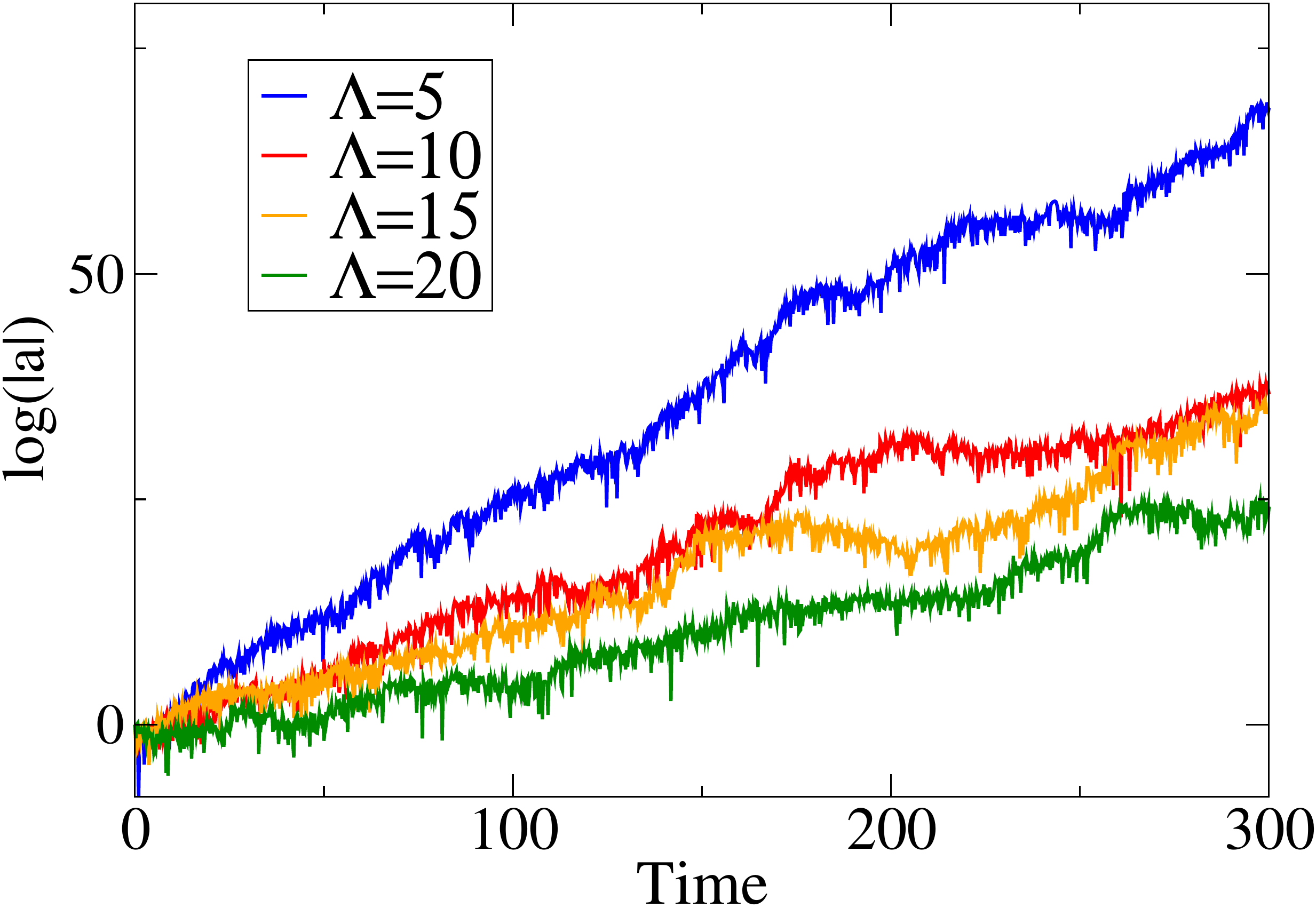}
\caption{\label{fig:one_field_1}Numeric result for $\log |a_o(t)|$ for a single real massless scalar field. It shows that as $\Lambda$ increases, the slope of  $\log |a_o(t)|$ decreases.}
\end{figure}

The main idea is to rewrite the time dependent frequency $\Omega(t)$ in phase space. (To see more details about this numeric method, please check Appendix \ref{appendix 2}. Here we only list the most crucial results.)  For a real massless scalar field, we have
 \begin{equation}
 \begin{split}
&\Omega^2(\{x_{\mathbf{k}}\},\{p_{\mathbf{k}}\},t)=\frac{8\pi}{3}\int \frac{d^3k d^3k'}{(2\pi)^3}x_{\mathbf{k}} x_{\mathbf{k'}}\omega\omega'\sin\omega t \sin\omega' t\\
&+p_{\mathbf{k}} p_{\mathbf{k'}}\cos\omega t \cos\omega' t-2x_{\mathbf{k}} p_{\mathbf{k'}}\omega\sin\omega t \cos
\omega' t. 
\end{split} 
\end{equation}
This is  the Weyl transformation of the operator $\hat{\Omega}^2(t)$. Here $\{x_{\mathbf{k}}, p_{\mathbf{k}}\}$ are phase space points of a particular  field mode with momentum $\mathbf{k}$. Approximately, for a particular choice of $ \{x_{\mathbf{k}}\},\{p_{\mathbf{k}}\}$, we can get an classic equation for $a$:

\begin{equation}\label{eq:a_k}
\ddot{a}(\{x_{\mathbf{k}}\},\{p_{\mathbf{k}}\},t)+\Omega^2 (\{x_{\mathbf{k}}\},\{p_{\mathbf{k}}\}, t)  a(\{x_{\mathbf{k}}\},\{p_{\mathbf{k}}\},t)  =0
\end{equation}
The observed value $a_o(t)$ is the average of $ a(\{x_{\mathbf{k}}\},\{p_{\mathbf{k}}\},t)$ over the Wigner pseudo distribution function $W(\{x_{\mathbf{k}}\},\{p_{\mathbf{k}}\},t)$, which is based on the wave function of the quantum field: 
\begin{equation}
a_{o}(t)=\int \left(\prod_{\mathbf{k}} dx_{\mathbf{k}} dp_{\mathbf{k}} \right) a(\{x_{\mathbf{k}}\},\{p_{\mathbf{k}}\},t)W(\{x_{\mathbf{k}}\},\{p_{\mathbf{k}}\},t).
\end{equation}
If the quantum field is in its ground state,  we have 
\begin{equation}
W(\{x_{\mathbf{k}}\},\{p_{\mathbf{k}}\},t)=\prod_{\mathbf{k}}  \frac{1}{\pi} e^{-\frac{p_{\mathbf{k}}^2}{\omega}-x_{\mathbf{k}}^2\omega}
\end{equation}
which means $ \{x_{\mathbf{k}}\},\{p_{\mathbf{k}}\}$ are all Gaussian variables. Based on this observation, our method to simulate this equation  is as following: 
i) at first we generate a set of  random Gaussian numbers for  $\{x_{\mathbf{k}}\},\{p_{\mathbf{k}}\}$ ; ii) we  solve the equation \eqref{eq:a_k} for this particular set of numbers; iii) then we repeat the process for another set of random numbers until a certain amount of repetitions; iv) The result $a_o(t)$ is the average over  all samples we have generated. We choose the  repetition amount to be  big enough for the results to converge. The result of a single scalar field case is illustrated in Fig. \ref{fig:one_field_1}.  

We can find that the slope of $\log |a_o(t)|\sim t$ is decreasing as we increase the cutoff $\Lambda$ as we expect. For the single field,  $\Omega^2(t)=\frac{8\pi}{3}\dot{\phi}^2$ repeatedly reaches zero since classically $\dot{\phi}$ is continuous and oscillates from positive to negative. Around these zero points the slow varying condition for $\Omega^2$ is not satisfied. But because the time duration of reaching zero is very short, this would not cause the adiabatic expanding scheme to breaks down. This point reveals in the numerical calculation.

\begin{figure}
\includegraphics[scale=0.35]{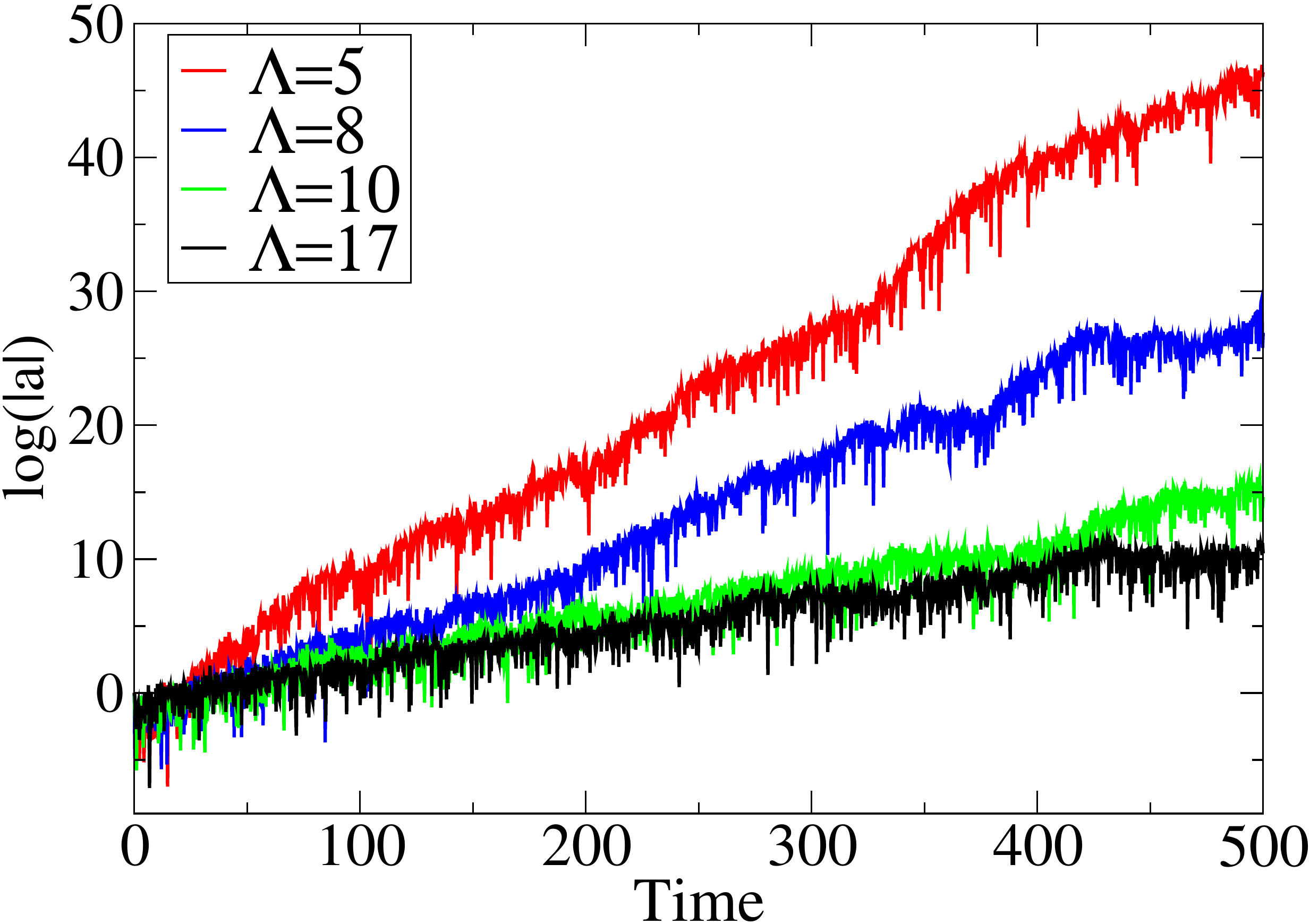}
\caption{\label{fig:two_field_1}Numeric result for $\log |a_o(t)|$ when two scalar fields are present and it shows that as $\Lambda$ increases, the slope of  $\log |a_o(t)|$ decreases. }
\end{figure}

The real Universe contains many different quantum fields. In Fig. \ref{fig:two_field_1}, we show the result when we include two independent massless scalar fields in which $\Omega^2(t)=\frac{8\pi}{3}(\dot{\phi}_1^2+\dot{\phi}_2^2)$. In this case $\Omega^2$ would not reach zero at almost all times since this can happen only when both $\dot{\phi}_1$ and $\dot{\phi}_2$ pass zero, which is unlikely to happen frequently.

\begin{figure}
\includegraphics[scale=0.35]{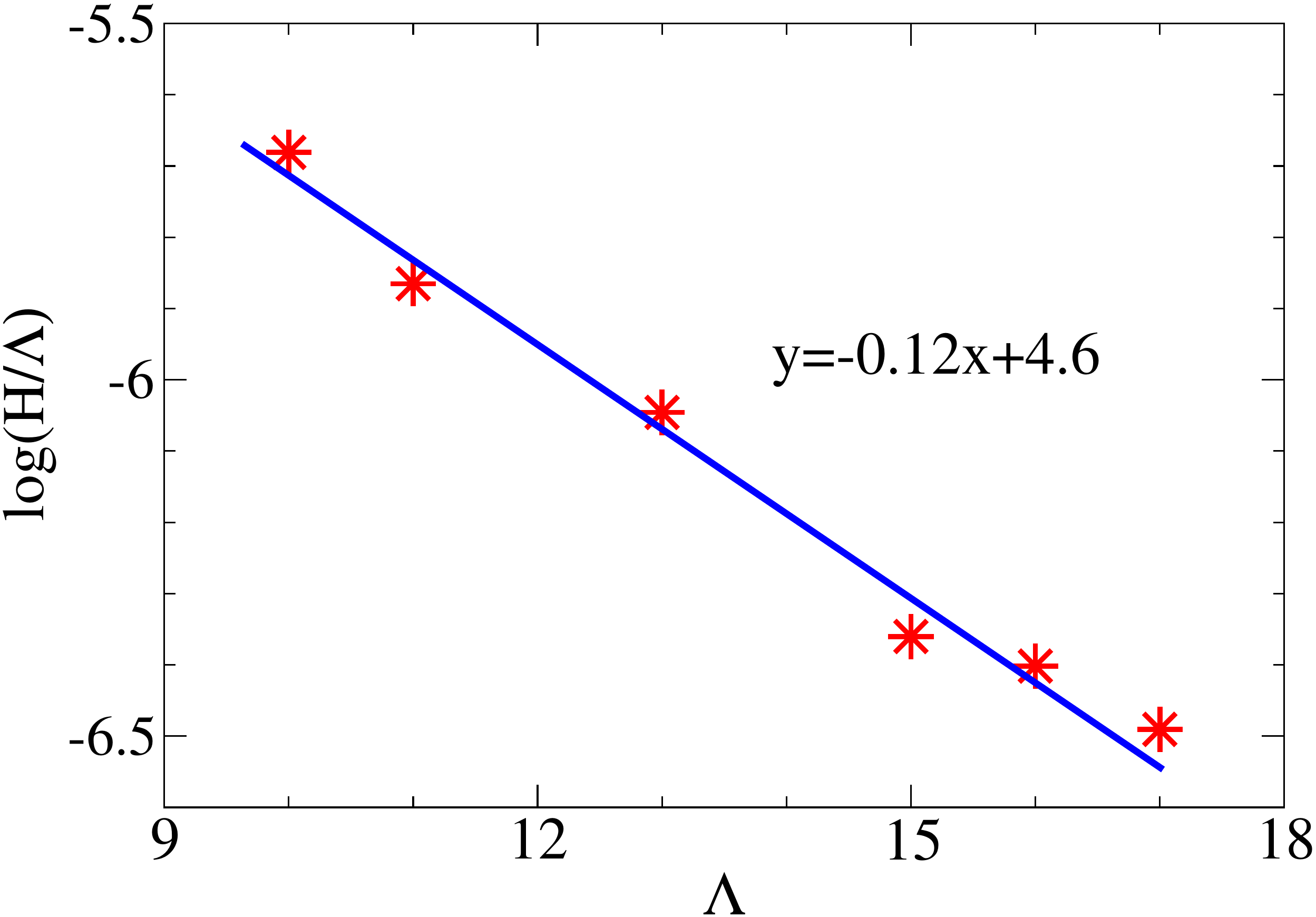}
\caption{\label{fig:two_field_fit}The plot of $\log (H/\Lambda)$ over $\Lambda$. The fitting result shows that $\alpha=e^{4.6}\approx100$ and $\beta=0.12$ in this two-field case. }
\end{figure}

In the two field case, we plot the $\log (H/\Lambda) \sim \Lambda$ graph to verify the quantitative relation  \eqref{dependence of H on Lambda}. The result is illustrated in Fig.\ref{fig:two_field_fit}. We can see that for $\Lambda\geq 10$, the result shows decent linearity, which is what we expected since the derivation of \eqref{dependence of H on Lambda} is only valid for large $\Lambda$. In this case, the two constants $\alpha=e^{4.6}\approx 100$ and $\beta=0.12$ according to the numeric calculation.

\section{Meaning of our results}\label{sec vi}
It is interesting to notice that both \eqref{cmc prediction} and \eqref{new accelerating expansion} give the exponential evolution and predict an accelerated expanding Universe. However, the underlying mechanisms are completely different, which leads to opposite results on the predicted magnitude of the observable Hubble expansion rate $H$.

The solution \eqref{cmc prediction} is based on the assumption that quantum vacuum energy density is constant all over the spacetime, which is a necessary requirement if one suppose that vacuum acts as a cosmological constant. This assumption leads to a huge Hubble expansion rate 
\begin{equation}
H=\sqrt{\frac{8\pi G\rho^{vac}}{3}}\propto \sqrt{G}\Lambda^2\to+\infty
\end{equation}
as taking the high energy cutoff $\Lambda$ to infinity.

Our proposal \eqref{new accelerating expansion} is based on the fact that quantum vacuum energy density is constantly fluctuating and extremely inhomogeneous all over the whole spacetime. This fact leads to a small Hubble expansion rate given by \eqref{dependence of H on Lambda} which goes to zero as taking the high energy cutoff $\Lambda$ to infinity.

If we can literally take the cutoff $\Lambda$ in \eqref{dependence of H on Lambda} to infinity, then $H=0$. In this sense, at least the ``old'' cosmological constant problem would be resolved.

In principle, this effective theory is valid only up to a large but finite cutoff $\Lambda$, which leads to a tiny but nonzero $H$. Since $H\to 0$ as $\Lambda\to+\infty$, there always exists a very large cutoff value of $\Lambda$ such that $H=\sqrt{\Omega_\lambda}H_0\approx 1.2\times 10^{-42} \,\mbox{GeV}$ to match the observation, where $H_0$ is current observed Hubble constant.

So our result suggests that there is no necessity to introduce the cosmological constant, which is required to be fine tuned to an accuracy of $10^{-120}$, or other forms of dark energy, which are required to have peculiar negative pressure, to explain the observed accelerating expansion of the Universe.

The exact value of $\Lambda$ cannot be determined since we do not know the values of the two dimensionless parameters $\alpha$ and $\beta$ in \eqref{dependence of H on Lambda}. In principle, we need the knowledge of all fundamental fields in the Universe to determine $\alpha$ and $\beta$, this deserves further investigations in the future and might provide some hint on elementary particle physics.

The value of $\Lambda$ should be on the order of Planck energy or higher. According to the numerical calculation in the last subsection, $\Lambda\sim 1000 E_P$ if we consider contributions to $\Omega^2$ from only two scalar fields. If more fundamental fields are included, we expect the value of $\beta$ would increase and thus decrease the value of $\Lambda$ needed. This is because that it increases the mean value of $\Omega$ and, as a consequence, reduces the ratio between the variation of $\Omega$ over its mean value $\langle \Omega \rangle$ that $\Omega$ varies slower. A slower $\Omega$ leads to smaller $H$ since the parametric resonance is weaker.

\section{The back reaction}\label{back reaction}
In this section, we investigate the back reaction effect by quantizing the field $\phi$ in the resulting curved spacetime to justify our method of using the quantized field expansion \eqref{field expansion} in Minkowski spacetime as an approximation.

The standard way to quantize the scalar field $\phi$ in a generic curved spacetime $g_{\mu\nu}$ is by first defining the following inner product on a spacelike hypersurface $\Sigma$ with induced metric $h_{ij}$ and unit normal vector $n^{\mu}$ (see e.g. \cite{carroll2004spacetime, 9780511622632}):
\begin{equation}\label{inner product}
(\phi_1, \phi_2)=-i\int_\Sigma (\phi_1\partial_{\mu}\phi_2^*-\phi_2^*\partial_{\mu}\phi_1)n^{\mu}\sqrt{h}d^3x,
\end{equation}
where $h=\det h_{ij}$ and $\phi_1$, $\phi_2$ are solutions to the equation \eqref{wave eqaution in curved spacetime}. The above inner product is independent of the choice of $\Sigma$.

One then choose a complete set of mode solutions $u_{\mathbf{k}}$ of \eqref{wave eqaution in curved spacetime} which are orthonormal in the product \eqref{inner product}:
\begin{eqnarray}\label{normalization condition}
(u_{\mathbf{k}}, u_{\mathbf{k}'})&=&\delta(\mathbf{k}-\mathbf{k}'),\\
(u_{\mathbf{k}}^*, u_{\mathbf{k}'}^*)&=&-\delta(\mathbf{k}-\mathbf{k}'),\\
(u_{\mathbf{k}}, u_{\mathbf{k}'}^*)&=&0.
\end{eqnarray}

Then the field $\phi$ may be expanded as
\begin{equation}\label{phi mode expansion}
\phi=\sum_{\mathbf{k}}\left(a_{\mathbf{k}}u_{\mathbf{k}}+a_{\mathbf{k}}^{\dag}u_{\mathbf{k}}^*\right).
\end{equation}

For the flat Minkowski spacetime, i.e. $g_{\mu\nu}=\eta_{\mu\nu}$, \eqref{wave eqaution in curved spacetime} reduces to the usual wave equation
\begin{equation}\label{homogeneous wave equation}
\ddot{\phi}-\nabla^2\phi=0.
\end{equation}
In this case, the mode solutions are usually chosen as
\begin{equation}\label{flat plane wave mode}
u_{\mathbf{k}}(t, \mathbf{x})=\frac{1}{(2\pi)^{3/2}}\frac{1}{\sqrt{2\omega}}e^{-i(\omega t-\mathbf{k}\cdot\mathbf{x})},
\end{equation}
where $\omega=|\mathbf{k}|$. Plugging \eqref{flat plane wave mode} into \eqref{phi mode expansion} just gives the usual quantum field expansion \eqref{field expansion}.

For our specific metric \eqref{inhomogenous FLRW coordinate}, \eqref{wave eqaution in curved spacetime} reduces to \eqref{eq:fullwave}. In this case, since the rate of accelerating expansion is extremely small, the back reaction effect due to the macroscopic expansion of the Universe is only important on large cosmological time scales. For this reason, we only worry about the back reaction due to the wildly fluctuating spacetime at small scales. i.e. we neglect the small exponential factor in \eqref{expected form of solution} and use the form of the $a$ based on the solution \eqref{wkb}:
\begin{equation}
a( t,\mathbf{x})=\frac{A_0}{\sqrt{\Omega(t,\mathbf{x})}}\cos\left(\Theta(t, \mathbf{x})\right),
\end{equation}
where 
\begin{equation}\label{phase angle}
\Theta(t,\mathbf{x})=\int_0^t\Omega(t',\mathbf{x}) dt'+\theta_{\mathbf{x}}.
\end{equation}
Then \eqref{eq:fullwave} becomes

\begin{widetext}
\begin{equation}\label{eq:fullexpansion}
\frac{A_0^2}{\Omega}\cos^2\Theta\ddot{\phi}-\nabla^2\phi-\frac{3A_0^2}{2}\left(\frac{\dot{\Omega}}{\Omega^2}\cos^2\Theta+\sin2\Theta\right)\dot{\phi}+\left(\frac{\nabla\Omega}{2\Omega}+\tan\Theta\nabla\Theta\right)\cdot\nabla\phi=0.
\end{equation}
\end{widetext}

In order to understand the effect from back reaction, we need to find out how the mode solutions of the above equation \eqref{eq:fullexpansion} in the resulting curved spacetime change from the mode solutions \eqref{flat plane wave mode} of the equation \eqref{homogeneous wave equation} in the flat Minkowski spacetime.

Physically, the correction to \eqref{flat plane wave mode} should be small for wave modes with frequencies lower than the cutoff frequency $\Lambda$. That is because the wave length of those field modes is larger than $2\pi/\Lambda$, while our spacetime fluctuates on the length scale $2\pi/\Omega\sim 1/(\sqrt{G}\Lambda^2)\ll 2\pi/\Lambda$. The relatively long wave length modes should not be sensitive to what is happening on small scales. This is analogous to the situation of sound waves traveling in the medium such as air or water or solids. The medium is constantly fluctuating at atomic scales, but this fluctuation does not affect the propagation of the sound wave whose wavelength is much larger than the atomic scale. Similarly, the propagation of the field modes in the ``medium''--the spacetime, which is constantly fluctuating on scales much smaller than the wavelength of the field modes, should also not be affected. 

Mathematical demonstration will be given in the following subsections.

\subsection{A simplified toy model}
It is complicated to obtain the mode solutions of \eqref{eq:fullexpansion} for a generic stochastic function $\Theta(t, \mathbf{x})$ whose stochastic property is determined by the quantum nature of the field $\phi$. To illustrate the underlying physical mechanism more clearly, we start with a simplified toy model by restricting the phase angle $\Theta(t, \mathbf{x})$ defined by \eqref{phase angle} to take the following form:
\begin{equation}\label{toy form phase angle}
\Theta(t, \mathbf{x})=\Omega t+\mathbf{K}\cdot\mathbf{x},
\end{equation}
where both $\Omega$ and $\mathbf{K}$ are constants and they have the same order of magnitude $\Omega\sim |\mathbf{K}|\sim\sqrt{G}\Lambda^2$.

Of course this toy model does not describe the real spacetime sourced by the quantum vacuum since the $\Omega$ is by no means a constant but always varying, although the varying is slow compared to it own magnitude. However, this toy model possesses the key property needed --- the spacetime is constantly fluctuating. It will be convenient for visualizing the back reaction effect from a fluctuating spacetime. 

After setting the $\Omega\equiv  Constant$ and the phase angle $\Theta(t, \mathbf{x})$ to be the form of \eqref{toy form phase angle}, the equation of motion \eqref{eq:fullexpansion} for $\phi$ becomes
\begin{widetext}
\begin{equation}\label{toy wave equation}
\left(1+\cos 2\left(\Omega t+\mathbf{K}\cdot\mathbf{x}\right)\right)\ddot{\phi}-\nabla^2\phi-3\Omega\sin2\left(\Omega t+\mathbf{K}\cdot\mathbf{x}\right)\dot{\phi}+\tan\left(\Omega t+\mathbf{K}\cdot\mathbf{x}\right)\mathbf{K}\cdot\nabla\phi=0, 
\end{equation}
where we have set $A_0=\sqrt{2\Omega}$ such that the average of the coefficient $\frac{A_0^2}{\Omega}\cos^2\Theta$ before $\ddot{\phi}$ is $1$ for convenience.

In the flat spacetime case \eqref{homogeneous wave equation}, each mode solution $u_{\mathbf{k}}$ in \eqref{flat plane wave mode} contains only one single frequency. However, for the above fluctuating spacetime case \eqref{toy wave equation}, high frequencies mixes with low frequencies and each mode solution must contain multiple frequencies. In fact, since \eqref{toy wave equation} describes a strictly periodic system with time period $\pi/\Omega$ and spatial period $\pi/|\mathbf{K}|$, each mode solution $u_{\mathbf{k}}$ must change from \eqref{flat plane wave mode} to the following form:
\begin{equation}\label{expected toy form of solution}
u_{\mathbf{k}}(t, \mathbf{x})=e^{-i\left(\omega t-\mathbf{k}\cdot\mathbf{x}\right)}\left(c_0+\sum_{\substack{m=-\infty\\ m\neq 0}}^{+\infty}c_m e^{i2m\left(\Omega t+\mathbf{K}\cdot\mathbf{x}\right)}\right),
\end{equation}
where $c_m$ are constants.

Inserting \eqref{expected toy form of solution} into \eqref{toy wave equation} and using the orthogonality of $e^{2im(\Omega t+\mathbf{K}\cdot\mathbf{x})}$, we obtain the following infinite system of linear equations:
\begin{eqnarray}\label{original infinite system of linear equations}
m\text{th equation:}\quad &&\sum_{n=-\infty}^{m-2}(-1)^{m+n}\mathbf{K}\cdot\left(\mathbf{k}+2n\mathbf{K}\right)c_n\nonumber\\
+&&\left[\frac{1}{2}\left(\omega-2\left(m-1\right)\Omega\right)^2-\frac{3}{2}\Omega\left(\omega-2\left(m-1\right)\Omega\right)-\mathbf{K}\cdot\left(\mathbf{k}+2\left(m-1\right)\mathbf{K}\right)\right]c_{m-1}\nonumber\\
+&&\left[\left(\omega-2m\Omega\right)^2-\left(\mathbf{k}+2m\mathbf{K}\right)^2\right]c_{m}\\
+&&\left[ \frac{1}{2}\left(\omega-2\left(m+1\right)\Omega\right)^2+\frac{3}{2}\Omega\left(\omega-2\left(m+1\right)\Omega\right)+\mathbf{K}\cdot\left(\mathbf{k}+2\left(m+1\right)\mathbf{K}\right)\right]c_{m+1}\nonumber\\
+&&\sum_{n=m+2}^{+\infty}(-1)^{m+n+1}\mathbf{K}\cdot\left(\mathbf{k}+2n\mathbf{K}\right)c_n\nonumber\\
=&&0, \quad\quad\quad\quad\quad\quad \quad\quad\quad\quad \quad m=0, \pm 1, \pm 2, \pm 3, \dots\nonumber
\end{eqnarray}
In the above calculations, we have used the Fourier series expansion
\begin{equation}\label{tanx fourier}
\tan x=-2\displaystyle\sum_{n=1}^{+\infty}(-1)^n\sin 2nx
\end{equation}
to expand the term $\tan(\Omega t+\mathbf{K}\cdot\mathbf{x})$ in \eqref{toy wave equation}.

For the equations of $m\leq-1$, we successively add the ($m+1$)th equation to the $m$th equation by the order from $m=-\infty$ to $m=-1$; and for the equations of $m\geq 1$, we successively add the ($m-1$)th equation to the $m$th equation by the order from $m=+\infty$ to $m=1$. Most terms can be eliminated by these elementary row operations and the above infinite system of linear equations \eqref{original infinite system of linear equations} becomes
\begin{eqnarray}
\text{if}\,\,m\leq-1,\quad&&\frac{1}{2}\left(\omega-2\left(m-1\right)\Omega\right)\left(\omega-\left(2m+1\right)\Omega\right)c_{m-1}\nonumber\\
+&&\left[\frac{3}{2}\left(\omega-2m\Omega\right)\left(\omega-\left(2m+1\right)\Omega\right)-\left(\mathbf{k}+2m\mathbf{K}\right)\cdot\left(\mathbf{k}+\left(2m+1\right)\mathbf{K}\right)\right]c_m \nonumber\\
+&&\left[\frac{3}{2}\left(\omega-2(m+1)\Omega\right)\left(\omega-\left(2m+1\right)\Omega\right)-\left(\mathbf{k}+2(m+1)\mathbf{K}\right)\cdot\left(\mathbf{k}+\left(2m+1\right)\mathbf{K}\right)\right]c_{m+1}\nonumber\\
+&&\frac{1}{2}\left(\omega-2\left(m+2\right)\Omega\right)\left(\omega-\left(2m+1\right)\Omega\right)c_{m+2}=0; \nonumber\\
\text{if}\,\,m=0,\quad\quad&& \sum_{n=-\infty}^{-2}(-1)^n\mathbf{K}\cdot\left(\mathbf{k}+2n\mathbf{K}\right)c_n\nonumber\\
+&&\left[\frac{1}{2}\left(\omega+2\Omega\right)\left(\omega-\Omega\right)-\mathbf{K}\cdot\left(\mathbf{k}-2\mathbf{K}\right)\right]c_{-1}\nonumber\\
+&&\left(\omega^2-\mathbf{k}^2\right)c_0\nonumber\\
+&&\left[\frac{1}{2}\left(\omega-2\Omega\right)\left(\omega+\Omega\right)+\mathbf{K}\cdot\left(\mathbf{k}+2\mathbf{K}\right)\right]c_1\nonumber\\
+&&\sum_{n=2}^{+\infty}(-1)^{n+1}\mathbf{K}\cdot\left(\mathbf{k}+2n\mathbf{K}\right)c_n=0; \nonumber\\
\text{if}\,\,m\geq 1, \quad\quad&&\frac{1}{2}\left(\omega-2\left(m-2\right)\Omega\right)\left(\omega-\left(2m-1\right)\Omega\right)c_{m-2}\nonumber\\
+&&\left[\frac{3}{2}\left(\omega-2(m-1)\Omega\right)\left(\omega-\left(2m-1\right)\Omega\right)-\left(\mathbf{k}+2(m-1)\mathbf{K}\right)\cdot\left(\mathbf{k}+\left(2m-1\right)\mathbf{K}\right)\right]c_{m-1}\nonumber\\
+&&\left[\frac{3}{2}\left(\omega-2m\Omega\right)\left(\omega-\left(2m-1\right)\Omega\right)-\left(\mathbf{k}+2m\mathbf{K}\right)\cdot\left(\mathbf{k}+\left(2m-1\right)\mathbf{K}\right)\right]c_{m}\nonumber\\
+&&\frac{1}{2}\left(\omega-2\left(m+1\right)\Omega\right)\left(\omega-\left(2m-1\right)\Omega\right)c_{m+1}=0. \label{simplified equations}
\end{eqnarray}

To characterize the property of the solutions of this system more clearly, we define the following parameters for convenience:
\begin{equation}\label{parameters definition}
\epsilon=\frac{\omega}{\Omega}, \quad \upsilon=\frac{|\mathbf{k}|}{\Omega}, \quad \delta=\frac{|\mathbf{K}|}{\Omega}, \quad \cos\gamma=\frac{\mathbf{K}\cdot\mathbf{k}}{|\mathbf{K}||\mathbf{k}|}.
\end{equation}

As mentioned before that our effective theory has a cutoff $\Lambda$ such that only modes with $\omega, |\mathbf{k}|\leq\Lambda$ are relevant, which are much smaller than $\Omega\sim |\mathbf{K}|\sim \sqrt{G}\Lambda^2$ as $\Lambda$ grows large. Therefore, we are only interested in the solutions of \eqref{original infinite system of linear equations} or \eqref{simplified equations} when $\omega, |\mathbf{k}|\ll\Omega$, i.e. when $\epsilon, \upsilon\to 0$.

Dividing both sides of \eqref{simplified equations} by $\Omega^2$ and doing some necessary algebraic manipulations, \eqref{simplified equations} can be rewritten as
\begin{eqnarray}
&&\text{if}\,\,m\leq-1,\nonumber\\
&& \left[\left(m-1\right)-\frac{\epsilon}{2}\right]c_{m-1}\nonumber\\
+&&\left[\left(3-2\delta^2\right)m-\frac{3\epsilon}{2}-2m\delta^2\sum_{n=1}^{+\infty}\left(\frac{\epsilon}{2m+1}\right)^n-\frac{\upsilon}{2m+1}\left(\left(4m+1\right)\delta\cos\gamma+\upsilon\right)\sum_{n=0}^{+\infty}\left(\frac{\epsilon}{2m+1}\right)^n\right]c_m \nonumber\\
+&&\left[\left(3-2\delta^2\right)(m+1)-\frac{3\epsilon}{2}-2(m+1)\delta^2\sum_{n=1}^{+\infty}\left(\frac{\epsilon}{2m+1}\right)^n-\frac{\upsilon}{2m+1}\left(\left(4m+3\right)\delta\cos\gamma+\upsilon\right)\sum_{n=0}^{+\infty}\left(\frac{\epsilon}{2m+1}\right)^n\right]c_{m+1} \nonumber\\
+&&\left[\left(m+2\right)-\frac{\epsilon}{2}\right]c_{m+2}=0;\nonumber\\
\nonumber\\
&&\text{if}\,\,m=0,\nonumber\\
&&\sum_{n=-\infty}^{-2}(-1)^n\left(2n\delta^2+\delta\upsilon\cos\gamma\right)c_n+\left[-1+2\delta^2+\frac{\epsilon}{2}-\delta\upsilon\cos\gamma+\frac{\epsilon^2}{2}\right]c_{-1}+\left(\epsilon^2-\upsilon^2\right)c_0\label{full series expansion equation}\\
+&&\left[-1+2\delta^2-\frac{\epsilon}{2}+\delta\upsilon\cos\gamma+\frac{\epsilon^2}{2}\right]c_1+\sum_{n=2}^{+\infty}(-1)^{n+1}\left(2n\delta^2+\delta\upsilon\cos\gamma\right)c_n=0;\nonumber\\
\nonumber\\
&&\text{if}\,\,m\geq 1,\nonumber\\
&&\left[\left(m-2\right)-\frac{\epsilon}{2}\right]c_{m-2}\nonumber\\
+&&\left[\left(3-2\delta^2\right)\left(m-1\right)-\frac{3\epsilon}{2}-2\left(m-1\right)\delta^2\sum_{n=1}^{+\infty}\left(\frac{\epsilon}{2m-1}\right)^n-\frac{\upsilon}{2m-1}\left(\left(4m-3\right)\delta\cos\gamma+\upsilon\right)\sum_{n=0}^{+\infty}\left(\frac{\epsilon}{2m-1}\right)^n\right]c_{m-1}\nonumber\\
+&&\left[\left(3-2\delta^2\right)m-\frac{3\epsilon}{2}-2m\delta^2\sum_{n=1}^{+\infty}\left(\frac{\epsilon}{2m-1}\right)^n-\frac{\upsilon}{2m-1}\left(\left(4m-1\right)\delta\cos\gamma+\upsilon\right)\sum_{n=0}^{+\infty}\left(\frac{\epsilon}{2m-1}\right)^n\right]c_{m}\nonumber\\
+&&\left[\left(m+1\right)-\frac{\epsilon}{2}\right]c_{m+1}=0.\nonumber
\end{eqnarray}

As $\epsilon, \upsilon\to 0$, the leading order asymptotic solution for $\{c_n\}$ of the above system of linear equations \eqref{full series expansion equation} depends only on the leading order of the coefficients before $\{c_n\}$. By keeping only the leading term for each coefficient, we obtain that the leading order solution of \eqref{full series expansion equation} for $\{c_n\}$ satisfies the following infinite system of linear equations:
\begin{equation}\label{asymptotic matrix}
\begin{pmatrix}
\ddots &\vdots &\vdots  & \vdots  &\vdots  &\vdots  &\vdots  &\vdots  & \iddots\\
\cdots  & -3(3-2\delta^2) & -2(3-2\delta^2) & -1 & 0 & 0 & 0 & 0  & \cdots\\
\cdots & -3 & -2(3-2\delta^2) & -(3-2\delta^2) & -\frac{\epsilon}{2} & 0 & 0 & 0 &\cdots \\
\cdots & 0 & -2 & -(3-2\delta^2)  & -\frac{3\epsilon}{2}-\delta\upsilon\cos\gamma & 1 & 0 & 0 &\cdots\\
\cdots   &6\delta^2 & -4\delta^2  & -1+2\delta^2 & \epsilon^2-\upsilon^2 & -1+2\delta^2 & -4\delta^2 & 6\delta^2  & \cdots\\
\cdots    &0 & 0 & -1 & -\frac{3\epsilon}{2}-\delta\upsilon\cos\gamma & 3-2\delta^2 & 2 & 0  & \cdots\\
\cdots  &0 & 0 & 0 & -\frac{\epsilon}{2} & 3-2\delta^2 & 2(3-2\delta^2) & 3 & \cdots\\
\cdots &0 & 0 & 0 & 0 & 1 & 2(3-2\delta^2) & 3(3-2\delta^2) & \cdots\\
\iddots  & \vdots  & \vdots &\vdots  &\vdots   &\vdots &\vdots  &\vdots  & \ddots
\end{pmatrix}
\begin{pmatrix}
 \vdots\\
 c_{-3}\\
 c_{-2}\\
  c_{-1} \\
  c_0 \\
  c_1 \\
  c_2\\
  c_3\\
  \vdots
  \end{pmatrix}
 =\begin{pmatrix}
 \vdots\\
  0 \\
   0 \\
    0 \\
  0 \\
  0 \\
  0 \\
   0 \\
  \vdots
 \end{pmatrix}.
\end{equation}
\end{widetext}

We will denote the infinite matrix in the above equation \eqref{asymptotic matrix} by $B$ and its elements by $b_{mn}$ with $-\infty<m, n<+\infty$. In order to have a nonzero solution, the determinant of $B$ must be zero. This gives us the dispersion relation that $\epsilon$ and $\upsilon$ must satisfy in the asymptotic regime $\epsilon, \upsilon\to 0$.

The determinant can be calculated by Laplace expansion:
\begin{equation}\label{Laplace expansion}
\operatorname{det}(B)=b_{00}M_{00}+\sum_{\substack{n=-\infty\\n\neq 0}}^{+\infty}(-1)^n b_{0n}M_{0n},
\end{equation}
where $M_{0n}$ is the $0, n$ minor of $B$, i.e. the infinite determinant that results from deleting the $0$th row and the $n$th column of B. Due to the symmetry property of $B$, we have that, for each $n\neq 0$,
\begin{equation}
b_{0n}=b_{0,-n}, \quad M_{0n}=-M_{0,-n},
\end{equation}
which implies that all the terms inside the summation symbol $\sum$ of \eqref{Laplace expansion} exactly cancel. Therefore, only the first term in \eqref{Laplace expansion} survive and thus we have that
\begin{equation}
\operatorname{det}(B)=M_{00}(\delta^2)\left(\epsilon^2-\upsilon^2\right)=0,
\end{equation}
which leads to
\begin{equation}
\epsilon^2=\upsilon^2,
\end{equation}
or equivalently
\begin{equation}
\omega^2=\mathbf{k}^2.
\end{equation}
This proves that the usual dispersion relation still holds for low frequency field modes. 

After setting $\epsilon^2=\upsilon^2$, we start solving the infinite system \eqref{asymptotic matrix}.

First, we rewrite \eqref{asymptotic matrix} as the following form:
\begin{equation}\label{moving c0 to right}
\sum_{\substack{n=-\infty\\n\neq 0}}^{+\infty}b_{mn}c_n=-b_{m0}c_0, \quad m=0, \pm1, \pm 2, \pm 3, \cdots
\end{equation}
Notice that the matrix elements of $B$ has the following symmetry properties:
\begin{eqnarray}
&&b_{mn}=-b_{-m, -n},\quad\text{if}\,m, n\neq 0\\
&&b_{m0}=b_{-m,0},\quad b_{0n}=b_{0,-n}.
\end{eqnarray}
The above symmetry properties leads to the following relation
\begin{equation}\label{antisymmetric solution}
c_n=-c_{-n}, \quad n\neq 0,
\end{equation}
which implies that we only need to solve $c_n$ for $n>0$ to solve the whole system.

For convenience, we define the following new variables $x_n$ by
\begin{equation}\label{xn definition}
c_n=\epsilon c_0 x_n, \quad n\neq 0.
\end{equation}
Then using the relation \eqref{antisymmetric solution}, the infinite system of linear equations \eqref{moving c0 to right} simplifies to the following infinite recurrence equations:
\begin{widetext}
\begin{eqnarray}
&&\left(4-2\delta^2\right)x_1+2x_2=\frac{3}{2}+\delta\cos\gamma,\label{first equation}\\
&&\left(3-2\delta^2\right)x_1+\left(3-2\delta^2\right)2x_2+3x_3=\frac{1}{2},\label{second equation}\\
&&(m-2)x_{m-2}+\left(3-2\delta^2\right)(m-1)x_{m-1}+\left(3-2\delta^2\right)mx_m+(m+1)x_{m+1}=0, \quad \text{if}\,\, m\geq 3, \label{recurrence equation}
\end{eqnarray}
\end{widetext}
where the dependence on $\epsilon$ in the equation \eqref{asymptotic matrix} or \eqref{moving c0 to right} has been eliminated by introducing the new variables $x_n, n\neq 0$ through \eqref{xn definition} and the solution for $x_n$ depends only on $\delta$.

In order to find the general formula for the sequence $\{x_m\}$, we define the following new variables:
\begin{equation}\label{ym definition}
y_m=(m-1)x_{m-1}+mx_m, \quad m\geq 3.
\end{equation}
Then the recurrence equations \eqref{recurrence equation} become
\begin{equation}\label{recurrence relation for y}
y_{m-1}+2(1-\delta^2)y_m+y_{m+1}=0, \quad  m\geq 3.
\end{equation}
Sequences satisfying \eqref{recurrence relation for y} must take the following form:
\begin{equation}\label{general formula for ym}
y_m=D\cos\left(m\vartheta+\psi\right), \quad  m\geq 3,
\end{equation}
where $D$ and $\psi$ are two constants and $\vartheta$ is determined by
\begin{equation}
\cos\vartheta=-1+\delta^2,\quad \sin\vartheta=\delta\sqrt{2-\delta^2}.
\end{equation}

Combining \eqref{general formula for ym} and \eqref{ym definition}, the general formula for $x_m$ can be obtained by iteration
\begin{widetext}
\begin{eqnarray}\label{xm solution}
x_m&&=\frac{1}{m}\left(D\sum_{n=3}^m (-1)^{m-n}\cos(n\vartheta+\psi)+(-1)^{m}2x_2\right)\\
&&=\frac{(-1)^m}{m}\left(-D\sec(\frac{\vartheta}{2})\sin\left(\frac{(m-2)\vartheta}{2}+\frac{m\pi}{2}\right)\sin\left(\frac{(m+3)\vartheta}{2}+\psi+\frac{m\pi}{2}\right)+2x_2\right), \quad m\geq 3.\nonumber
\end{eqnarray}

Replacing the $c_m$ in \eqref{expected toy form of solution} by $x_m$ through \eqref{xn definition} we obtain that, as $\epsilon\to 0$, the mode solution $u_{\mathbf{k}}(t, \mathbf{x})$ is asymptotic to
\begin{equation}\label{toy model asymptotic mode solution}
u_{\mathbf{k}}(t, \mathbf{x})=c_0 e^{-i\left(\omega t-\mathbf{k}\cdot\mathbf{x}\right)}\left(1+\epsilon\sum_{\substack{m=-\infty\\m\neq 0}}^{+\infty}x_me^{i2m\left(\Omega t+\mathbf{K}\cdot\mathbf{x}\right)}\right),
\end{equation}
where $x_m$ is determined by \eqref{antisymmetric solution}, \eqref{xn definition}, \eqref{first equation}, \eqref{second equation}, \eqref{recurrence equation} and \eqref{xm solution}.
\end{widetext}

Using the orthogonality of $e^{i2m\left(\Omega t+\mathbf{K}\cdot\mathbf{x}\right)}$, the relative magnitude of the correction to $u_{\mathbf{k}}$ from the usual plane wave mode $e^{-i(\omega t-\mathbf{k}\cdot\mathbf{x})}$ in Minkowski spacetime can be characterized by applying Parseval's identity:
\begin{equation}\label{parseval}
|\Delta u_{\mathbf{k}}(t, \mathbf{x})|=\epsilon\left(\sum_{\substack{m=-\infty\\m\neq 0}}^{+\infty}x_m^2\right)^{\frac{1}{2}}.
\end{equation}
From the solution \eqref{xm solution} we know that as $m\to\infty$,
\begin{equation}
x_m^2\sim\frac{1}{m^2}.
\end{equation}
Thus the summation inside the bracket of \eqref{parseval} converges and the correction
\begin{equation}\label{Delta u}
|\Delta u_{\mathbf{k}}(t, \mathbf{x})|\sim\epsilon\to 0, \quad\text{as}\,\,\, \epsilon\to 0.
\end{equation}
Thus we have demonstrated that the low frequency wave modes ($\omega\leq\Lambda$) are almost not affected by the fluctuating spacetime with much higher frequency ($\Omega\sim\sqrt{G}\Lambda^2$).

\subsection{General case}
The methods used and results obtained in the last subsection for the particular simplified toy model \eqref{toy wave equation} can be generalized to the generic case \eqref{eq:fullexpansion}. To start, we rewrite \eqref{eq:fullexpansion} to the following form:
\begin{equation}\label{quasiperiodic wave equation}
\left(1+f_1\right)\ddot{\phi}-\nabla^2\phi-\Omega_0f_2\dot{\phi}+K_0\mathbf{f}_3\cdot\nabla\phi=0,
\end{equation}
where
\begin{eqnarray}
&&f_1=\frac{A_0^2}{\Omega}\cos^2\Theta-1,\label{f1 definition}\\
&&f_2=\frac{3A_0^2}{2}\left(\frac{\dot{\Omega}}{\Omega^2}\cos^2\Theta+\sin2\Theta\right)/\Omega_0,\label{f2 definition}\\
&&\mathbf{f}_3=\left(\frac{\nabla\Omega}{2\Omega}+\tan\Theta\nabla\Theta\right)/K_0,\label{f3 definition}\\
&&\Omega_0=\left\langle\Omega\right\rangle, \quad K_0=\left\langle|\nabla\Theta|\right\rangle.
\end{eqnarray}
For convenience, we choose the constant $A_0$ such that the average of $f_1$
\begin{equation}\label{setting A0}
\left\langle f_1(t, \mathbf{x})\right\rangle=0.
\end{equation}

Unlike the toy model \eqref{toy wave equation} we used in the last subsection, \eqref{quasiperiodic wave equation} is not strictly periodic. However, \eqref{quasiperiodic wave equation} is quasiperiodic and its quasiperiod is the same as the period of \eqref{toy wave equation}. This property is reflected in the Fourier transforms $f_1(\omega, \mathbf{k})$, $f_2(\omega, \mathbf{k})$ and $\mathbf{f}_3(\omega, \mathbf{k})$ of the functions $f_1(t, \mathbf{x})$, $f_2(t, \mathbf{x})$ and $\mathbf{f}_3(t, \mathbf{x})$ respectively which are defined by
\begin{eqnarray}
&&f_1(t, \mathbf{x})=\int d\omega d^3k\,f_1(\omega, \mathbf{k})e^{i(\omega t+\mathbf{k}\cdot\mathbf{x})},\label{f1 fourier transform}\\
&&f_2(t, \mathbf{x})=\int d\omega d^3k\,f_2(\omega, \mathbf{k})e^{i(\omega t+\mathbf{k}\cdot\mathbf{x})},\label{f2 fourier transform}\\
&&\mathbf{f}_3(t, \mathbf{x})=\int d\omega d^3k\,\mathbf{f}_3(\omega, \mathbf{k})e^{i(\omega t+\mathbf{k}\cdot\mathbf{x})}.\label{f3 fourier transform}
\end{eqnarray}

For the function $f_1(t, \mathbf{x})$ defined by \eqref{f1 definition}, after setting the constant $A_0$ by \eqref{setting A0} and considering the slow varying property of $\Omega(t, \mathbf{x})$ and $\Theta(t, \mathbf{x})$ in both temporal and spatial directions, its leading order goes as
\begin{equation}\label{f1 sim}
f_1(t, \mathbf{x})\sim\cos2\Theta,
\end{equation}
which implies that the Fourier transform $f_1(\omega, \mathbf{k})$ would have two peaks centered at
\begin{equation}\label{f1 center}
\omega=\pm2\Omega_0,\quad |\mathbf{k}|=2K_0.
\end{equation}

For the function $f_2(t, \mathbf{x})$ defined by \eqref{f2 definition}, the second term which includes the factor $\sin2\Theta$ is dominant since the first term which includes the factor $\dot{\Omega}/\Omega^2$ goes as $\sim1/\Lambda\to 0$ due to the slow varying condition described by \eqref{omega expectation value} and \eqref{d omega expectation value}. Thus, its leading order goes as
\begin{equation}\label{f2 sim}
f_2(t, \mathbf{x})\sim 3\sin2\Theta,
\end{equation}
which implies that the Fourier transform $f_2(\omega, \mathbf{k})$ would also have two peaks centered at
\begin{equation}\label{f2 center}
\omega=\pm2\Omega_0,\quad |\mathbf{k}|=2K_0.
\end{equation}

Similarly, for the function $\mathbf{f}_3(t, \mathbf{x})$ defined by \eqref{f3 definition}, the second term which includes the factor $\tan\Theta$ is dominant since the absolute value of the first term which includes the factor $\nabla\Omega/(\Omega K_0)$ also goes as $\sim 1/\Lambda\to 0$ due to the slow varying property of $\Omega$ in spatial directions. Thus, its leading order goes as
\begin{equation}\label{f3 sim}
\mathbf{f}_3(t, \mathbf{x})\sim\tan\Theta\frac{\nabla\Theta}{K_0}.
\end{equation}
Then using the Fourier series expansion \eqref{tanx fourier} for $\tan\Theta$, we know that the Fourier transform $\mathbf{f}_3(\omega, \mathbf{k})$ would have infinitely many peaks centered at
\begin{equation}\label{f3 center}
\omega=\pm2n\Omega_0, \quad|\mathbf{k}|=2nK_0,\quad n=1, 2, 3, \cdots.
\end{equation}
(For a rough calculation of the above Fourier transforms, see Appendix \ref{appendix 3})

In addition, we have the zero frequency component (see \eqref{zero frequency} in Appendix \ref{appendix 3})
\begin{equation}\label{zero for zero frequency}
f_i(\omega= 0, \mathbf{k}= 0)\sim 0, \quad i=1, 2, 3.
\end{equation}

In summary, the system described by \eqref{quasiperiodic wave equation} is very similar to the system described by the simplified toy model \eqref{toy wave equation}. The only difference is that the Fourier transforms of the coefficients $f_1$, $f_2$, and $\mathbf{f}_3$ in \eqref{quasiperiodic wave equation} spread around center points given by \eqref{f1 center}, \eqref{f2 center} and \eqref{f3 center} while the Fourier transforms of the corresponding coefficients in \eqref{toy wave equation} are ideal delta functions exactly located at same points given by \eqref{f1 center}, \eqref{f2 center} and \eqref{f3 center}.

Therefore, the mode solution of \eqref{quasiperiodic wave equation} would take the form similar to \eqref{expected toy form of solution}:
\begin{widetext}
\begin{equation}\label{general form of solution}
u_{\mathbf{k}}(t, \mathbf{x})=e^{-i(\omega t-\mathbf{k}\cdot\mathbf{x})}\left(c_0+\int_{\substack{\omega'\neq 0\\
\mathbf{k}'\neq\mathbf{0}}} d\omega'd^3k' \, u_{\mathbf{k}}(\omega', \mathbf{k}')e^{i(\omega' t+\mathbf{k}'\cdot\mathbf{x})}\right),
\end{equation}
where $u_{\mathbf{k}}(\omega', \mathbf{k}')$ is non-negligible only when $\omega', \mathbf{k}'$ are taking values around the centers given by \eqref{f1 center}, \eqref{f2 center} and \eqref{f3 center}.

Inserting \eqref{general form of solution} into \eqref{quasiperiodic wave equation} and replacing the coefficients $f_1(t, \mathbf{x})$, $f_2(t, \mathbf{x})$ and $\mathbf{f}_3(t, \mathbf{x})$ in \eqref{quasiperiodic wave equation} by the equations \eqref{f1 fourier transform}, \eqref{f2 fourier transform} and \eqref{f3 fourier transform} and then using the orthogonality of $e^{i(\omega' t+\mathbf{k}'\cdot\mathbf{x})}$, we obtain the following uncountably infinite system of linear equations which are similar to \eqref{original infinite system of linear equations}:
\begin{eqnarray}\label{general original infinite system of linear equations}
(\omega', \mathbf{k}')\text{th equation}:\quad&&\left[\left(\omega-\omega'\right)^2-\left(\mathbf{k}+\mathbf{k}'\right)^2\right]u_{\mathbf{k}}\left(\omega', \mathbf{k}'\right)\nonumber\\
+&&\int d\omega''d^3k''\Big[\left(\omega-\left(\omega'-\omega''\right)\right)^2f_1\left(\omega'', \mathbf{k}''\right)-i\Omega_0\left(\omega-\left(\omega'-\omega''\right)\right)f_2\left(\omega'', \mathbf{k}''\right)\nonumber\\
-&&iK_0\left(\mathbf{k}+\left(\mathbf{k}'-\mathbf{k}''\right)\right)\cdot\mathbf{f}_3\left(\omega'', \mathbf{k}''\right)\Big]u_{\mathbf{k}}\left(\omega'-\omega'', \mathbf{k}'-\mathbf{k}''\right)=0,
\end{eqnarray}
where we have defined the notation $u_{\mathbf{k}}(0, \mathbf{0})=c_0\delta(0, \mathbf{0})$ for convenience. 

To characterize the property of the solutions of this system more clearly, we define the following parameters similar to \eqref{parameters definition} for convenience:
\begin{equation}
\epsilon=\frac{\omega}{\Omega_0}, \quad \upsilon=\frac{|\mathbf{k}|}{\Omega_0}, \quad \delta=\frac{K_0}{\Omega_0}, \quad\cos\gamma=\frac{\mathbf{k}\cdot\mathbf{k}'}{|\mathbf{k}||\mathbf{k}'|}, \quad \cos\mu=\frac{\mathbf{k}\cdot\mathbf{f}_3}{|\mathbf{k}||\mathbf{f}_3|}, \quad \cos\mu'=\frac{\mathbf{k}'\cdot\mathbf{f}_3}{|\mathbf{k}'||\mathbf{f}_3|},\quad \cos\mu''=\frac{\mathbf{k}''\cdot\mathbf{f}_3}{|\mathbf{k}''||\mathbf{f}_3|}.
\end{equation}

Dividing both sides of \eqref{general original infinite system of linear equations} by $\Omega_0^2$ gives
\begin{eqnarray}\label{general system divide by Omega square}
(\omega', \mathbf{k}')\text{th equation}:\quad&&\left[\left(\epsilon-\frac{\omega'}{\Omega_0}\right)^2-\left(\upsilon^2+\frac{\mathbf{k}'^2}{\Omega_0^2}+2\upsilon\frac{|\mathbf{k}'|}{\Omega_0}\cos\gamma\right)\right]u_{\mathbf{k}}\left(\omega', \mathbf{k}'\right)\nonumber\\
+&&\int d\omega''d^3k''\Bigg[\left(\epsilon-\left(\frac{\omega'}{\Omega_0}-\frac{\omega''}{\Omega_0}\right)\right)^2f_1\left(\omega'', \mathbf{k}''\right)-i\left(\epsilon-\left(\frac{\omega'}{\Omega_0}-\frac{\omega''}{\Omega_0}\right)\right)f_2\left(\omega'', \mathbf{k}''\right)\nonumber\\
-&&i\delta\left(\upsilon\cos\mu+\left(\frac{|\mathbf{k}'|}{\Omega_0}\cos\mu'-\frac{|\mathbf{k}''|}{\Omega_0}\cos\mu''\right)\right)|\mathbf{f}_3\left(\omega'', \mathbf{k}''\right)|\Bigg]u_{\mathbf{k}}\left(\omega'-\omega'', \mathbf{k}'-\mathbf{k}''\right)=0.
\end{eqnarray}

Similar to the toy model case, as $\epsilon, \upsilon\to 0$, the leading order solution of \eqref{general system divide by Omega square} for $u_{\mathbf{k}}(\omega', \mathbf{k}')$ satisfies the following uncountably infinite system of linear equations:
\begin{eqnarray}
\text{if}\,\,  (\omega', \mathbf{k}')=(0, \mathbf{0}):\quad&&\left(\epsilon^2-\upsilon^2\right)\delta\left(0, \mathbf{0}\right)c_0\nonumber\\
+&&\int d\omega''d^3k''\Bigg[\left(\frac{\omega''}{\Omega_0}\right)^2f_1\left(\omega'', \mathbf{k}''\right)-i\left(\frac{\omega''}{\Omega_0}\right)f_2\left(\omega'', \mathbf{k}''\right)\nonumber\\
+&&i\delta\left(\frac{|\mathbf{k}''|}{\Omega_0}\cos\mu''\right)|\mathbf{f}_3\left(\omega'', \mathbf{k}''\right)|\Bigg]u_{\mathbf{k}}\left(-\omega'', -\mathbf{k}''\right)=0,\nonumber\\
\text{if}\,\,  (\omega', \mathbf{k}')\neq(0, \mathbf{0}): \quad
&&\left(-i\epsilon f_2(\omega', \mathbf{k}')-i\delta\upsilon\cos\mu |\mathbf{f}_3(\omega', \mathbf{k}')| \right)c_0\nonumber\\
+&&\left[\left(\frac{\omega'}{\Omega_0}\right)^2-\left(\frac{\mathbf{k}'}{\Omega_0}\right)^2\right]u_{\mathbf{k}}\left(\omega', \mathbf{k}'\right)\nonumber\\
+&&\int_{\substack{\omega''\neq\omega'\\\mathbf{k}''\neq\mathbf{k}'}} d\omega''d^3k''\Bigg[\left(\frac{\omega'}{\Omega_0}-\frac{\omega''}{\Omega_0}\right)^2f_1\left(\omega'', \mathbf{k}''\right)+i\left(\frac{\omega'}{\Omega_0}-\frac{\omega''}{\Omega_0}\right)f_2\left(\omega'', \mathbf{k}''\right)\nonumber\\
-&&i\delta\left(\frac{|\mathbf{k}'|}{\Omega_0}\cos\mu'-\frac{|\mathbf{k}''|}{\Omega_0}\cos\mu''\right)|\mathbf{f}_3\left(\omega'', \mathbf{k}''\right)|\Bigg]u_{\mathbf{k}}\left(\omega'-\omega'', \mathbf{k}'-\mathbf{k}''\right)=0, \label{general aysmptotic equation}
\end{eqnarray}
where we have used the property \eqref{zero for zero frequency} in obtaining \eqref{general aysmptotic equation} from \eqref{general system divide by Omega square}.

The above uncountably infinite system of linear equations \eqref{general aysmptotic equation} can also be written formally in matrix form similar to \eqref{asymptotic matrix}. We use similar notations that denoting the matrix here by $B$ and its elements by $b_{(\omega', \mathbf{k}'), (\omega'', \mathbf{k}'')}$ for convenience.

In order to have nonzero solutions, the determinant of the uncountably infinite matrix $B$ has to be zero, which gives the dispersion relations that $\epsilon$ and $\upsilon$ must be satisfied in the asymptotic region $\epsilon, \upsilon\to 0$.

The ``determinant'' of $B$ can be formally calculated through Laplace expansion similar to \eqref{Laplace expansion}:
\begin{equation}\label{generalized Laplace expansion}
\operatorname{det}B=b_{(0, \mathbf{0}), (0, \mathbf{0})}M_{(0, \mathbf{0}), (0, \mathbf{0})}+\int_{\substack{\omega''\neq 0\\\mathbf{k}''\neq 0}} d\omega''d^3k'' (-1)^{(\omega'', \mathbf{k}'')} b_{(0, \mathbf{0}), (\omega'', \mathbf{k}'')}M_{(0, \mathbf{0}), (\omega'', \mathbf{k}'')},
\end{equation}
where $M_{(0, \mathbf{0}), (\omega'', \mathbf{k}'')}$ is the $(0, \mathbf{0}), (\omega'', \mathbf{k}'')$ minor of $B$, i.e. the `determinant' resulting from deleting the $(0, \mathbf{0})$th row and $(\omega'', \mathbf{k}'')$th column of $B$.

Notice that since $f_1(t, \mathbf{x})$, $f_2(t, \mathbf{x})$ and $\mathbf{f}_3(t, \mathbf{x})$ are all real, their Fourier transforms $f_1(\omega, \mathbf{k})$, $f_2(\omega, \mathbf{k})$ and $\mathbf{f}_3(\omega, \mathbf{k})$ defined by \eqref{f1 fourier transform}, \eqref{f2 fourier transform} and \eqref{f3 fourier transform} must satisfy the following relations:
\begin{equation}\label{symmetric relation between fourier components}
f_1(\omega, \mathbf{k})=f_1(-\omega, -\mathbf{k})^*, \quad f_2(\omega, \mathbf{k})=f_2(-\omega, -\mathbf{k})^*, \quad \mathbf{f}_3(\omega, \mathbf{k})=\mathbf{f}_3(-\omega, -\mathbf{k})^*,
\end{equation}
where the $*$ means complex conjugate.

The above symmetry property \eqref{symmetric relation between fourier components} leads to
\begin{equation}
b_{(0, \mathbf{0}), (\omega'', \mathbf{k}'')}=b_{(0, \mathbf{0}), (-\omega'', -\mathbf{k}'')}, \quad M_{(0, \mathbf{0}), (\omega'', \mathbf{k}'')}=-M_{(0, \mathbf{0}), (-\omega'', -\mathbf{k}'')}, \quad\text{if}\quad (-\omega'', -\mathbf{k}'')\neq(0, \mathbf{0}),
\end{equation}
which implies that all the terms inside the integral symbol $\int$ of \eqref{generalized Laplace expansion} exactly cancel. Therefore, only the first term in \eqref{generalized Laplace expansion} survives and thus we have
\begin{equation}
\operatorname{det}B=M_{(0, \mathbf{0}), (0, \mathbf{0})}(\epsilon^2-\upsilon^2)=0,
\end{equation}
which gives again the usual dispersion relation
\begin{equation}\label{general dispersion relation}
\epsilon^2=\upsilon^2  \quad\text{or}\quad \omega^2=\mathbf{k}^2.
\end{equation}

After setting the dispersion relation \eqref{general dispersion relation}, we only need to solve the $(\omega', \mathbf{k}')\neq(0, \mathbf{0})$th equations in \eqref{general aysmptotic equation} since $\operatorname{det}B=0$ implies that the $(\omega', \mathbf{k}')=(0, \mathbf{0})$th equation is redundant.

For convenience, we define new variables $x_{\mathbf{k}}(\omega', \mathbf{k}')$ similar to the $x_n$ defined in \eqref{xn definition}:
\begin{equation}\label{xk definition}
u_{\mathbf{k}}(\omega', \mathbf{k}')=\epsilon c_0 x_{\mathbf{k}}(\omega', \mathbf{k}'), \quad  (\omega', \mathbf{k}')\neq(0, \mathbf{0}).
\end{equation}
Then \eqref{general aysmptotic equation} can be rewritten as
\begin{eqnarray}
\text{if}\,\,  (\omega', \mathbf{k}')\neq(0, \mathbf{0}): \quad &&\left[\left(\frac{\omega'}{\Omega_0}\right)^2-\left(\frac{\mathbf{k}'}{\Omega_0}\right)^2\right]x_{\mathbf{k}}\left(\omega', \mathbf{k}'\right)\nonumber\\
+&&\int_{\substack{\omega''\neq\omega'\\\mathbf{k}''\neq\mathbf{k}'}} d\omega''d^3k''\Bigg[\left(\frac{\omega'}{\Omega_0}-\frac{\omega''}{\Omega_0}\right)^2f_1\left(\omega'', \mathbf{k}''\right)+i\left(\frac{\omega'}{\Omega_0}-\frac{\omega''}{\Omega_0}\right)f_2\left(\omega'', \mathbf{k}''\right)\nonumber\\
-&&i\delta\left(\frac{|\mathbf{k}'|}{\Omega_0}\cos\mu'-\frac{|\mathbf{k}''|}{\Omega_0}\cos\mu''\right)|\mathbf{f}_3\left(\omega'', \mathbf{k}''\right)|\Bigg]x_{\mathbf{k}}\left(\omega'-\omega'', \mathbf{k}'-\mathbf{k}''\right) \nonumber\\
=&& i f_2(\omega', \mathbf{k}')+i\delta\cos\mu |\mathbf{f}_3(\omega', \mathbf{k}')|.\label{general equation without epsilon}
\end{eqnarray}

Replacing the $u_{\mathbf{k}}(\omega', \mathbf{k}')$ in \eqref{general form of solution} by $x_{\mathbf{k}}(\omega', \mathbf{k}')$ through \eqref{xk definition} we obtain that, as $\epsilon\to 0$, the mode solution $u_{\mathbf{k}}(t, \mathbf{x})$ is asymptotic to
\begin{equation}\label{general uk}
u_{\mathbf{k}}(t, \mathbf{x})=c_0e^{-i(\omega t-\mathbf{k}\cdot\mathbf{x})}\left(1+\epsilon\int_{\substack{\omega'\neq 0\\\mathbf{k}'\neq 0}}d\omega'd^3k'\, x_{\mathbf{k}}(\omega', \mathbf{k}')e^{i(\omega' t+\mathbf{k}'\cdot\mathbf{x})}\right),
\end{equation}
where $x_{\mathbf{k}}(\omega', \mathbf{k}')$ is determined by \eqref{general equation without epsilon}.

\end{widetext}

Analogous to \eqref{xm solution} in the simplified toy model, $x_{\mathbf{k}}(\omega', \mathbf{k}')$ would also go as
\begin{equation}\label{general xm asymptotics}
x_{\mathbf{k}}(\omega', \mathbf{k}')\sim\frac{1}{m},
\end{equation}
when $\omega', \mathbf{k}'$ taking values around the centers
\begin{equation}
\omega'\sim \pm 2m\Omega_0, \quad |\mathbf{k}'|\sim 2mK_0, \quad   m=1, 2, 3, \cdots
\end{equation}
($x_{\mathbf{k}}(\omega', \mathbf{k}')$ is negligible if $\omega', \mathbf{k}'$ is far away from these centers).

Due to Parseval's theorem, \eqref{general xm asymptotics} implies that the integral inside the bracket of \eqref{general uk} converges which is similar to \eqref{parseval} and thus the correction to $u_{\mathbf{k}}(t, \mathbf{x})$ also goes as $\epsilon$.

Therefore, when we quantize the scalar field $\phi$ in our wildly fluctuating spacetime by expanding it in terms of the annihilation and creation operators according to \eqref{phi mode expansion}, the leading order would still be the form of the Minkowski quantum field expansion \eqref{field expansion}. The correction to the dispersion relation $\omega^2=\mathbf{k}^2$ and the plane wave mode $e^{-i(\omega t-\mathbf{k}\cdot\mathbf{x})}$ are on the order $\sim\epsilon$. In addition, the extra wave modes which mixing in \eqref{toy model asymptotic mode solution} or \eqref{general uk} are all modes with frequencies higher than $\Omega_0\sim\sqrt{G}\Lambda^2$, which is much larger than our effective QFT's cutoff $\Lambda$. These extremely high frequency modes beyond the cutoff are irrelevant to our low energy physics. This also explains why the ordinary QFT works by assuming fixed Minkowski spacetime. The small scale structure averages out in its effect on the long wavelength low energy fields.

In summary, we have argued that although our spacetime sourced by the quantum vacuum is highly curved and wildly fluctuating, the back reaction of the resulting spacetime on the quantum field sitting on it is small. This justifies our method of neglecting back reaction and using the quantum field expansion \eqref{field expansion} in Minkowski spacetime at the beginning.

\section{the more general inhomogeneous metrics}\label{general coordinate}\label{sec viii}
In previous sections we assume the simplest inhomogeneous metric \eqref{inhomogenous FLRW coordinate} to describe the spacetime resulting from the inhomogeneous vacuum. In this section, we try to generalize the result to more general inhomogeneous metrics.

The quantum fluctuations of the vacuum is not completely arbitrary, the magnitude of the fluctuations must be the same everywhere and in every direction, i.e. the spacetime is still stochastically homogeneous and isotropic. Thus we can always choose a special gauge and construct the following general synchronous coordinate:
\begin{equation}\label{synchronized coordinate}
ds^2=-dt^2+h_{ab}(t,\mathbf{x})dx^adx^b,\quad a, b=1,2,3.
\end{equation}

For the above metric \eqref{synchronized coordinate}, we employ the initial value formulation of general relativity. In this formulation, the Einstein equation is equivalent to six equations for the evolution of the second fundamental form
\begin{equation}\label{evolution equation}
\begin{split}
\dot{k}_{ab}=&-R_{ab}^{(3)}-(tr k)k_{ab}+2k_{ac}k^c_b\\
+&4\pi G\rho h_{ab}+8\pi G\left(T_{ab}-\frac{1}{2}h_{ab}tr T\right),
\end{split}\end{equation}
plus the usual four constraint equations,
\begin{equation}\label{Hamiltonian constraint}
R^{(3)}+(trk)^2-k_{ab}k^{ab}=16\pi G\rho,
\end{equation}
\begin{equation}\label{momentum constraint}
D_ak^a_{b}-D_b(trk)=8\pi G j_b,
\end{equation}
where $k_{ab}=\frac{1}{2}\dot{h}_{ab}$, $k^{ab}=h^{ac}h^{bd}k_{cd}$, $tr k=h^{ab}k_{ab}$, $\rho=T_{00}$, $j_b=h_b^aT_{0a}$, $trT=h^{ab}T_{ab}$, $R^{(3)}$ is the 3-dimensional spatial curvature and $D_a$ is the derivative operator associated with $h_{ab}$.

Taking trace on both sides of \eqref{evolution equation} and then combining with \eqref{Hamiltonian constraint} gives:
\begin{equation}\label{dynamical equaiton}
h^{ab}\dot{k}_{ab}-k_{ab}k^{ab}=-4\pi G\left(\rho+trT\right).
\end{equation}
It is interesting to notice that there are no spatial derivatives included on the left hand of the above equation \eqref{dynamical equaiton}. The key evolution equation \eqref{hmo} for $a(t, \mathbf{x})$ we used in previous sections is just the special case of the above equation \eqref{dynamical equaiton}.

Direct calculation using the expression \eqref{stress energy tensor for a massless scalar field} shows that, the contribution from a real massless scalar field to the right-hand side of \eqref{dynamical equaiton} is
\begin{equation}\label{genearal rho plus 3p}
\rho+trT=2\dot{\phi}^2,
\end{equation}
where all the spatial derivatives of $\phi$ and all the explicit dependence on the metric $g_{\mu\nu}$ in the definition of stress energy tensor \eqref{stress energy tensor for a massless scalar field} are canceled. It is also interesting to notice that the above exact expression \eqref{genearal rho plus 3p} is exactly the same with the corresponding expression \eqref{massless field contribution} for the simplest inhomogeneous metric \eqref{inhomogenous FLRW coordinate} case. 

We first consider the following special case:
\begin{equation}\label{metric 2}
h_{ab}(t,\mathbf{x}) = 
 \begin{pmatrix}
  a^2(t,\mathbf{x}) & 0 & 0 \\
  0 & b^2(t,\mathbf{x}) & 0 \\
  0 & 0 & c^2(t,\mathbf{x})
 \end{pmatrix}.
\end{equation}
The spacetime described by the above coordinate \eqref{metric 2} possesses more freedoms than \eqref{inhomogenous FLRW coordinate} and thus would exhibit richer structures. In this case, the expansion rate at the same point becomes directionally dependent. Along the three principle axes $\hat{x}$, $\hat{y}$ and $\hat{z}$, which are eigenvectors of the symmetric matrix $h_{ab}$ in \eqref{metric 2}, the expansion rates $\dot{a}/a$, $\dot{b}/b$ and $\dot{c}/c$ can be different. This means that, at one same point, the space can be expanding in one or two directions and contracting on the other two or one directions. 

Under the coordinate system \eqref{metric 2}, equation \eqref{dynamical equaiton} becomes
\begin{equation}\label{generalized hmo}
\frac{\ddot{a}}{a}+\frac{\ddot{b}}{b}+\frac{\ddot{c}}{c}=-4\pi G\left(\rho+trT\right),
\end{equation}
which is a generalization of the key evolution equation \eqref{hmo} we used in previous sections.

Let 
\begin{equation}
\frac{\ddot{a}}{a}=-\Omega_1^2(t,\mathbf{x}), \quad\frac{\ddot{b}}{b}=-\Omega_2^2(t,\mathbf{x}),\quad \frac{\ddot{c}}{c}=-\Omega_3^2(t,\mathbf{x}),
\end{equation}
then \eqref{generalized hmo} immediately leads to
\begin{equation}
\Omega_1^2(t,\mathbf{x})+\Omega_2^2(t,\mathbf{x})+\Omega_3^2(t,\mathbf{x})=4\pi G\left(\rho+trT\right).
\end{equation}
As the functions $a$, $b$ and $c$ are alternately symmetric, their expectation values must be equal
\begin{equation}
\left\langle\Omega_i^2(t,\mathbf{x})\right\rangle=\frac{4\pi G}{3}\left\langle\rho+trT\right\rangle, \quad i=1,2,3.
\end{equation}

Unlike equation \eqref{hmo}, here $\Omega_i^2(t,\mathbf{x})$ does not necessarily go exactly the same as $\frac{4\pi G}{3}\left(\rho+trT\right)$. Since $4\pi G(\rho+trT)$ is slowly varying, $\Omega_i^2$ must also be slowly varying functions. Otherwise we would have two or three fast varying functions sum together and precisely cancel each other to give a slowly varying function, which is almost impossible in the system with such huge quantum fluctuations. Thus the evolution of $a$, $b$ and $c$ are still adiabatic processes and the conclusion we obtained in previous sections still holds. We can obtain solutions similar to \eqref{expected form of solution} that
\begin{eqnarray}
a(t, \mathbf{x})&\simeq& e^{\int_0^t H_{1\mathbf{x}}(t')dt'}P_1(t, \mathbf{x}),\label{a}\\
b(t, \mathbf{x})&\simeq& e^{\int_0^t H_{2\mathbf{x}}(t')dt'}P_2(t, \mathbf{x}),\label{b}\\
c(t, \mathbf{x})&\simeq& e^{\int_0^t H_{3\mathbf{x}}(t')dt'}P_3(t, \mathbf{x}),\label{c}
\end{eqnarray}
and on average
\begin{equation}\label{Hi dependence Lambda}
H_i=\alpha\Lambda e^{-\beta\sqrt{G}\Lambda},\quad  i=1,2,3,
\end{equation}
where
\begin{equation}
H_i=\frac{1}{t}\int_0^t H_{i\mathbf{x}}(t')dt'.
\end{equation}
Therefore, the observable physical volume is,
\begin{equation}
V(t)=\int\sqrt{h(t, \mathbf{x})}d^3x=V(0)e^{3Ht},
\end{equation}
where $h=\det h_{ab}=a^2b^2c^2$.

Next we investigate the most general case
\begin{equation}\label{most genearl metric}
h_{ab}(t,\mathbf{x}) = 
 \begin{pmatrix}
  a^2(t,\mathbf{x}) & d(t,\mathbf{x}) & e(t,\mathbf{x}) \\
  d(t,\mathbf{x}) & b^2(t,\mathbf{x}) & f(t,\mathbf{x}) \\
  e(t,\mathbf{x}) & f(t,\mathbf{x}) & c^2(t,\mathbf{x}) 
 \end{pmatrix}.
\end{equation}

In this case, the three orthogonal eigenvectors of the symmetric matrix $h_{ab}$ can rotate in space. This gives more freedom and structure to the spacetime evolution than in the case described by the coordinate system \eqref{metric 2}. For example, an initial sphere will distort toward an ellipsoid with principle axes given by eigenvectors of $h_{ab}$, with rates given by time derivatives $\dot{\lambda}_i/\lambda_i$ of the corresponding eigenvalues $\lambda_i^2(t, \mathbf{x}), i=1,2,3$.

Expanding the dynamic equation \eqref{dynamical equaiton} using the metric \eqref{most genearl metric} gives
\begin{widetext}
\begin{equation}\label{genearl metric expansion}
\frac{a^2h^{\ast}_{11}}{h}\frac{\ddot{a}}{a}+\frac{b^2h^{\ast}_{22}}{h}\frac{\ddot{b}}{b}+\frac{c^2h^{\ast}_{11}}{h}\frac{\ddot{c}}{c}+\frac{dh^{\ast}_{12}}{h}\frac{\ddot{d}}{d}+\frac{e h^{\ast}_{13}}{h}\frac{\ddot{e}}{e}+\frac{f h^{\ast}_{23}}{h}\frac{\ddot{f}}{f}+F(h_{ab},\dot{h}_{ab})=-4\pi G(\rho+trT),
\end{equation}
\end{widetext}
where $h=\mathrm{det}(h_{ab})$ is the determinant of the matrix \eqref{most genearl metric}, $h^{\ast}_{ab}$ is the matrix's $(a,b)$ cofactor and $F$ is a nonlinear function of the metric components $h_{ab}$ and their first time derivatives $\dot{h}_{ab}$. \eqref{genearl metric expansion} is difficult to handle. Further investigations are needed in the future.

However, the results we obtained for the coordinate \eqref{metric 2} suggest that, for the most general case \eqref{most genearl metric}, the eigenvalues $\lambda_i^2(t, \mathbf{x})$ should also evolve adiabatically similar to $a^2$, $b^2$ and $c^2$. In other words, we expect that the results \eqref{a}, \eqref{b}, \eqref{c} and \eqref{Hi dependence Lambda} can be generalized to $\lambda_i$ in the most general case and the physical volume of space would expand as
\begin{eqnarray}
V(t)=&&\int \sqrt{h(t, \mathbf{x})}d^3x\nonumber\\
=&&\int \sqrt{\lambda_1^2\lambda_2^2\lambda_3^2} d^3x\nonumber\\
=&&V(0)e^{3Ht},
\end{eqnarray}
where $H$ is determined by \eqref{dependence of H on Lambda}.

\section{Discussion}\label{sec ix}
So far, we have presented a new mechanism of vacuum gravitation and showed that it leads to a slow accelerating expansion instead of a catastrophic explosion of the Universe. In this section, we discuss some questions raised and a couple of new concepts suggested by this different way of vacuum gravitating.

\subsection{Lorentz invariant cutoffs}
A potential concern is that the cutoff $\Lambda$ we are using is not Lorentz invariant. However, the results would not change if using Lorentz invariant cutoffs instead. This is because the $\Lambda$ is just used for comparing the magnitude of different infinities, whose leading order dependencies on $\Lambda$ can also be obtained just by dimensional analysis. In previous sections, we have taken this simple non-Lorentz invariant cutoff $\Lambda$ for convenience. In this subsection, we use more complicated but Lorentz invariant Pauli-Villars type cutoffs to show directly that the results do not change.

First, we calculate the two point function
\begin{equation}\label{two point function}
\left\langle\phi(t_1, \mathbf{x})\phi(t_2, \mathbf{x})\right\rangle=\int\frac{d\omega d^3k}{(2\pi)^4}\frac{i}{\omega^2-\mathbf{k}^2+i\epsilon}e^{-i\omega\Delta t},
\end{equation}
where $\Delta t=t_1-t_2$. We then replace the photon propagator
\begin{equation}
\frac{1}{\omega^2-\mathbf{k}^2+i\epsilon}
\end{equation}
in \eqref{two point function} by
\begin{equation}
\frac{1}{\omega^2-\mathbf{k}^2+i\epsilon}-\frac{1}{\omega^2-\mathbf{k}^2-\Lambda_1^2+i\epsilon},
\end{equation}
where $\Lambda_1$ can be thought of as a fictitious heavy photon, which can serve as a Lorentz invariant cutoff.

Then \eqref{two point function} becomes
\begin{equation}\label{first pauli villars}
-i\Lambda_1^2\int\frac{d\omega d^3k}{(2\pi)^4}e^{-i\omega\Delta t}\int_0^1d\alpha_1\left(\frac{1}{\omega^2-\mathbf{k}^2-\alpha_1\Lambda_1^2+i\epsilon}\right)^2,
\end{equation}
where we have used the identity
\begin{equation}\label{integration identity}
\frac{1}{AB}=\int_0^1\frac{d\alpha}{\left(A+(B-A)\alpha\right)^2}.
\end{equation}

\eqref{first pauli villars} is still logarithmically divergent when setting $\Delta t=0$. To make it converge, we employ another Pauli-Villars type cutoff $\Lambda_2$ by replacing the
\begin{equation}\label{pauli villars technique1}
\frac{1}{\omega^2-\mathbf{k}^2-\alpha_1\Lambda_1^2+i\epsilon}
\end{equation}
in \eqref{first pauli villars} with
\begin{equation}\label{pauli villars technique2}
\frac{1}{\omega^2-\mathbf{k}^2-\alpha_1\Lambda_1^2+i\epsilon}-\frac{1}{\omega^2-\mathbf{k}^2-\alpha_1\Lambda_1^2-\Lambda_2^2+i\epsilon}.
\end{equation}

Then \eqref{first pauli villars} becomes
\begin{widetext}
\begin{equation}\label{second pauli villars}
-i\Lambda_1^2\Lambda_2^4\int\frac{d\omega d^3k}{(2\pi)^4}e^{-i\omega\Delta t}\int_0^1d\alpha_1\int_0^1d\alpha_2\int_0^1d\alpha_2'\frac{1}{(\omega^2-\mathbf{k}^2-\alpha_1\Lambda_1^2-\alpha_2\Lambda_2^2+i\epsilon)^2(\omega^2-\mathbf{k}^2-\alpha_1\Lambda_1^2-\alpha_2'\Lambda_2^2+i\epsilon)^2},
\end{equation}
where we have used the identity \eqref{integration identity} again in obtaining \eqref{second pauli villars}.

Performing the integration $\int d^3k$ first, \eqref{second pauli villars} becomes
\begin{eqnarray}\label{third pauli villars}
&&\frac{\Lambda_1^2\Lambda_2^4}{16\pi^2}\int_0^1d\alpha_1\int_0^1d\alpha_2\int_0^1d\alpha_2'\int d\omega e^{-i\omega\Delta t}\\&&\cdot\frac{1}{\sqrt{\omega^2-\tilde{\Lambda}^2(\alpha_1, \alpha_2)+i\epsilon}}\cdot\frac{1}{\sqrt{\omega^2-\tilde{\Lambda}'^2(\alpha_1, \alpha_2')+i\epsilon}}\cdot\frac{1}{\left(\sqrt{\omega^2-\tilde{\Lambda}^2(\alpha_1, \alpha_2)+i\epsilon}+\sqrt{\omega^2-\tilde{\Lambda}'^2(\alpha_1, \alpha_2')+i\epsilon}\right)^3},\nonumber
\end{eqnarray}
where
\begin{equation}
\tilde{\Lambda}^2(\alpha_1, \alpha_2)=\alpha_1\Lambda_1^2+\alpha_2\Lambda_2^2, \quad\quad \tilde{\Lambda}'^2(\alpha_1, \alpha_2')=\alpha_1\Lambda_1^2+\alpha_2'\Lambda_2^2.
\end{equation}

Then let $\omega=\tilde{\Lambda}(\alpha_1, \alpha_2)u$, \eqref{third pauli villars} becomes
\begin{equation}\label{fourth pauli villars}
\frac{\Lambda_1^2\Lambda_2^4}{16\pi^2}\int_0^1d\alpha_1\int_0^1d\alpha_2\int_0^1d\alpha_2'\frac{1}{\tilde{\Lambda}^4(\alpha_1, \alpha_2)}\int du\, e^{-i\tilde{\Lambda}(\alpha_1, \alpha_2)\Delta t u}I_0(u, \alpha_1, \alpha_2, \alpha_2'),
\end{equation}
where the integrand is,
\begin{eqnarray}
&&I_0(u, \alpha_1, \alpha_2, \alpha_2')\\
=&&\frac{1}{\sqrt{u^2-1+\frac{i\epsilon}{\tilde{\Lambda}^2(\alpha_1, \alpha_2)}}}\cdot\frac{1}{\sqrt{u^2-\frac{\tilde{\Lambda}'^2(\alpha_1, \alpha_2')}{\tilde{\Lambda}^2(\alpha_1, \alpha_2)}+\frac{i\epsilon}{\tilde{\Lambda}^2(\alpha_1, \alpha_2)}}}\cdot\frac{1}{\left(\sqrt{u^2-1+\frac{i\epsilon}{\tilde{\Lambda}^2(\alpha_1, \alpha_2)}}+\sqrt{u^2-\frac{\tilde{\Lambda}'^2(\alpha_1, \alpha_2')}{\tilde{\Lambda}^2(\alpha_1, \alpha_2)}+\frac{i\epsilon}{\tilde{\Lambda}^2(\alpha_1, \alpha_2)}}\right)^3}.\nonumber
\end{eqnarray}

 If $\alpha_2 \neq\alpha_2'$,  the integrand is multivalued in complex plane and has four branch points 
 \begin{equation}
 u_{1,2,3,4}=\pm \sqrt{1-\frac{i\epsilon}{\tilde{\Lambda}^2}},\;\pm \frac{\tilde{\Lambda}'}{
 \tilde{\Lambda}}\sqrt{1-\frac{i\epsilon}{\tilde{\Lambda}'^2}}
 \end{equation}. 

Assuming $\Delta t >0$, the integral contour goes around the lower half plane. Without loss of generality assuming $\alpha_2 >\alpha_2'$,  we can choose the branch cut being the line connecting $u_1=\sqrt{1-\frac{i\epsilon}{\tilde{\Lambda}^2}}$ and $u_3=\frac{\tilde{\Lambda}'}{
\tilde{\Lambda}}\sqrt{1-\frac{i\epsilon}{\tilde{\Lambda}'^2}}$. The integral $I_0$ is determined by the integral along this branch cut.  Since $I_0\rightarrow \frac{1}{\sqrt{u-u_{1,3}}}$ when $u\rightarrow u_{1,3}$, we can easily see that the integral along the branch cut converges to a finite number.  
 
If $\alpha_2 =\alpha_2'$, $I_0$ becomes $\left( u^2-1+\frac{i\epsilon}{\tilde{\Lambda}^2(\alpha_1, \alpha_2)}\right)^{-5/2}/8$. The integral can be solved easily in this case, i.e. $\int_{0}^{\infty} du I_0=\frac{1}{12}$ when $\epsilon\rightarrow 0$ , which is also finite. 

Therefore, after setting $\Delta t=0$, we have
\begin{equation}
\left\langle\phi^2\right\rangle=\frac{\Lambda_1^2\Lambda_2^4}{16\pi^2}\int_0^1d\alpha_1\int_0^1d\alpha_2\int_0^1d\alpha_2'\frac{1}{\tilde{\Lambda}^4(\alpha_1, \alpha_2)}\int du I_0(u, \alpha_1, \alpha_2, \alpha_2')\sim\Lambda^2, \quad \text{as} \quad \Lambda\sim\Lambda_1\sim\Lambda_2\to+\infty.
\end{equation}
Similarly, we have
\begin{eqnarray}\label{lorentz invariant phi dot}
&&\left\langle\dot{\phi}^2\right\rangle=\lim_{\Delta t\to 0}\frac{d}{dt_1}\frac{d}{dt_2}\left\langle\phi(t_1, \mathbf{x})\phi(t_2, \mathbf{x})\right\rangle\nonumber\\
=&&\frac{\Lambda_1^2\Lambda_2^4}{16\pi^2}\int_0^1d\alpha_1\int_0^1d\alpha_2\int_0^1d\alpha_2'\frac{1}{\tilde{\Lambda}^2(\alpha_1, \alpha_2)}\int du\, u^2I_0(u, \alpha_1, \alpha_2, \alpha_2')\sim\Lambda^4, \quad \text{as} \quad \Lambda\sim\Lambda_1\sim\Lambda_2\to+\infty,
\end{eqnarray}
which gives the same result as \eqref{omega square expectation value}.
\end{widetext}

When calculating the quantity $\left\langle\ddot{\phi}^2\right\rangle$ from \eqref{fourth pauli villars} using the same technique as in \eqref{lorentz invariant phi dot}, we would get the integral $\int du\, u^4I_0$, which is logarithmically divergent. To make it converge, we can employ another two Pauli-Villars type cutoffs $\Lambda_3$ and $\Lambda_4$ just as what we did from \eqref{pauli villars technique1} to \eqref{pauli villars technique2}. The exact dependence of $\left\langle\ddot{\phi}^2\right\rangle$ on $\Lambda_1$, $\Lambda_2$, $\Lambda_3$ and $\Lambda_4$ is complicated, but the result for its leading order goes as
\begin{equation}\label{lorentz invariant phi ddot}
\left\langle\ddot{\phi}^2\right\rangle\sim\Lambda^6,
\end{equation}
which gives the same result as \eqref{omega time derivative expectation value}.

So far, we have obtained that the leading order dependencies of $\left\langle\dot{\phi}^2\right\rangle$ and $\left\langle\ddot{\phi}^2\right\rangle$ on the non-Lorentz invariant sharp cutoff $\Lambda$ and on the Lorentz invariant Pauli-Villars type cutoffs $\Lambda_1$, $\Lambda_2$, ($\Lambda_3$, $\Lambda_4$) are the same. As we mentioned in the beginning of this subsection, these results are natural since they could have been guessed by dimensional analysis. Therefore, the slow varying condition \eqref{slow varying condition} is still satisfied by using Lorentz invariant cutoffs that our results would not change.

\subsection{The singularities at $a(t, \mathbf{x})=0$}
In our way of vacuum gravitating, the space is alternatively expanding and contracting at each spatial point, and, during each such cycle, the expansion outweighs the contraction a little bit due to the weak parametric resonance effect. This process gives a slowly increasing amplitude $A(t, \mathbf{x})$ of the scale factor $a(t, \mathbf{x})$, whose observable effect is just the accelerating expansion of our Universe. 

Probably one of the biggest concerns about this physical picture is the appearance of singularities at points $a(t, \mathbf{x})=0$---according to the solution \eqref{wkb}, the scale factor $a(t, \mathbf{x})$ must go through zero whenever the space at $\mathbf{x}$ switches from contraction phase to expansion phase. 

Singularities are a generic feature of the solution of Einstein field equations under rather general energy conditions (e.g. strong, weak, dominant etc.), which is guaranteed by Penrose-Hawking singularity theorems \cite{PhysRevLett.14.57, Hawking511, Hawking490, Hawking187, Hawking:1973uf, Hawking:1969sw}. In this paper, since we investigate the gravitational property of quantum vacuum without modifying either QFT or GR, the appearance of singularities is inevitable---QFT predicts a huge vacuum energy, and according to GR, huge energy must collapse to form singularity. 

It is usually thought that the Einstein field equations break down at singularities and thus the spacetime evolution will stop once the singularity is formed. However, it is not the case for our solution to the key dynamic evolution equation \eqref{hmo}. \eqref{hmo} describes the oscillating motion a harmonic oscillator. It is natural for a harmonic oscillator to pass its equilibrium point $a(t, \mathbf{x})=0$ at maximum speed without stopping. So in our solution, the singularity immediately disappears after it forms and the spacetime continues to evolve without stopping. Singularities just serve as the turning points at which the space switches from contraction phase to expansion phase.

\subsubsection{Resolving singularity by multiplying $a$}

In order to understand why in our solution the singularity is not the end of spacetime evolution, it is helpful to review one crucial step in deriving \eqref{hmo} from \eqref{00+ii}. Rigorously speaking, we can only obtain the following equation from \eqref{00+ii}:
\begin{equation}\label{intermidiate step}
-\frac{\ddot{a}}{a}=\Omega^2(t, \mathbf{x}),
\end{equation}
which is not equivalent to \eqref{hmo}. To get \eqref{hmo}, we need one more step---multiply both sides of \eqref{intermidiate step} by $a$.

Mathematically, $a$ is not allowed to be zero in \eqref{intermidiate step} since it is in the denominator. In fact, when writing down the Einstein field equations \eqref{Einstein equations 00}, \eqref{Einstein equations ii}, \eqref{Einstein equations 0i} and \eqref{Einstein equations ij}, it has been presumed that $a\neq 0$ since if $a=0$, the metric would become degenerate ($g=\operatorname{det}(g_{\mu\nu})=-a^6=0$), the curvature would become infinite and the Einstein tensor are simply not defined there.

But, after the inequivalent algebraic manipulation of multiplying both sides of \eqref{intermidiate step} by $a$, $a$ is allowed to evolve to zero in the resulting equation \eqref{hmo} since there is nothing wrong for a harmonic oscillator to go through its equilibrium point. In this sense, we have smoothly extended the solution beyond the singularity by the mathematical operation of multiplying both sides of \eqref{intermidiate step} by $a$ (or more generally by some power of the metric determinant).

The idea of resolving a singularity by mulptiplying Einstein equations with some power of the determinant of the metric is not new. Einstein himself had proposed this idea with his collaborator Rosen in 1935 (for which they credited this idea to Mayer) \cite{PhysRev.48.73}. Ashtekar used a similar trick in his method of ``new variables'' to develop an equivalent Hamiltonian formulation of GR \cite{PhysRevD.36.1587}. It is also proposed by Stoica that the equations obtained after multiplying the usual Einstein equations by some power of the metric determinant are actually more fundamental than the usual Einstein equations \cite{Stoica:2013wx, doi:10.1142/S0219887814500418, Stoica:2012my, Stoica:2012gb, Stoica:2011xf, Stoica:2011nm, Stoica:2014tpa, Stoica:2015yfa, Stoica:2015hba}. In this sense, we argue that our spacetime with singularities due to the metric becoming degenerate ($a=0$) is a legitimate solution of GR. 

\subsubsection{Singularities do not cause problems}
While singularities are natural and inevitable in solutions to Einstein's equations, we must discuss the consequences they bring to this calculation.

Will the singularities cause serious problems? At least in our case we do not feel they cause problems. To see this, we investigate how the singularities affect the propagation of the field modes in our toy model \eqref{toy wave equation}.

In this toy model, the spacetime have singularities at the hypersurfaces
\begin{equation}\label{singularity points}
\Omega t+\mathbf{K}\cdot\mathbf{x}=(n+\frac{1}{2})\pi, \quad n=0, \pm1, \pm2, \pm3, \cdots
\end{equation}
Using the relation $x_m=-x_{-m}$, which is evident from \eqref{antisymmetric solution} and \eqref{xn definition}, the asymptotic mode solution \eqref{toy model asymptotic mode solution} becomes
\begin{widetext}
\begin{equation}\label{alternative uk expression}
u_{\mathbf{k}}(t, \mathbf{x})=c_0 e^{-i\left(\omega t-\mathbf{k}\cdot\mathbf{x}\right)}\left(1+2i\epsilon\sum_{m=1}^{+\infty}x_m \sin{2m\left(\Omega t+\mathbf{K}\cdot\mathbf{x}\right)}\right).
\end{equation}
\end{widetext}
At the singularities \eqref{singularity points}, the terms $\sin 2m(\Omega t+\mathbf{K}\cdot\mathbf{x})$ of \eqref{alternative uk expression} are all zero and thus we have
\begin{equation}
u_{\mathbf{k}}(t, \mathbf{x})=c_0 e^{-i\left(\omega t-\mathbf{k}\cdot\mathbf{x}\right)}
\end{equation}
So $u_{\mathbf{k}}$ is normal at singularities which shows that the field can naturally pass the singularities without problems.

One might still worry about the divergences of the time or spatial derivatives of $u_{\mathbf{k}}$ at the singularities \eqref{singularity points}. However, these divergences arise from those small high frequency corrections (terms inside the summation symbol $\Sigma$ of \eqref{alternative uk expression}) with frequencies $2m\Omega$ which are much higher than our effective QFT's cutoff $\Lambda$. When looking at low energy scales ($\leq\Lambda$), $u_{\mathbf{k}}$ behaves the same as the mode solution \eqref{flat plane wave mode} when the background spacetime is flat; only when looking at high energy scales ($\geq\sqrt{G}\Lambda^2$) which are much higher than the cutoff scale $\Lambda$, those small high frequency corrections are noticeable.

In this sense, the singularities do not cause problems at the observable low energy regime---after all, the singularities only appear (and immediately disappear) above Planck energy scales, which should not affect the low energy physics whose energy scale is far below Planck.

\subsection{Similarity of effects of vacuum energy in non-gravitational system and gravitational system}\label{section iv}
Vacuum fluctuations and their associated vacuum energies are direct consequences of the Heisenberg's uncertainty principle of quantum mechanics. Although it is still controversial \cite{PhysRevD.72.021301}, various observable effects are often ascribed to the existence of vacuum energies and have been experimentally verified, which strongly suggests the reality of vacuum fluctuations. These vacuum fluctuation effects include the spontaneous emission \cite{scully1997quantum}, the Lamb shift \cite{PhysRev.72.241}, the anomalous magnetic moment of the electron \cite{PhysRev.73.416, PhysRevA.83.052122} and the Casimir effect \cite{Casimir:1948dh, PhysRevLett.78.5, PhysRevLett.81.4549, PhysRevLett.88.041804}. The reality of the vacuum energy associated to the spontaneous symmetry breaking of electroweak theory has also been confirmed by the discovery of the Higgs boson at the LHC \cite{Aad20121, Chatrchyan201230}.

If we assume that the vacuum fluctuations do exist as evidenced by the above listed observable effects, then according to the equivalence principle, the associated vacuum energies would gravitate as well as all other forms of energy. This has been experimentally demonstrated by, for example, the gravitational test of Lamb shift energy \cite{Polchinski:2006gy, Masso:2009zd, Braginskii:1971tn}. The gravitational property of Casimir energy has not been tested experimentally, but has been demonstrated theoretically with the conclusion that it does gravitate according to equivalence principle \cite{Fulling:2007xa, Milton:2007ar, Milton:2007hd, Milton:2008ks}.

However, in the literature, the value of vacuum energy density is usually thought to play a different role in non-gravitational systems and in gravitational systems. The actual value of the vacuum energy density is generally regarded as irrelevant in non-gravitational contexts based on the argument that only energy differences from the vacuum are measurable; while when gravity is present, the actual value of the energy matters, not just the differences, since the source for the gravitational field is the entire energy momentum tensor that its large value may be potentially disastrous.

We argue differently in this section with the following points: (i) the value of vacuum energy density can also be relevant in non-gravitational contexts; (ii) the huge value of vacuum energy density is not a direct observable and that it is not disastrous in a theory of gravity. Moreover, there is essentially no difference between the roles played by vacuum energy in non-gravitational systems and in gravitational systems. In other words, although technically more complicated when gravity is included, the gravitational effect of the vacuum energy on spacetime metric is intrinsically the same as its effect on material bodies when gravity is excluded.

\subsubsection{Value of vacuum energy is relevant in Casimir effect}
Let us first consider the Casimir effect. The Casimir force is usually derived by calculating the change in vacuum energy due to the presence of the conducting plates, which acts as mirrors to reflect electromagnetic waves (We will call them mirrors in the following). This derivation is straightforward, but loses some important physical details about what is going on in the system \cite{GONZALEZ1985228, PhysRevA.38.1621}. Due to quantum fluctuations, the zero point fields constantly impinge on both sides of the mirror and then reflect back, which transmit momentum to the mirror and thus result in forces on both sides of the mirror. The Casimir stress (force per unit area) is just the difference between the pressure exerted by the electromagnetic field vacuum from inside and outside
\begin{equation}\label{casimr stress}
S(t, x, y)=T_{zz}^{\mathrm{inside}}-T_{zz}^{\mathrm{outside}},
\end{equation}
where we have set that the two parallel mirrors are normal to the $z$ axis. Since the vacuum fluctuations between the two mirrors are different from the vacuum fluctuations outside, the expectation values of $T_{zz}^{\mathrm{inside}}$ and $T_{zz}^{\mathrm{outside}}$ would be different and thus gives a net average force. Although both $T_{zz}^{\mathrm{inside}}$ and $T_{zz}^{\mathrm{outside}}$ are divergent, this average force is finite since the quartic divergent Minkowski zero point fluctuations are canceled after the subtraction in \eqref{casimr stress} and one obtains the well known Casimir stress \cite{PhysRevA.38.1621}
\begin{equation}\label{expectation value of Casimir stress}
\left\langle S\right\rangle=-\frac{\pi^2}{240 d^4}.
\end{equation}
Thus the effect of the value of zero point energy disappears in the calculations. It is for this reason that although the Casimir effect is usually regarded as evidence of the reality of zero point energy, the actual value of its energy density is thought to be irrelevant in this effect.

However, the value of zero point energy density does have an effect. Note that \eqref{expectation value of Casimir stress} only gives the expectation value of the Casimir stress $S$, but $S$ is never a constant, it fluctuates. That's because the amount of momentum carried by the zero point fields which impinge on both sides of the mirror is constantly fluctuating due to the fact that the vacuum is not an eigenstate of the $zz$ component of the stress energy tensor $T_{zz}$. The magnitude of the fluctuation of each $T_{zz}$ is large and diverges as the same order  of the vacuum energy density $\left\langle T_{00}\right\rangle$. For a perfect mirror, since the fields on the two sides fluctuate independently of each other, the mean-squared stresses on the two sides simply add, resulting in the magnitude of the fluctuation of the net stress also diverges as 
\begin{equation}\label{perfect mirror stress fluctuation}
\left\langle \Delta S^2\right\rangle=\left\langle (S-\left\langle S\right\rangle)^2\right\rangle\sim\left\langle T_{00}\right\rangle^2\to\infty.
\end{equation}
For more realistic imperfect mirrors which become transparent for frequencies higher than its plasma frequency $\Lambda$, the $\left\langle T_{00}\right\rangle$ in \eqref{perfect mirror stress fluctuation} contains contributions only from field modes of frequencies lower than $\Lambda$ and the mean squared value of the net stress $S$ goes as
\begin{equation}\label{imperfect mirror stress fluctuation}
\left\langle \Delta S^2\right\rangle\sim\left\langle T_{00}\right\rangle^2\sim\Lambda^8.
\end{equation}
The plasma frequency $\Lambda$ in \eqref{imperfect mirror stress fluctuation} acts as an effective cutoff which depends on the microstructure of the mirror. It is similar but distinct from the effective QFT's cutoff $\Lambda$ in \eqref{dependence of H on Lambda}, which depends on the microstructure of spacetime.

Therefore, the value of zero point energy density is still physically significant even in non-gravitational system. Its value appears in \eqref{perfect mirror stress fluctuation} and \eqref{imperfect mirror stress fluctuation} to characterize the strength of Casimir stress fluctuation, which implies that the net Casimir stress is constantly fluctuating with huge magnitudes around its small mean value \eqref{expectation value of Casimir stress}. Due to this huge fluctuation, at almost any instant, the magnitude of the stress at each single point of the mirror is as large as the value of the zero point energy density. 

However, this effect is strong only at small scales. Its measurable effect becomes small at larger scales. In practice, the measurements must be taken over some finite time interval $T$ and some finite surface area of order $l^2$. More precisely, what the force detector measures is the time and surface average
\begin{equation}
\bar{S}=\int dtdxdy f(t,x,y)S(t,x,y),
\end{equation}
where the averaging function $f$ satisfies
\begin{equation}
\int dtdxdy f(t,x,y)=1.
\end{equation}
The exact shape of the averaging function depends on the measuring apparatus. On physical grounds one can choose $f$ to be a single peak over a time interval $T$ comparable to the experimental resolving time and over a spatial region of area $l^2$ comparable to the resolution of the measuring device. Although the magnitude of the fluctuations of the net stress $S$ is formally infinite as shown in \eqref{perfect mirror stress fluctuation}, the magnitude of the measurable fluctuations of its average $\bar{S}$ is finite. This is because the effect of the vacuum fluctuations at small scales is significantly weakened when averaging over larger scales. The calculations have been done by G Barton in \cite{0305-4470-24-5-014} with the conclusion that, for the realistic case where $l \ll cT$, the mean squared deviation
\begin{equation}\label{Barton fluctuation}
\left\langle \Delta\bar{S}^2\right\rangle=\left\langle \left(\bar{S}-\left\langle\bar{S}\right\rangle\right)^2\right\rangle=\frac{\mathrm{constant}}{T^8},
\end{equation}
where the ``$\mathrm{constant}$'' here is a pure number as could have been foreseen on dimensional grounds. The above equation \eqref{Barton fluctuation} shows that $\left\langle \Delta\bar{S}^2\right\rangle$ increases as $T$ decreases, which means that the better the measuring device, the stronger fluctuation due to the effect of the value of the zero point energy density can be measured. And in principle, using a perfect instantaneous measuring device ($T\to 0$), one can measure the infinite fluctuations of the Casimir stress on a perfect mirror due to the infinite value of zero point energy density. In practice, however, $\left\langle \Delta\bar{S}^2\right\rangle$ is too small to be measured for a real force detector whose resolving time $T$ is too large \cite{0305-4470-24-5-014}.

\subsubsection{Effect of vacuum energy on the motion of mirrors}

The value of zero point energy density also has effects on the dynamic motion of small material bodies. Imagine that we place a single mirror of very small size in the vacuum and then release it. The mirror would experience a fluctuating force exerted by the quantum field vacuum and starts  to move. The equation of motion of the mirror, which is called quantum Langevin equation, can be generally described by
\begin{equation}\label{equation of motion of mirror}
\ddot{X}=F\left(t, X, \dot{X}, \phi, \dot{\phi}, \dots\right),
\end{equation}
where $X$ is the mirror's position, $\phi$ represents the field interacting with the mirror which is usually taken to be a scalar field for simplicity and we have set the mirror's mass $M=1$ for convenience. The average force in this case would be zero because of symmetry
\begin{equation}\label{zero average force}
\left\langle F\right\rangle=0,
\end{equation}
and similar to the Casimir stress fluctuation \eqref{perfect mirror stress fluctuation}, the force here also undergoes wild fluctuations with a magnitude 
\begin{equation}\label{magnitude of force fluctuation}
\left\langle F^2\right\rangle\propto\left\langle T_{00}\right\rangle^2\to\infty.
\end{equation}

The mathematically infinite fluctuating force $F$ gives infinite instantaneous accelerations of the mirror through \eqref{equation of motion of mirror}. Similar to the case of infinite Casimir stress fluctuation \eqref{perfect mirror stress fluctuation}, this infinite fluctuating force and infinite instantaneous acceleration make sense since they are also only significant at very small scales and will not result in infinite fluctuation of the mirror's position at observable larger scales. In fact, the mirror would oscillate back and forth with very high speeds, but its range of motion is still small \cite{PhysRevD.89.085009, PhysRevD.92.063520, Gour:1998my, Jaekel:1992ef, Stargen:2016euf}.

More precisely, suppose that the mirror is initially located at $X(0)=0$ with velocity $\dot{X}(0)=0$ and is then released at $t=0$. The magnitude of its acceleration $\ddot{X}(t)$ and velocity $\dot{X}(t)$, which can be characterized by the quantity $\left\langle \ddot{X}^2(t)\right\rangle$ and $\left\langle \dot{X}^2(t)\right\rangle$, is large. But, the magnitude of the range of the mirror's fluctuating motion, which can be characterized by the observable mean squared displacement $\left\langle X^2(t)\right\rangle$, is still small. 

In this sense, the value of vacuum energy density is still relevant even in non-gravitational physics. This value appears in the equation \eqref{magnitude of force fluctuation} to characterize the strength of the force fluctuations acting on the mirror at small scales and it may have small observable effects at larger scales such as diffusions predicted in \cite{Gour:1998my, Jaekel:1992ef, Stargen:2016euf}.

\subsubsection{Analogies between the motion of mirror and the motion of $a(t, \mathbf{x})$}
Although technically more complicated in gravity, the basic dynamic equation of motion \eqref{hmo} satisfied by the scale factor $a(t, \mathbf{x})$ is in fact very similar to the equation of motion \eqref{equation of motion of mirror} satisfied by the mirror's position $X(t)$. Consider only the contribution from the massless scalar field $\phi$, \eqref{hmo} is just the following same form as the equation \eqref{equation of motion of mirror}
\begin{equation}
\ddot{a}=F\left(a, \dot{\phi}\right),
\end{equation}
where
\begin{equation}
F\left(a, \dot{\phi}\right)=-\frac{8\pi G}{3}\dot{\phi}^2a.
\end{equation}
Also, the average of the fluctuating force $F\left(a, \dot{\phi}\right)$ is zero due to symmetry
\begin{equation}\label{zero average force for a}
\left\langle F\left(a, \dot{\phi}\right)\right\rangle=0,
\end{equation}
and its magnitude of fluctuation
\begin{equation}\label{huge fluctuation force for a}
\left\langle F^2\left(a, \dot{\phi}\right)\right\rangle\propto\left\langle T_{00}\right\rangle^2\to\infty.
\end{equation}
The above two statistical properties \eqref{zero average force for a} and \eqref{huge fluctuation force for a} satisfied by the ``force'' driving the ``motion'' of the scale factor $a$ are the same with the statistical properties \eqref{zero average force} and \eqref{magnitude of force fluctuation} satisfied by the force driving the motion of the mirror. In this sense, the role played by the value of the vacuum energy density in gravitational system is similar to its role in non-gravitational system.

Concretely speaking, the vacuum energy density results in large instantaneous acceleration $\ddot{X}$ and velocity $\dot{X}$ of the mirror, but the observable position fluctuations of the mirror, which can be characterized by the quantity $\left\langle X^2\right\rangle$, is not large. Analogously, the vacuum energy density results in the large instantaneous ``acceleration'' $\ddot{a}$ and ``velocity'' $\dot{a}$ of the scale factor, but the observable physical distance defined by \eqref{length definition}, whose value is determined by the quantity $\left\langle a^2\right\rangle$, is also not large. These properties about $a(t, \mathbf{x})$ are evident from the solutions \eqref{expected form of solution}, \eqref{wkb} and \eqref{dependence of H on Lambda}, from which we can see that the quantities $\left\langle\ddot{a}^2\right\rangle$ and $\left\langle \dot{a}^2\right\rangle$ are as large as $\left\langle T_{00}\right\rangle^2$ and $\left\langle T_{00}\right\rangle$ respectively, while the magnitude of the quantity $\left\langle a^2\right\rangle$ is on the order $1$. 

In this sense, the role played by vacuum energy in gravitational system is similar to its role in the non-gravitational mirror systems---it appears both at \eqref{magnitude of force fluctuation} and  \eqref{huge fluctuation force for a} to show the strongness of vacuum fluctuations at microscopic scales (for mirrors, microscopic means atomic scale; for gravity, microscopic means Planck scale) and their observable effects are both small at macroscopic scales.

By this same kind of mechanism, the violent gravitational effect produced by the vacuum energy density is confined to Planck scales, and its effect at macroscopic scales---the accelerating expansion of the Universe, due to the weak parametric resonance is so small that, it is only observable after accumulations on the largest scale---the cosmological scale.

\section*{Acknowledgements}
We would like to thank Andrei Barvinsky, Gordon W. Semenoff, Mark Van Raamsdonk, Philip C. E. Stamp, Daniel Carney, Yin-Zhe Ma, Daoyan Wang, Michael Desrochers, Giorgio Torrieri, Niayesh Afshordi, Roland de Putter, J\'{e}r\^{o}me Gleyzes and Olivier Dor\'{e} for helpful discussions and criticisms. We especially thank Michael Desrochers for helping revising the manuscript. This research is partially supported by the Natural Sciences and Engineering Research Council of Canada (NSERC), the Templeton Foundation, and the Canadian Institute for Advanced Research. Zhen Zhu would also like to thank Philip C. E. Stamp for financial support.

\appendix
\section{Real Massless Scalar Field}\label{appendix 1}
In this appendix we give the calculation details about how the quantum vacuum fluctuates all over the spacetime by using the massless scalar field \eqref{field expansion} as an example.

We first define the covariance of the energy density operator at two spacetime points $x=(t,\mathbf{x})$ and $x'=(t',\mathbf{x}')$
\begin{eqnarray}\label{covariance definition}
&&\operatorname{Cov}\big(T_{00}(x),T_{00}(x')\big)\nonumber\\
=&&\langle\left\{\big(T_{00}(x)-\left\langle T_{00}(x)\right\rangle\big)\big(T_{00}(x')-\left\langle T_{00}(x')\right\rangle\big)\right\}\rangle, 
\end{eqnarray}
where the curly bracket $\{\}$ in \eqref{covariance definition} is the symmetrization operator which is defined as, for any two operators $A$ and $B$,
\begin{equation}\label{symmetrization def}
\left\{AB\right\}=\frac{1}{2}\left(AB+BA\right).
\end{equation}

Inserting \eqref{field expansion} and \eqref{energy density definition} into \eqref{covariance definition} gives the following result
\begin{equation}\begin{split}\label{correlation function 0}
&\operatorname{Cov}\big(T_{00}(x),T_{00}(x')\big)\\
=&\frac{1}{2}\int\frac{d^3kd^3k'}{(2\pi)^6}\frac{\left(\omega\omega'+\mathbf{k}\cdot\mathbf{k'}\right)^2}{2\omega 2\omega'}\\
&\cdot\cos\Big((\omega+\omega') \Delta t-(\mathbf{k}+\mathbf{k'})\cdot\mathbf{\Delta x}\Big),
\end{split}\end{equation}
where $\Delta t=t-t'$ and $\mathbf{\Delta x}=\mathbf{x-x'}$ are time and space separation of the two spacetime points $x$ and $x'$.

If $x$ and $x'$ are timelikely separated, we can find a reference frame to set $\Delta x=|\mathbf{\Delta x}|=0$. In this case, evaluation of the integral in \eqref{correlation function 0} for a high frequency cutoff $|\mathbf{k}|=\Lambda$ gives
\begin{widetext}
\begin{eqnarray}\label{timelike correlation}
&&\operatorname{Cov}\big(T_{00}(x),T_{00}(x')\big)=\frac{1}{24\pi^4{\Delta t}^8}\Bigg(\left[-(\Lambda\Delta t)^6+21(\Lambda\Delta t)^4-72(\Lambda\Delta t)^2+36\right]\cos(2\Lambda\Delta t)\\
&&+6\left[(\Lambda\Delta t)^5-8(\Lambda\Delta t)^3+12\Lambda\Delta t\right]\sin(2\Lambda\Delta t)+12\left[(\Lambda\Delta t)^3-6\Lambda\Delta t\right]\sin(\Lambda\Delta t)+36\left[(\Lambda\Delta t)^2-2\right]\cos(\Lambda\Delta t)+36\Bigg).\nonumber
\end{eqnarray}

If $x$ and $x'$ are spacelikely separated, we can find a reference frame to set $\Delta t=0$. In this case, evaluation of the integral in \eqref{correlation function} for a high frequency cutoff $|\mathbf{k}|=\Lambda$ gives
\begin{eqnarray}
\label{spacetime correlation}
\operatorname{Cov}\big(T_{00}(x),T_{00}(x')\big)=&&\frac{1}{32\pi^4{\Delta x}^8}\Bigg(\left[2(\Lambda\Delta x)^4-34(\Lambda\Delta x)^2+33\right]\cos(2\Lambda\Delta x)-\left[12(\Lambda\Delta x)^3-50\Lambda\Delta x\right]\sin(2\Lambda\Delta x)\nonumber\\
&&+16\left[(\Lambda\Delta x)^2-6\right]\cos(\Lambda\Delta x)-64\Lambda\Delta x\sin(\Lambda\Delta x)+63\Big)
\end{eqnarray}
\end{widetext}

As $\Delta t$ and $\Delta x$ goes to $0$, both \eqref{timelike correlation} and \eqref{spacetime correlation} reduces to the variance of the energy density,
\begin{equation}\label{magnitude of mean squared energy density}
\left\langle\big(T_{00}-\left\langle T_{00}\right\rangle\big)^2\right\rangle=\frac{2}{3}\left(\frac{\Lambda^4}{16\pi^2}\right)^2
=\frac{2}{3}\left\langle T_{00}\right\rangle^2
\end{equation}

We then investigate the Pearson product-moment correlation coefficient
\begin{equation}
\rho_{x,x'}=\frac{\operatorname{Cov}\big(T_{00}(x),T_{00}(x')\big)}{\sigma_x\sigma_{x'}},
\end{equation}
where
\begin{equation}\label{standard deviation}
\sigma_x=\sqrt{\left\langle\left(T_{00}(x)-\left\langle T_{00}(x)\right\rangle\right)^2\right\rangle}.
\end{equation}

The correlation coefficient $\rho_{x,x'}$ shows by its magnitude the strength of correlation between two random variables. $\rho_{x,x'}$ is positive if the energy density $T_{00}$ at $x$ and $x'$ are most possibly lying on the same side of the vacuum expectation value $\left\langle T_{00}\right\rangle=\Lambda^4/(16\pi^2)$. Thus a positive correlation coefficient $\rho_{x,x'}$ implies the energy density at $x$ and $x'$ tend to be simultaneously greater than, or simultaneously less than the expectation value. Similarly, a negative $\rho_{x,x'}$ implies the energy density tend to lie on opposite sides of the expectation value. We will call the energy density $T_{00}$ at $x$ and $x'$ are positively correlated if $\rho_{x,x'}>0$ or negatively correlated (anticorrelation) if $\rho_{x,x'}<0$.

\begin{figure}
\includegraphics[scale=0.65]{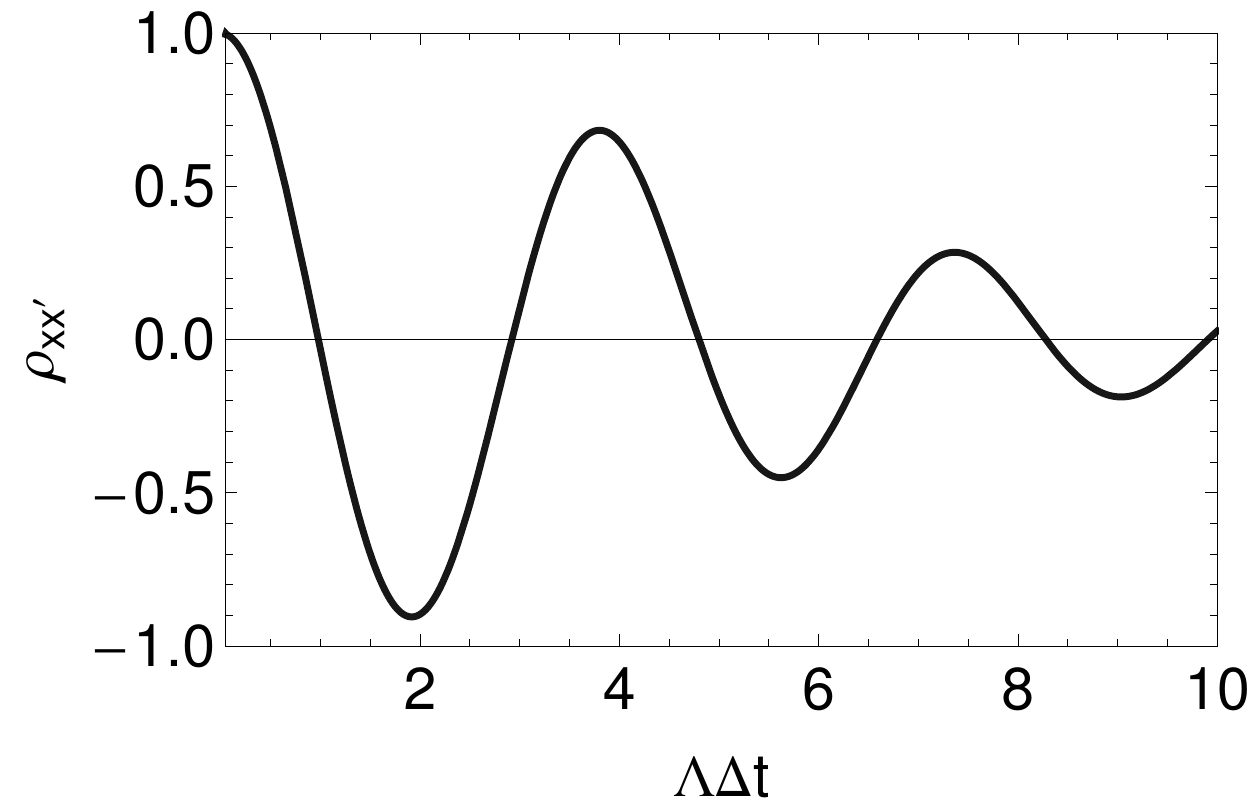}
\caption{\label{time correlation coefficients} Plot of correlation coefficient $\rho_{x,x'}$ as a function of time separation $\Lambda\Delta t$ in the case $\Delta x=0$.}
\end{figure}

\begin{figure}
\includegraphics[scale=0.65]{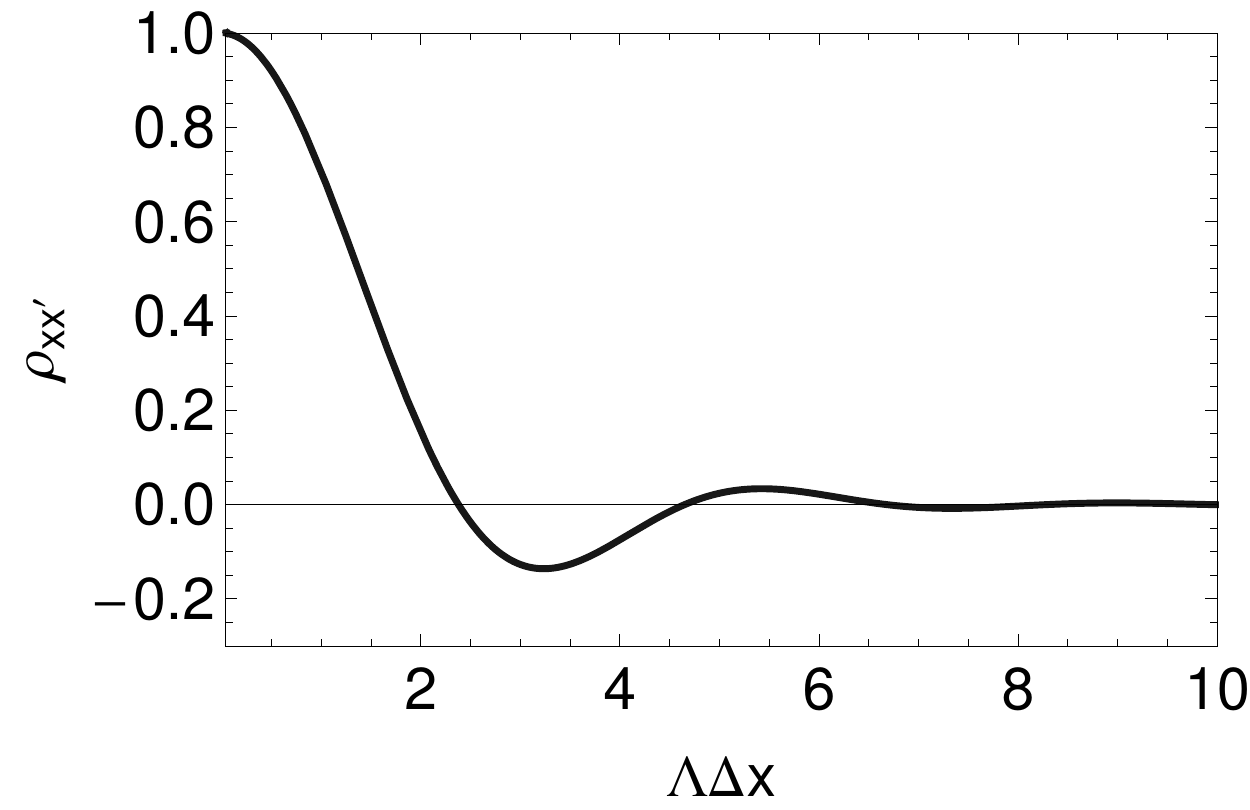}
\caption{\label{space correlation coefficients} Plot of correlation coefficient $\rho_{x,x'}$ as a function of spatial separation $\Lambda\Delta x$ in the case $\Delta t=0$.}
\end{figure}

Because of transnational invariance, $\rho_{x,x'}$ is only dependent on the temporal and spatial separation $\Delta t=t-t', \mathbf{\Delta x}=\mathbf{x}-\mathbf{x}'$. For the real massless scalar field \eqref{field expansion}, the behavior of the correlation coefficient $\rho_{x, x'}$ as a function of temporal separation $\Lambda\Delta t$ for the case of $\Delta x=0$ and as a function of spatial separation $\Lambda\Delta x$ for the case of $\Delta t=0$ are plotted in FIG. \ref{time correlation coefficients} and \ref{space correlation coefficients} respectively.

In the temporal direction, i.e. the case of $\Delta x=0$ (Fig. \ref{time correlation coefficients}), the correlation coefficient goes quickly from $1$ down to around $-0.9$ in a time scale around $\Delta t=1.9/\Lambda$ and then goes up to $0.7$ in a time scale around $\Delta t=3.8/\Lambda$ and then goes down and up alternatively from positive values to negative values with decreasing amplitudes. It roughly oscillates as $-\cos(2\Lambda\Delta t)/(\Lambda\Delta t)^2$ with a period $\pi/\Lambda$ as $\Delta t$ is large. Thus at the extremely small time scales $\Delta t\sim 1.9/\Lambda$, ($\Lambda\to+\infty$), the energy density are strongly anticorrelated. In other words, if at some time the value of the energy density is larger than its expectation value, for example, by an amount of $0.82\left\langle T_{00}\right\rangle$, after a short time $\Delta t=1.9/\Lambda$, its value is most likely to be smaller than the expectation value, for example, by an amount of $0.74\left\langle T_{00}\right\rangle$. The difference is $1.56\left\langle T_{00}\right\rangle$ only after such a short time.

In the spatial direction, i.e. the case of $\Delta t=0$ (Fig. \ref{space correlation coefficients}), the correlation coefficient goes quickly from $1$ down to around $-0.14$ in a length scale around $\Delta x=3.24/\Lambda$ and then goes up to $0.03$ in a length scale around $\Delta x=5.4/\Lambda$ and then goes down and up alternatively from positive values to negative values with decreasing amplitudes. Compared to the temporal direction, the decay in the oscillation amplitude of the correlation coefficient is faster in spatial direction. It roughly oscillates as $2\cos(2\Lambda\Delta x)/(\Lambda\Delta x)^4$ with a period $\pi/\Lambda$ as $\Delta x$ is large. These properties show that the strength of the correlation between energy densities at close range in spatial direction is not as strong as in the temporal direction. For larger spatial separations, $\rho_{x, x'}$ approaches zero and the vacuum energy density $T_{00}$ at different $x$ and $x'$ fluctuate independently. These properties result in extreme spatial inhomogeneities of the quantum vacuum which can be characterized by the quantity $\Delta\rho^2$ defined by \eqref{spatial inhomogeneous rho} in section \ref{sec iii}.

The quantity $\Delta\rho^2$ is related to $\rho_{x, x'}$ by
\begin{equation}
\Delta\rho^2=1-\rho_{x, x'}.
\end{equation}
The behavior of $\Delta\rho^2$ has been plotted in FIG. \ref{density difference}.

Next we calculate the $\chi(\Delta t)$ defined by \eqref{cov def} in section \ref{The accelerating expansion from weak parametric resonance}. Wick expansion of \eqref{cov def} gives
\begin{equation}\label{cov}
\chi(\Delta t)=\frac{\left\langle\dot{\phi}(t_1,\mathbf{x})\dot{\phi}(t_2,\mathbf{x})\right\rangle^2+\left\langle\dot{\phi}(t_2,\mathbf{x})\dot{\phi}(t_1,\mathbf{x})\right\rangle^2}{2\left\langle\dot{\phi}^2(t, \mathbf{x})\right\rangle^2},
\end{equation}
where the correlation function can be calculated directly by inserting \eqref{field expansion}
\begin{equation}\label{correlation function}
\left\langle\dot{\phi}(t_1,\mathbf{x})\dot{\phi}(t_2,\mathbf{x})\right\rangle=\frac{1}{4\pi^2}\int_0^{\Lambda} k^3e^{-ik\Delta t}dk.
\end{equation}
Plugging \eqref{correlation function} into \eqref{cov} gives the following result 
\begin{widetext}
\begin{eqnarray}\label{phi dot square covariance}
\chi(\Delta t)&=&\frac{16}{\Lambda^8\Delta t^8}\Big(36\left(-2+\Lambda^2\Delta t^2\right)\cos(\Lambda\Delta t)+\left(36-72\Lambda^2\Delta t^2+21\Lambda^4\Delta t^4-\Lambda^6\Delta t^6\right)\cos(2\Lambda\Delta t)\nonumber\\
&+&6\left(6+2\Lambda\Delta t\left(-6+\Lambda^2\Delta t^2\right)\sin(\Lambda\Delta t)+\Lambda\Delta t\left(12-8\Lambda^2\Delta t^2+\Lambda^4\Delta t^4\right)\sin(2\Lambda\Delta t)\right)\Big).
\end{eqnarray}
The behavior of $\chi(\Delta t)$ has been plotted in FIG. \ref{omega covariance}. It is closely related to the correlation coefficient $\rho_{x, x'}$ as a function of time difference $\Delta t$ in the case $\Delta x=0$ (FIG. \ref{time correlation coefficients}).

Next we derive the equation \eqref{big Omega} in section \ref{The accelerating expansion from weak parametric resonance}. First, $\Omega^2(t, \mathbf{0})$ can be expanded as
\begin{eqnarray}\label{phi dot square expansion}
\Omega^2(t, \mathbf{0})=\frac{8\pi G}{3}\int_{\omega, \omega'\leq\Lambda}\frac{d^3kd^3k'}{(2\pi)^3}&&\frac{\sqrt{\omega\omega'}}{2}\bigg[\left(a_{\mathbf{k}}a_{\mathbf{k}'}^{\dag}+a_{\mathbf{k}}^{\dag}a_{\mathbf{k}'}\right)\cos\left(\omega-\omega'\right)t+i\left(-a_{\mathbf{k}}a_{\mathbf{k}'}^{\dag}+a_{\mathbf{k}}^{\dag}a_{\mathbf{k}'}\right)\sin\left(\omega-\omega'\right)t\nonumber\\
+&&\left(-a_{\mathbf{k}}a_{\mathbf{k}'}-a_{\mathbf{k}}^{\dag}a_{\mathbf{k}'}^{\dag}\right)\cos\left(\omega+\omega'\right)t+i\left(a_{\mathbf{k}}a_{\mathbf{k}'}-a_{\mathbf{k}}^{\dag}a_{\mathbf{k}'}^{\dag}\right)\sin\left(\omega+\omega'\right)t\bigg].
\end{eqnarray}
\end{widetext}
Specially, the vacuum state $|0\rangle$ is an eigenstate of the operator coefficients of the first two terms in the above expression \eqref{phi dot square expansion}. If $\mathbf{k}\neq\mathbf{k}'$, the eigenvalues of the operator coefficients of the first two terms are zero. Thus in this case, the first two terms have to both take zero values. If $\mathbf{k}=\mathbf{k}'$, the second term is zero since in this case $\omega=\omega'$ and thus the factor $\sin(\omega-\omega')t=0$. So only the first term survives and gives the expectation value of $\Omega^2(t, \mathbf{0})$:
\begin{equation}
\Omega_0^2=\left\langle \Omega^2\right\rangle=\frac{8\pi G}{3}\int_{\omega\leq\Lambda}\frac{d^3k}{(2\pi)^3}\frac{\omega}{2}=\frac{G\Lambda^4}{6\pi}.
\end{equation}

For the operator coefficients of the last two terms in the expression \eqref{phi dot square expansion}, the vacuum state $|0\rangle$ is not an eigenstate. So the last two terms are constantly fluctuating, and the time varying of $\Omega^2$ comes from these two terms.

After some algebraic manipulations, \eqref{phi dot square expansion} can be rewritten as the form of \eqref{big Omega} for the vacuum state $|0\rangle$, where
\begin{equation}\label{f gamma}
f(\gamma)d\gamma=\frac{16\pi^2}{\Lambda^4}\int_{\gamma\leq\omega+\omega'\leq\gamma+d\gamma}\frac{d^3kd^3k'}{(2\pi)^3}\frac{\sqrt{\omega\omega'}}{2}\left(-a_{\mathbf{k}}a_{\mathbf{k}'}-a_{\mathbf{k}}^{\dag}a_{\mathbf{k}'}^{\dag}\right),
\end{equation}
\begin{equation}\label{g gamma}
g(\gamma)d\gamma=\frac{16\pi^2}{\Lambda^4}\int_{\gamma\leq\omega+\omega'\leq\gamma+d\gamma}\frac{d^3kd^3k'}{(2\pi)^3}\frac{\sqrt{\omega\omega'}}{2}i\left(a_{\mathbf{k}}a_{\mathbf{k}'}-a_{\mathbf{k}}^{\dag}a_{\mathbf{k}'}^{\dag}\right).
\end{equation}

Evaluating the above integrals gives the expectation values
\begin{equation}
\left\langle f(\gamma)d\gamma\right\rangle=\left\langle g(\gamma)d\gamma\right\rangle=0,
\end{equation}
and their fluctuations
\begin{widetext}
\begin{equation}\label{power spectrum calculation}
 \left\langle \left(f(\gamma)d\gamma\right)^2\right\rangle = \left\langle \left(g(\gamma)d\gamma\right)^2\right\rangle=
  \begin{cases} 
   \frac{4}{35}\left(\frac{\gamma}{\Lambda}\right)^7\frac{d\gamma}{2\Lambda}, & \text{if }   0\leq\gamma \geq \Lambda, \\
   -\frac{4}{35}\left(40-140\frac{\gamma}{\Lambda}+168\left(\frac{\gamma}{\Lambda}\right)^2-70\left(\frac{\gamma}{\Lambda}\right)^3+\left(\frac{\gamma}{\Lambda}\right)^7\right) \frac{d\gamma}{2\Lambda},      & \text{if }   \Lambda\leq\gamma \geq 2\Lambda.
  \end{cases}
\end{equation}
\end{widetext}
The above expression \eqref{power spectrum calculation} gives the power spectrum density of the varying part of $\Omega^2(t, \mathbf{0})$ (except the constant $\Omega_0^2$ part), which has been plotted in FIG. \ref{power spectrum}.

\section{Wigner-Weyl Description of Quantum Mechanics and Numeric simulations}\label{appendix 2}
This chapter explain the principle of the numeric calculations in the main text. Same as the numeric part in the  main text, we set $G=1$ in this section. 
Wigner functions and Weyl transforms of operators offer a formulation of quantum mechanics that is equivalent to the standard approach given by the Schr\"{o}dinger equation. The Wigner distribution function is a quasi distribution function in the phase space. For a  particular quantum wave function  $\psi(x)$, its Wigner function is defined as
\begin{equation}
W(x,p)=\int d y e^{-i p y} \psi(x+\frac{y}{2}) \psi^{*}(x-\frac{y}{2})
\end{equation} 
The Weyl transform of an quantum operator $\hat{A}$ is defined as 
\begin{equation}
A(x,p)=\int   d y e^{-i p y} \langle x+\frac{y}{2}|\hat{A}|x-\frac{y}{2} \rangle
\end{equation}
Then the expectation value of the operator $\hat{A}$ under the state $\psi(x)$ can be written as 
\begin{equation}
\langle \hat{A}\rangle=\int \int dx dp W(x,p )A(x,p) 
\end{equation}
These two transformations give the Wigner-Weyl discription for quantum mechanics. The expectation values of
physical quantities are obtained by averaging their Weyl transforms over phase space.

For a harmonic oscillator with frequency $\omega$ and $m=1$, the ground state Wigner function is a Gaussian distribution function for both $x$ and $p$
\begin{equation}
W_0(x, p)=\frac{1}{\pi} e^{-\frac{p^2}{\omega}-x^2\omega}
\end{equation}
We can easily check that the Weyl transform of an operator $H(\hat{x})$ ( or $H(\hat{p})$) is simply  replaced the operator $\hat{x}$ by $x$ (or $\hat{p}$ by $p$). Other than that, another particular transform we are going to use in this write-up is 
\begin{equation}
\hat{x}\hat{p}\rightarrow x p+\frac{i}{2};\quad \hat{p}\hat{x}\rightarrow x p-\frac{i}{2}
\end{equation}
We can see that the transform of the  product does not necessarily equal to the product of transforms. In the following part we are going to get the general expression for the transform of the product. 

Before that we notice that Weyl transform  can be used to construct the original operator , i.e. 
\begin{equation}\label{eq:harmonicgs}
\langle x |\hat{A}| y \rangle =\frac{1}{2\pi}\int dp A(\frac{x+y}{2}, p) e^{i p (x-y)}
\end{equation}
Using this formula we can construct the transform of  product of two states: 
\begin{equation}
\begin{split}
&\int dy \langle x+\frac{y}{2} |\hat{A}\hat{B}|x-\frac{y}{2}\rangle e^{-ipy} \\
&=\int d z  dy \langle x+\frac{y}{2} |\hat{A}|z\rangle \langle z|\hat{B}|x-\frac{y}{2}\rangle e^{-ipy} \\
&=\frac{1}{4\pi^2}\int d z dy dp_1 dp_2 e^{i p_1(x+\frac{y}{2}-z)}e^{-i p_2(x-\frac{y}{2}-z)} e^{-ipy}\\
&\quad\quad \cdot A(\frac{ x+y/2+z}{2}, p_1) B(\frac{ x-y/2+z}{2}, p_2)\\
&=\frac{1}{4\pi^2} \int dz_1 dz_2 dp_1 dp_2 e^{i z_1 (p_2-p)} e ^{i z_2 (p-p_1)}\\
&\quad\quad \cdot A(x+\frac{z_1}{2}, p_1) B(x+\frac{z_2}{2}, p_2)\\
\end{split}
\end{equation}
Here we define 
\begin{eqnarray}
z_1 &=&\frac{y}{2}+z-x\\
z_2 &=&-\frac{y}{2}+z-x
\end{eqnarray}
We Taylor-expand $ A(x+\frac{z_1}{2}, p_1) $ and $B(x+\frac{z_2}{2}, p_2)$ around $x$ and have 
\begin{eqnarray}
A(x+\frac{z_1}{2}, p_1) =\sum_{n=0}^{\infty} \frac{1}{n!}A^{(n)}(x,p_1) (z_1/2)^n\\
B(x+\frac{z_2}{2}, p_2) =\sum_{n=0}^{\infty} \frac{1}{n!} B^{(n)}(x,p_2) (z_2/2)^n
\end{eqnarray}
and use the facts 
\begin{equation}
\frac{1}{2\pi}\int dx x^n e^{i x y}=  (-i)^{n}\delta^{(n)}(y)
\end{equation}
and
\begin{equation}
\int dy \delta^{(n)}(y) f(y)=(-1)^{n} f^{(n)}(0)
\end{equation}
Therefore, we can write the Weyl transform of operator $\hat{A}\hat{B}$ as
\begin{equation}\label{eq:prod}
\sum_{n,m} \frac{i^{n}(-i)^{m}}{2^{n+m}n!m!}A^{(n,m)}(x,p) B^{(m,n)}(x,p)
\end{equation}
The generalized FRW scale factor $a$ satisfies the equation
\begin{equation}
\ddot{a}+\Omega^2(t) a=0
\end{equation}
in which 
\begin{equation}
\Omega^2(t)=\frac{8\pi}{3} \dot{\phi}^{2}(t)
\end{equation}
Now we replace all the quantities by operators, assuming that operators still satisfy the previous equation
\begin{equation}
\ddot{\hat{a}}+\hat{\Omega}(t)^2 \hat{a}=0
\end{equation}
with 
\begin{equation}
\hat{\Omega}^2(t)=\frac{8\pi}{3} \dot{\hat{\phi}}^{2}(t)
\end{equation}
For a massless real scalar field, we can write it as
\begin{equation}
\hat{\phi}=\int\frac{d^3k}{(2\pi)^{3/2}} (\hat{x}_k\cos(\omega_k t)+\frac{1}{\omega_k}\hat{p}_k\sin(\omega_k t))
\end{equation}
 in which
 \begin{eqnarray}
 \hat{x}_k &=&\sqrt{\frac{1}{2\omega_k}}(b_k^{\dag}+b_k)\\
 \hat{p}_k &=&i\sqrt{\frac{\omega_k}{2}}(b_k^{\dag}-b_k) 
\end{eqnarray}
 are the generalized $\hat{x}$ $\hat{p}$ operators for each field modes. 

We can write the Weyl transformation of the $\hat{\Omega}(t)$
 \begin{equation}
 \begin{split}
&\Omega(\{x_k\},\{p_k\},t)^2=\frac{8\pi}{3}\iint \frac{d^3k d^3k'}{(2\pi)^3}x_k x_{k'}\omega_k\omega_{k'}\sin\omega_k t \sin\omega_{k'}t\\
&+p_k p_{k'}\cos\omega_k t \cos\omega_{k'}t-2x_k p_{k'}\omega_k\sin\omega_k t \cos
\omega_{k'}t
\end{split} 
\end{equation}
This expression is quadratic in  $x_k$ and $p_k$, so if we apply it to \eqref{eq:prod}, only $m+n\leq 2$ terms survive. Assuming $a(\{x_k\},\{p_k\},t)$ is the Weyl transform of operator $\hat{a}$, we have the equation for $a$ as
\begin{widetext}
\begin{equation}\label{eq:a_full}
\ddot{a}+\Omega^2 a +\frac{i}{2}\sum_k \left(\frac{\partial \Omega^2}{\partial x_k}   \frac{\partial a }{\partial p_k} -\frac{\partial  \Omega^2}{\partial p_k}  \frac{\partial a}{\partial x_k} \right)-\frac{1}{8}\sum_{k, k'} \left(\frac{\partial^2 \Omega^2}{\partial x_k \partial x_{k'}}\frac{\partial^2 a}{\partial p_k \partial p_{k'}}+\frac{\partial^2 \Omega^2}{\partial p_k \partial p_{k'}}\frac{\partial^2 a}{\partial x_k \partial x_{k'}}-2\frac{\partial^2 \Omega^2}{\partial x_k \partial p_{k'}}\frac{\partial^2 a}{\partial p_k \partial x_{k'}}\right)=0
\end{equation}
\end{widetext}
The observed value $a$ is the average over Wigner function $W(\{x_k\},\{p_k\},t)$
\begin{equation}
a_{o}(t)=\int \left(\prod_k dx_k dp_k \right) a(\{x_k\},\{p_k\},t)W(\{x_k\},\{p_k\},t)
\end{equation} 
If the quantum field is in the ground state, then by \eqref{eq:harmonicgs}
\begin{equation}
W(\{x_k\},\{p_k\},t)=\prod_k  \frac{1}{\pi} e^{-\frac{p_k^2}{\omega_k}-x_k^2\omega_k}
\end{equation}

\paragraph{Local approximation}Generally the equation \eqref{eq:a_full} depends on  not only the value of $\Omega$ and $a$ on a particular phase space point $(x,p)$, but also on the neighboring values (i.e. derivatives).  If our solution $a$ is "smooth" enough in the phase space then we can neglect the last two derivative terms in the \eqref{eq:a_full}. It can be simplified to 
\begin{equation}
\ddot{a}+\Omega^2 a=0.
\end{equation}
Assuming the length of the Universe is $L$. We can  replace the integral by summations. For simplicity, we define
\begin{eqnarray}
\tilde{t}&\rightarrow & \frac{2\pi t}{L}\\
\tilde{x}_n &\rightarrow & \sqrt{2\omega_{\frac{2\pi n}{L}}} x_{\frac{2\pi n}{L}}\\
\tilde{p}_n &\rightarrow & \sqrt{\frac{2}{\omega_{\frac{2\pi n}{L}}}}p_{\frac{2\pi n}{L}}
\end{eqnarray}
The equation can be written as 
\begin{equation}\label{eq:a_o}
\ddot {a} +\frac{4}{3L^2} \Omega(\tilde{t})^2 a=0
\end{equation}
with 
\begin{equation}
\begin{split}
\Omega(\tilde{t})^2=&\sum_{\vec{n},\vec{n'}} \sqrt{n n'}\big(\tilde{x}_{\vec{n}} \tilde{x}_{\vec{n'}} \sin n \tilde{t}\sin n' \tilde{t}\\
&+\tilde{p}_{\vec{n}} \tilde{p}_{\vec{n'}} \cos n \tilde{t}\cos n' \tilde{t}\\
&-\tilde{x}_{\vec{n}} \tilde{p}_{\vec{n'}} \sin n \tilde{t}\cos n' \tilde{t})\big)\\
=&\left[ \sum_{\vec{n}}\sqrt{n}(\tilde{x}_{\vec{n}} \sin n \tilde{t}-\tilde{p}_{\vec{n}} \cos n \tilde{t}) \right]^2
\end{split}
\end{equation}
Here $\vec{n}=(n_1, n_2, n_3), n_{1,2,3}\in \mathbb{Z}$ and $n=|\vec{n}|$. $\{\tilde{x}_{\vec{n}}\}$ $\{\tilde{p}_{\vec{n}}\}$ are random Gaussian variables with unit standard deviation.  We can solve the equation for a randomly generated set of  $\{\tilde{x}_{\vec{n}}\}$ and  $\{\tilde{p}_{\vec{n}}\}$, and repeat. The result $a_o(t)$ is the average over  all solutions as long as our sample size is big enough.

\section{Fourier transforms of the coefficients in \eqref{quasiperiodic wave equation}}  \label{appendix 3}
In this appendix, we demonstrate the property of the spectrum of the coefficients in \eqref{quasiperiodic wave equation} given by \eqref{f1 center}, \eqref{f2 center} and \eqref{f3 center}. Observing that the $\cos 2\Theta$, $\sin2\Theta$ and $\tan\Theta$ in \eqref{f1 sim}, \eqref{f2 sim} and \eqref{f3 sim} respectively can all be decomposed as Fourier series sum of the form $e^{i2n\Theta}$, where $n=\pm1, \pm2, \cdots$, we only need to analyze the spectrum of $e^{i2n\Theta}$.

For simplicity, we only analyze the time component Fourier transform of $e^{i2n\Theta}$. The spatial part has similar property. The phase angle $\Theta$ is determined by $\Omega$ through \eqref{phase angle} while $\Omega$ is determined by \eqref{big Omega}. The power spectrum of $\Omega^2$ is given by \eqref{power spectrum calculation} (illustrated in FIG. \ref{power spectrum}).

Calculation of the Fourier transform of $e^{i2n\Theta}$ exactly based on \eqref{big Omega} is complicated. For simplicity, we assume that $\Omega$ taking the following simple form which is similar to \eqref{small omega}
\begin{equation}
\Omega=\Omega_0(1+h\cos\gamma t),
\end{equation}
where $\gamma$ take the peak value of the power spectrum \eqref{power spectrum calculation} which is around $\sim 1.7\Lambda$ (see FIG. \ref{power spectrum}) and $h<1$ to make sure that $\Omega>0$.

Then we have
\begin{equation}
\Theta=\Omega_0t+\frac{h\Omega_0}{\gamma}\sin\gamma t.
\end{equation}
Using the Jacobi-Anger expansion we have
\begin{equation}\label{Jacobi-Anger}
e^{i2n\Theta}=\sum_{m=-\infty}^{+\infty}J_m(\frac{2nh\Omega_0}{\gamma})e^{i(2n\Omega_0+m\gamma) t},
\end{equation}
where $J_m$ is the $m$th Bessel function of the first kind.

As $|m|\to\infty$, we have
\begin{equation}
|J_m(\frac{2nh\Omega_0}{\gamma})|\sim\frac{1}{m!}\left(\frac{nh\Omega_0}{\gamma}\right)^{|m|},
\end{equation}
which drops faster than the exponential. Therefore, the Fourier transform of $e^{i2n\Theta}$ is centered around $2n\Omega_0$.

To estimate the magnitude of the Fourier coefficients of $e^{i2n\Theta}$ around zero frequency, we evaluate the Bessel function for 
\begin{equation}
m\sim -2n\Omega_0/\gamma\sim\sqrt{G}\Lambda\to\infty.
\end{equation}
In this case, the zero component Fourier coefficient is asymptotic to (see \cite{watson1995treatise})
\begin{equation}\label{zero frequency}
|J_m(-hm)|\sim\frac{e^{-(\nu-\tanh\nu)|m|}}{\sqrt{2\pi|m|\tanh\nu}}\to 0,
\end{equation}
since $\nu$ is determined by $h=\operatorname{sech}\nu<1$ that we always have $\nu-\tanh\nu>0$.

When calculating the Fourier transform of $e^{i2n\Theta}$ exactly based on \eqref{big Omega}, the spectrum becomes continuous instead of discrete. But the distribution of the spectrum should be similar.

\bibliographystyle{unsrt}
\bibliography{how_vacuum_gravitates}

\end{document}